\def\prd{Phys. Rev. D}

\def\apj{Astrophys. J.}

\def\mnras{Mon. Not. R. Astr. Soc.}
\def\aap{Astr. Astrophys.}

\def\procspie{Proc. SPIE}
\def\araa{Annual Review of Astronomy and Astrophysics}
\def \<{\langle}
\def \>{\rangle}

\newcommand{\ra}{\;\raise1.0pt\hbox{$'$}\hskip-6pt\partial\;}
\newcommand{\lo}{\;\overline{\raise1.0pt\hbox{$'$}\hskip-6pt\partial}\;}

\newcommand{\degree}{^\circ}

\newcommand{\Abs}{\abstract}
\newcommand{\Ack}{\acknowledgments}
\newcommand{\mktt}{\maketitle}

\documentclass[a4paper,11pt]{article}
\usepackage{jcappub,graphicx,epsfig,natbib,color,times,bm,amsmath,multirow,hyperref}
\usepackage[ddmmyyyy,hhmmss]{datetime}

\begin{document}
\title{ E and B families of the Stokes parameters in the polarized synchrotron and thermal dust foregrounds}

\author[a,b]{Hao Liu,}\emailAdd{liuhao@nbi.dk}
\author[a]{James Creswell}\emailAdd{dgz764@alumni.ku.dk}
\author[a]{and Pavel Naselsky}\emailAdd{naselsky@nbi.dk}

\affiliation[a]{The Niels Bohr Institute \& Discovery Center, Blegdamsvej 17, DK-2100 Copenhagen, Denmark}
\affiliation[b]{Key laboratory of Particle and Astrophysics, Institute of High Energy Physics, CAS, 19B YuQuan Road, Beijing, China}

\Abs{

Better understanding of Galactic foregrounds is one of the main obstacles to
detection of primordial gravitational waves through measurement of the B mode
in the polarized microwave sky. We generalize the method proposed in
\cite{2011PhRvD..83h3003B} and decompose the polarization signals into the E
and B families directly in the domain of the Stokes $Q$, $U$ parameters as
$(Q,U)\equiv(Q_E, U_E)+(Q_B,U_B)$. This also enables an investigation of the
morphology and the frequency dependence of these two families, which has been
done in the WMAP K, Ka (tracing synchrotron emission) and Planck 2015 HFI maps
(tracing thermal dust).  The results reveal significant differences in spectra
between the E and B families. The spectral index of the E family fluctuates
less across the sky than that of the B family, and the same tendency occurs
for the polarization angles of the dust and synchrotron channels. The new
insight from WMAP and Planck data on the North Polar Spur and BICEP2 zones
through our method clearly indicates that these zones are characterized by
very low polarization intensity of the B family compared to the E family. We
have detected global structure of the B family polarization angles at high
Galactic latitudes which cannot be attributed to the cosmic microwave
background or instrumental noise. However, we cannot exclude instrumental
systematics as a partial contributor to these anomalies.

}

\mktt

\section{Introduction}\label{sec:intro}

The next generation of cosmic microwave background (CMB) experiments will
focus on detection of the B mode of polarization, which is a unique indicator
of the presence of cosmological gravitational waves in the universe
\citep{2012SPIE.8442E..19H, 2016arXiv161002743A, 2011arXiv1110.2101K,
quijote2012}. The theory of the generation of the CMB polarization predicts
that in the absence of cosmological gravitational waves, the B mode is equal
to zero except for the lensing effect. The gravitational waves in a
cosmological plasma are usually characterized by the so called
tensor-to-scalar ratio $r$, which is constrained to $r< 0.07$ at a $95\%$
confidential level in~\cite{2015PhRvL.114j1301B}. This constraint reflects the
present status of our knowledge of Galactic and extra-galactic foregrounds,
including synchrotron radiation, thermal dust emission (TDE), free--free
radiation, anomalous dust emission (AME), as well as contamination from
systematic effects. Bearing in mind the target of $r\sim 10^{-4}$ to
$10^{-3}$, we should understand the properties of the strongly polarized
foregrounds (synchrotron and TDE) at least 2--3 orders of magnitude better
than now. Besides, we should not forget the potentially weak, but not
negligible polarization of the free--free and AME components, which are poorly
understood.

Better understanding of the polarized sky emission requires more detailed
analysis of the morphology of the foregrounds, including local and global
features of the synchrotron and TDE signals. Efficient removal of the
foregrounds after cleaning depends on the method applied and is limited not
only by the absolute value of polarized amplitudes, but also by the variation
of spectral indices. This problem is closely related to the proper usage of
sky masks, which are implemented in WMAP~\cite{WMAPdata:online} and
Planck~\cite{Planckdata:online} data analysis. These masks are designed to
remove the brightest zones of the polarization intensity from the analysis, in
order to minimize the leakage from the foregrounds to the derived polarized
CMB.

Implementation of the E--B decomposition for the pure cosmological CMB signal
is practically useful, since in absence of primordial gravitational waves the
B mode vanishes. However, in reality the polarized CMB sky consists of a mix
of the primordial CMB, the Galactic and extra-galactic foregrounds, the E mode
lensing, and residuals of systematic effects (uncertainties of the antenna
beam shape, possible bandpass leakage from temperature anisotropy and the
foregrounds, etc.). This is why \emph{a priori} it is not clear what kind of
representation (Q--U or E--B) will be most suitable for cleaning of the data
for different frequency bands in order to extract the cosmological signal.

In this paper we would like to highlight a novel method based on decomposition
of the Stokes parameters $Q$ and $U$ into the $(Q_E, U_E)$ and $(Q_B,U_B)$
families, associated with the E and B modes respectively,\footnote{We use
``family'' for the $(Q_E, U_E)$ and $(Q_B, U_B)$ vector maps to distinguish
them from the scalar E or B components of polarization.} which satisfy
$(Q,U)=(Q_E, U_E)+(Q_B,U_B)$. Based on this representation we will
investigate the polarization intensity maps for the E family
and the B family,
\begin{equation}
    P_E=\sqrt{Q^2_E+U^2_E}, \quad P_B=\sqrt{Q^2_B+U^2_B},
\end{equation}
in combination with the corresponding polarization angles $\theta_E = 0.5
\arctan (U_E,Q_E)$ and $\theta_B = 0.5 \arctan (U_B,Q_B)$. We show that these
estimators reveal new properties of the polarized synchrotron emission and TDE
in the WMAP K and Ka maps and the Planck 217 and 353 GHz
maps.\footnote{Obviously, 217 and 353 GHz are not the best tracers of TDE. It
would be much better to use 353 and 545 GHz maps, or even include 857 GHz.
However, only the 217 and 353 GHz maps are presented with polarization data in
the Planck public data release.}

We will show that  both the E and B family foregrounds have a relatively
stable structure of polarization angles across the K, Ka and 217, 353 GHz
frequencies, as well as confirm their existence against random noise with high
significance. For the whole sky the global structure of the synchrotron and
TDE polarization angles, $\theta^{\mathrm{sync}}_E$ and
$\theta^{\mathrm{dust}}_E$, reveals the existence of large angular structure,
which correlates with the Galactic and interstellar magnetic field.

Namely, we introduce the ratio $\rho(\textbf n)=P_E(\textbf n)/P_B(\textbf n)$
in each sky direction $\textbf n$ and look at the distribution functions
$F(\rho)$ for the whole sky in the WMAP K, Ka and Planck 30, 44 GHz maps. We
find remarkable similarity of $F(\rho)$ for the WMAP and Planck polarization
maps when $\rho\gg 1$, which allows us to determine the specific zones of the
sky where $P_E(\textbf n)\gg P_B(\textbf n)$. These zones are localized along
the northern part of the Loop I, and include the BICEP2 zone.

We will investigate the frequency dependence of the E and B families via their
spectral indices and identify some distinct patterns of variation of $\beta$,
$\beta_E$, and $\beta_B$ corresponding to the total intensity $P$, $P_E$, and
$P_B$. The $\beta_E$ index is significantly less varied across the full sky,
and especially in the NPS and BICEP2 zones. For the B family the variations of
the spectral index in the same angular domains are much more pronounced. This
makes the removal of the E family foregrounds (either synchrotron or dust)
more stable, unlike the B family foregrounds. This is also potentially
important for investigation of the E to B excess
problem~\cite{2016A&A...586A.133P, 2017ApJ...839...91C}.

Our method is sensitive enough for detection of the systematic anomalies of
the Planck 30, 44 GHz maps compared to the low frequency WMAP K, Ka maps.
These anomalies manifest themselves as a systematic shift of the distribution
function $F(\rho)$ for $\rho\ge 2$ in both the 30 and 44 GHz maps related to
the bandpass leakage correction of the Planck polarization.

In this work, we use $N_{\mathrm{side}}=128$ resolution and 2$\degree$
smoothing by default. The outline of our paper is the following: In
Section~\ref{sec:EB decompose} we will define the E and B-families of the
Stokes parameters as a result of linear convolution of $Q$ and $U$ through the
projection operator $W(\textbf n,\textbf n')$ and present the relationship
$Q_E,U_E\rightarrow E$, $Q_B,U_B\rightarrow B$. Section~\ref{sec:EB families
in foreground} is devoted to investigation of the morphology of the E and B
families of the synchrotron and the thermal dust foregrounds. We discuss the
angular distribution of the polarization angles and the corresponding
intensity across the sky, and identify the peculiar zones with low ratio
$\rho$. In Section~\ref{sec:EB ratio} we analyzed the frequency dependency of
the E and B family foreground and show that E family is characterized by
lesser variation of the spectral indexes in respect to the B family, and in
Section~\ref{sec:EB-beta} the frequency dependency is represented in form of
spectrum index, and we focus on two particular zones of the sky, related to
North Polar Spur and BICEP2 zone and derive the angular distribution and
variations of spectral indexes in them. We summarized our results in
Section~\ref{sec:discussion}.

\section{The E and B families of the Q and U Stokes parameters}\label{sec:EB decompose}

\subsection{Stokes parameters and basic definitions}

\begin{figure}[!hbtp]
 \centering
 \includegraphics[width=0.8\textwidth]{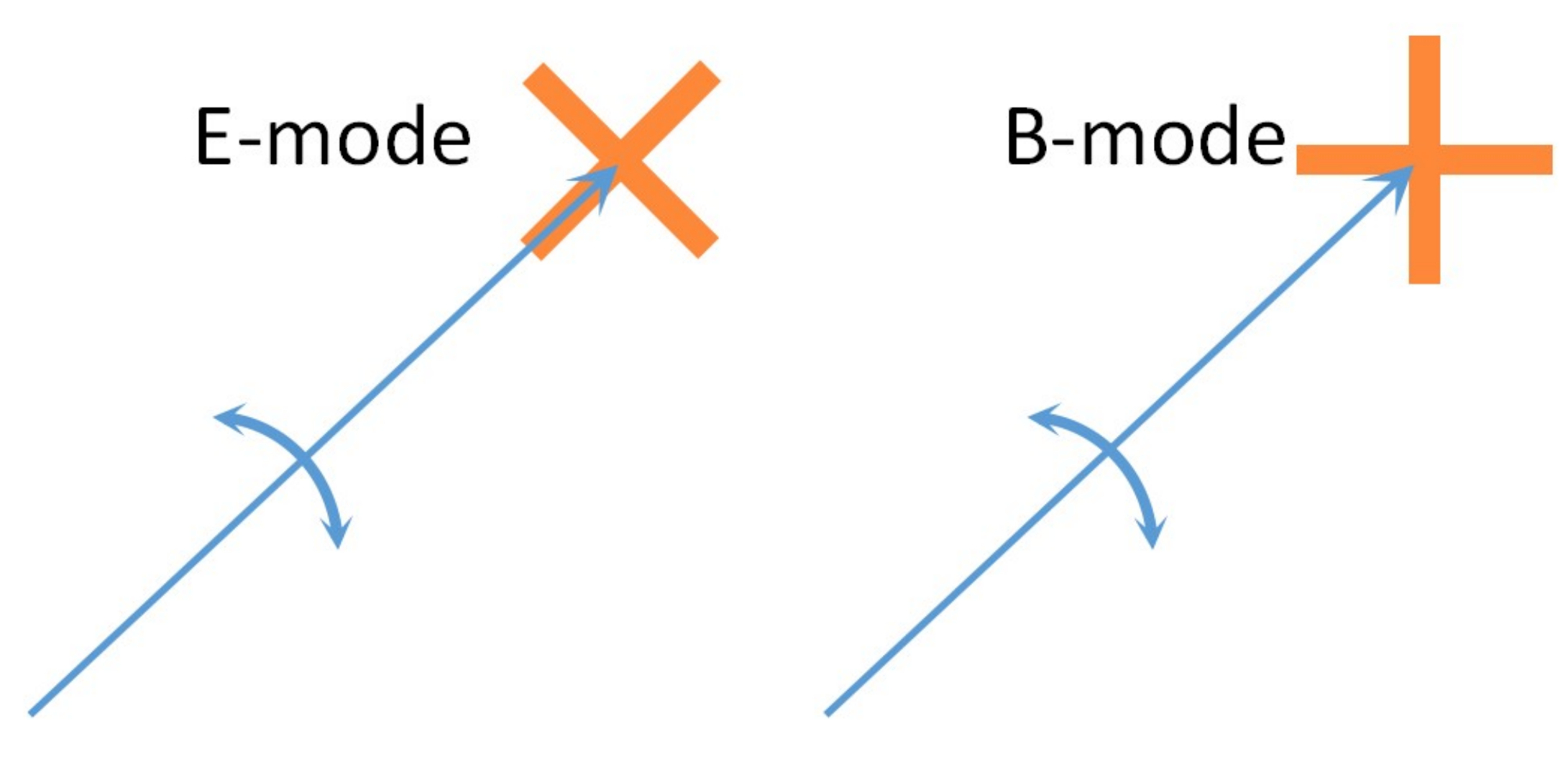}
 \caption{ An illustration of the E and B modes. For each mode, only two
  polarization directions (arms of the cross) are allowed in respect to the
  normal vector (arrow). Only one radial direction is plotted for example, and
  a complete E or B mode is made up of all its rotations. Therefore both E and
  B modes are rotationally invariant by design.}
 \label{fig:EB in real space}
\end{figure}

The state of polarization is described by the Stokes parameters $Q$ and $U$.
Since Thompson scattering does not generate circular polarization, $Q$ and $U$
are sufficient to describe the CMB polarization~\cite{2003moco.book.....D}. In
a sky observation, the $Q$ and $U$ Stokes parameters are separated by a
relative and constant $45\degree$ rotation around the line of sight (LOS), but
the reference frame is allowed to rotate freely around the LOS. However, if
one chooses to bind the reference frame with a family of normal vectors of a
given rotation, then $Q$ and $U$ are directly related to the E and B modes, as
illustrated in figure~\ref{fig:EB in real space} in the pixel domain.

Rotation of $Q$ and $U$ by an angle $\psi$ on the plane
perpendicular to the direction of $\mathbf{\hat n}$ is given
by~\cite{PhysRevD.55.1830, 0004-637X-503-1-1}
\begin{equation} \label{Q'U'}
(Q\pm i U)'(\mathbf{\hat{n}}) = e^{\mp 2i\psi}(Q \pm i U)(\mathbf{\hat{n}}).
\end{equation}
The separation into E and B modes can be derived from the Stokes
parameters~\cite{1997PhRvD..55.7368K, 1997PhRvL..78.2058K,
2016ARA&A..54..227K, 2010A&A...519A.104K}. Below we briefly review the
standard approach.

In the context of rotation, $Q$ and $U$ can be decomposed into spin $\pm2$
spherical harmonics~\cite{PhysRevD.55.1830, 0004-637X-503-1-1} as follows:
\begin{equation} \label{Q_lm+iU_lm}
    Q(\mathbf{\hat{n}})\pm i U(\mathbf{\hat{n}}) = \sum_{l,m} a_{\pm2,lm}\;{}_{\pm2}Y_{lm}(\mathbf{\hat n}),
\end{equation}
where $_{\pm2}Y_{lm}(\mathbf{\hat{n}})$ are the spin $\pm2$ spherical
harmonics, and the coefficients $a_{\pm2,lm}$ are given by
\begin{equation} \label{a2lm}
    a_{\pm2,lm}=\int \left(Q(\mathbf{\hat{n}})\pm i U(\mathbf{\hat{n}})\right) \,{}_{\pm2}Y^*_{lm}(\mathbf{\hat{n}})\,\mathrm d \mathbf{\hat{n}}.
\end{equation}
The E and B modes in harmonic space are then formed by
\begin{align} \label{equ:alm-eb} a_{E,lm} &= -(a_{2,lm} + a_{-2,lm})/2,
    \nonumber \\ a_{B,lm} &= i(a_{2,lm} - a_{-2,lm})/2,
\end{align}
and the real space representations of the scalar E and B modes are
\begin{align} \label{equ:EB-original}
    E(\mathbf {\hat n}) &= \sum\sqrt{\frac{(l+2)!}{(l-2)!}} a_{E,lm}\,Y_{lm}(\mathbf {\hat n}), \nonumber \\
    B(\mathbf {\hat n}) &= \sum\sqrt{\frac{(l+2)!}{(l-2)!}} a_{B,lm}\,Y_{lm}(\mathbf {\hat n}).
\end{align}

As we have pointed out in section \ref{sec:intro}, it is not clear \emph{a
priori} what kind of representation of the polarized signal ($Q, U$ or $E, B$)
will be most suitable for cleaning of the data in different frequency bands in
order to extract the cosmological component. The properties of the E and B
modes derived from the following decomposition can help to improve cleaning
using multi-frequency methods.

\subsection{Polarization intensity of the E and B modes}

For analysis of the polarized foregrounds, it is particularly important to
study the polarization intensity
\begin{equation} \label{equ:p}
    P = \sqrt{Q^2+U^2} 
\end{equation}
(for convenience, $\mathbf{\hat{n}}$ is suppressed from now on). Consequently,
it is also important to study the polarization intensity due to the E and B
modes respectively. However, the EB decomposition provided in
eq.~(\ref{equ:EB-original}) can not be directly transformed to a polarization
intensity, thus an alternative way is introduced as follows.

Eq.~(\ref{Q_lm+iU_lm}) can be solved for $Q$ and $U$, giving
\begin{align} \label{equ:qu from spin alm}
    Q &= \frac{1}{2} \left( \sum_{l,m} a_{2,lm}\;{}_{2}Y_{lm} + \sum_{l,m} a_{-2,lm}\;{}_{-2}Y_{lm} \right), \nonumber \\
    U &= \frac{1}{2i}\left( \sum_{l,m} a_{2,lm}\;{}_{2}Y_{lm} - \sum_{l,m} a_{-2,lm}\;{}_{-2}Y_{lm} \right).
\end{align}
In the particular case when $a_{B,lm}=0$ or $a_{E,lm}=0$, one gets the
contribution to $Q$ and $U$ from only the E or B mode respectively.
According to eq.~(\ref{equ:alm-eb}), when $a_{B,lm}=0$, one gets
\begin{equation} \label{equ:b is zero}
    -a_{E,lm} = a_{2,lm} = a_{-2,lm},
\end{equation}
and when $a_{E,lm}=0$, one gets
\begin{equation} \label{equ:e is zero}
    -ia_{B,lm} = a_{2,lm} = -a_{-2,lm}.
\end{equation}
We define $F_{+,lm}$ and $F_{-,lm}$ as
\begin{equation} \label{equ:define F}
    F_{+,lm} = -\frac{1}{2} \left({}_{2}Y_{lm} + {}_{-2}Y_{lm} \right), \quad F_{-,lm} = -\frac{1}{2i} \left({}_{2}Y_{lm} - {}_{-2}Y_{lm} \right).
\end{equation}
Then, by setting $a_{B,lm}=0$ and combining eqs.~(\ref{equ:qu from spin alm})
and (\ref{equ:b is zero}), one gets the $Q$ and $U$ parameters due to the E
mode:
\begin{equation} \label{equ: qe ue}
    Q_E = \sum_{l,m} a_{E,lm} F_{+,lm}, \quad U_E = \sum_{l,m} a_{E,lm} F_{-,lm}.
\end{equation}
Similarly, the Stokes parameters that are only from the B mode are
\begin{equation} \label{equ: qb ub}
    -Q_B = \sum_{l,m} a_{B,lm} F_{-,lm}, \quad U_B = \sum_{l,m} a_{B,lm} F_{+,lm}.
\end{equation}
Since the E and B modes are components of linear decomposition of the
input map, the corresponding Stokes parameters satisfy
\begin{equation} \label{equ:Qe Qb Ue Ub} 
    Q = Q_E + Q_B, \quad U = U_E + U_B.
\end{equation}
Therefore, one can calculate the polarization intensity that is only from the
E or B mode:
\begin{equation} \label{equ:pe pb} 
    P_E = \sqrt{Q_E^2+U_E^2}, \quad P_B = \sqrt{Q_B^2+U_B^2}.
\end{equation}

The E and B decomposition for the $Q$ and $U$ Stokes parameters is linear
and orthogonal while the polarization intensity is quadratic and consequently
nonlinear. Combining eqs.~(\ref{equ:p}),~(\ref{equ:Qe Qb Ue Ub})
and~(\ref{equ:pe pb}), one gets
\begin{align}
    P^2 = Q^2 + U^2 
        = P_B^2 \left(1 + \rho^2 + 2 \rho \cos(2\theta_E - 2\theta_B) \right),
\label{215}
\end{align}
where $\rho = P_E/P_B$, and
\begin{equation} \label{angle}
    \theta_{E,B} = \frac{1}{2}\arctan(U_{E,B},Q_{E,B})
\end{equation}
are the polarization angles for the E and B families. Defining a quadratic
difference of the polarization intensity as
\begin{equation}
    \Delta = P^2-(P_E^2+P_B^2)=2(Q_E Q_B + U_E U_B),
\end{equation}
and substituting eqs.~(\ref{equ: qe ue}--\ref{equ: qb ub}), one gets:
\begin{equation}
    \Delta = 2\sum_{lml'm'} { a_{E,lm} a^*_{B,l'm'} G_{lml'm'}},
\end{equation}
where
\begin{equation}
    G_{lml'm'} = F_{-,lm}F^*_{+,l'm'}-F_{+,lm}F^*_{-,l'm'}.
\end{equation}
Therefore, $\Delta$ is completely determined by the cross quadratic term
$a_{E,lm} a^*_{B,l'm'}$ between the E and B modes.

\subsection{E/B family decomposition in the pixel domain}\label{sub:EB from real space}

The method presented above for decomposing $Q$ and $U$ into the E and B
families $Q_E,U_E$ and $Q_B,U_B$ is applicable for the full sky analysis.
However, for data sets with partial sky coverage or defects (stripes, missing
data in the pixels, and others), the E and B families can be defined locally
in the pixel domain through linear convolution of the Stokes parameters. In
this subsection we will derive the corresponding approach from the full sky
approach, using some general relations presented in the previous section.
Then, the final representation of the linear filter in the pixel domain can be
easily generalized for the incomplete sky model.

By combining equations~(\ref{a2lm})--(\ref{equ:alm-eb}) with eq.~(\ref{equ:define
F}), we get
\begin{align}
    a_{E,lm}&=-(a_{2,lm} + a_{-2,lm})/2 = \int \left(Q(\mathbf {\hat n}) F^*_{+,lm}(\mathbf {\hat n}) + U(\mathbf {\hat n}) F^*_{-,lm}(\mathbf {\hat n})\right) \,\mathrm d \mathbf {\hat n}, \nonumber \\ 
    a_{B,lm}&=i(a_{2,lm} - a_{-2,lm})/2 =\int \left(-Q(\mathbf {\hat n}) F^*_{-,lm}(\mathbf {\hat n}) + U(\mathbf {\hat n}) F^*_{+,lm}(\mathbf {\hat n})\right) \,\mathrm d \mathbf {\hat n}.
\end{align}
Then we substitute eqs.~(\ref{equ: qe ue})--(\ref{equ: qb ub}), giving
\begin{align} \label{equ:QU-EB in real space}
    Q_E(\mathbf{\hat n}) &= \int \left(Q(\mathbf {\hat n}') G_{1}(\mathbf{\hat n},\mathbf{\hat n}') + U(\mathbf {\hat n}') G_{2}(\mathbf{\hat n},\mathbf{\hat n}') \right) \,\mathrm d \mathbf {\hat n}', \nonumber \\
    U_E(\mathbf{\hat n}) &= \int \left(Q(\mathbf {\hat n}') G_{3}(\mathbf{\hat n},\mathbf{\hat n}') + U(\mathbf {\hat n}') G_{4}(\mathbf{\hat n},\mathbf{\hat n}') \right)  \,\mathrm d \mathbf {\hat n}', \nonumber \\
    Q_B(\mathbf{\hat n}) &= \int \left(Q(\mathbf {\hat n}') G_{4}(\mathbf{\hat n},\mathbf{\hat n}') - U(\mathbf {\hat n}') G_{3}(\mathbf{\hat n},\mathbf{\hat n}') \right)  \,\mathrm d \mathbf {\hat n}', \nonumber \\
    U_B(\mathbf{\hat n}) &= \int \left(-Q(\mathbf {\hat n}') G_{2}(\mathbf{\hat n},\mathbf{\hat n}') + U(\mathbf {\hat n}') G_{1}(\mathbf{\hat n},\mathbf{\hat n}') \right) \,\mathrm d \mathbf {\hat n}',
\end{align}
where we have defined
\begin{align}
    G_{1}(\mathbf{\hat n},\mathbf{\hat n}') &= \sum_{l,m} F_{+,lm}(\mathbf{\hat n})F^*_{+,lm}(\mathbf{\hat n}'), \quad
    G_{2}(\mathbf{\hat n},\mathbf{\hat n}') = \sum_{l,m} F_{+,lm}(\mathbf{\hat n})F^*_{-,lm}(\mathbf{\hat n}'), \nonumber\\
    G_{3}(\mathbf{\hat n},\mathbf{\hat n}') &= \sum_{l,m} F_{-,lm}(\mathbf{\hat n})F^*_{+,lm}(\mathbf{\hat n}'), \quad
    G_{4}(\mathbf{\hat n},\mathbf{\hat n}') = \sum_{l,m} F_{-,lm}(\mathbf{\hat n})F^*_{-,lm}(\mathbf{\hat n}'),
\end{align}
which are pure real-space functions, whose physical meaning is the combined
two-point correlations for all spin-2 harmonics. A more simple matrix form of
eq.~(\ref{equ:QU-EB in real space}) can then be given as:
\begin{align}\label{equ:EB-QU_final simple}
\left[\begin{matrix}
Q_E \\
U_E
\end{matrix}\right](\mathbf{\hat n})&= \int
\left[\begin{matrix}
G_1 , & +G_2 \\
+G_3 , & G_4
\end{matrix}
\right](\mathbf{\hat n},\mathbf{\hat n}')
\left[\begin{matrix}
Q \\
U
\end{matrix}\right](\mathbf{\hat n}')\,\mathrm d \mathbf {\hat n}' \\ \nonumber
\left[\begin{matrix}
Q_B \\
U_B
\end{matrix}\right](\mathbf{\hat n})&= \int
\left[\begin{matrix}
G_4 , & -G_3 \\
-G_2 , & G_1
\end{matrix}
\right](\mathbf{\hat n},\mathbf{\hat n}')
\left[\begin{matrix}
Q \\
U
\end{matrix}\right](\mathbf{\hat n}')\,\mathrm d \mathbf {\hat n}'
\end{align}

According to eqs.~(\ref{equ:QU-EB in real space}) and~(\ref{equ:EB-QU_final
simple}), the E/B family decomposition of the Stokes parameters can be done
purely in the pixel space over the area of the data available. This provides
an important direction for a pure real space determination of the E and B
modes from partial sky coverage. However, in the case of an incomplete sky,
there is inevitably the problem of E/B leakage, and the decomposition is
partially ambiguous. In order to alleviate E/B leakage, the window function
for the cut sky can be smoothed along its edge (see, for example,
\cite{2010A&A...519A.104K}).

\section{Morphology of the E and B foreground families}\label{sec:EB families in foreground}

\begin{figure*}[!htb]
  \centering
  \includegraphics[width=0.32\textwidth]{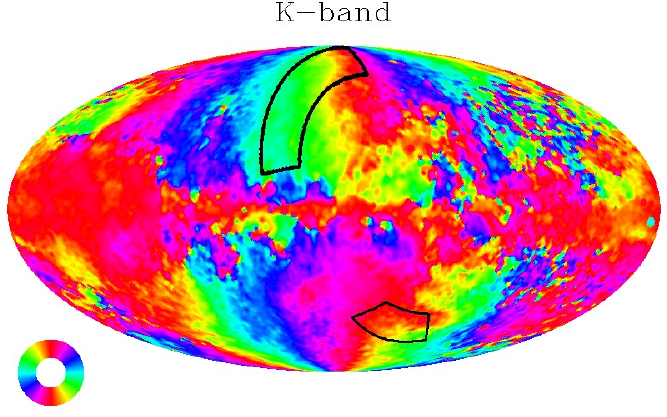}
  \includegraphics[width=0.32\textwidth]{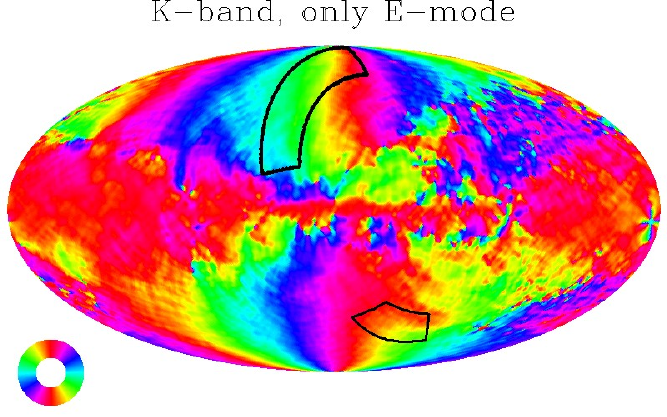}
  \includegraphics[width=0.32\textwidth]{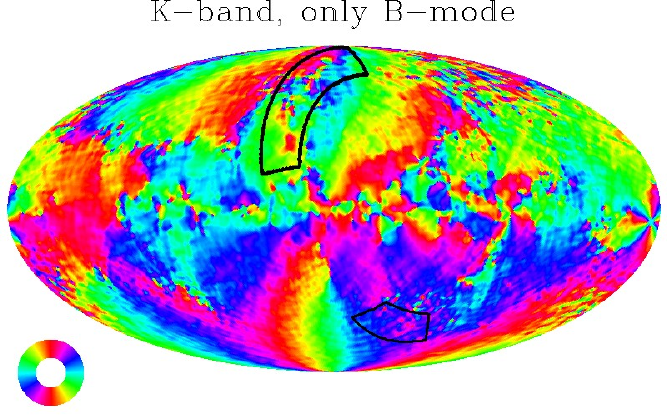} 
  
  \includegraphics[width=0.32\textwidth]{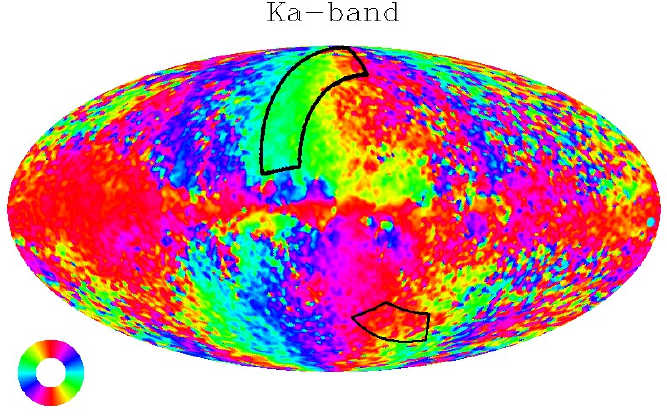}
  \includegraphics[width=0.32\textwidth]{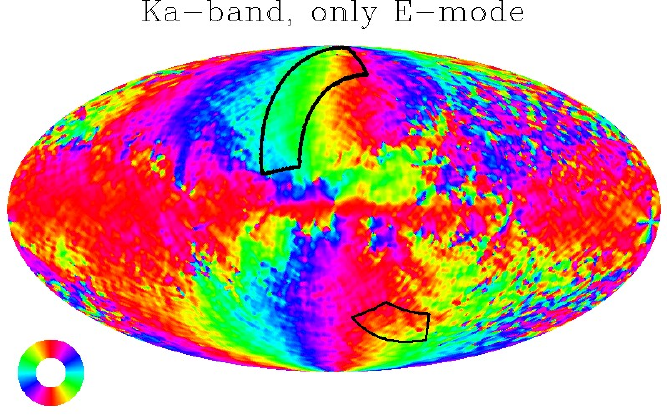}
  \includegraphics[width=0.32\textwidth]{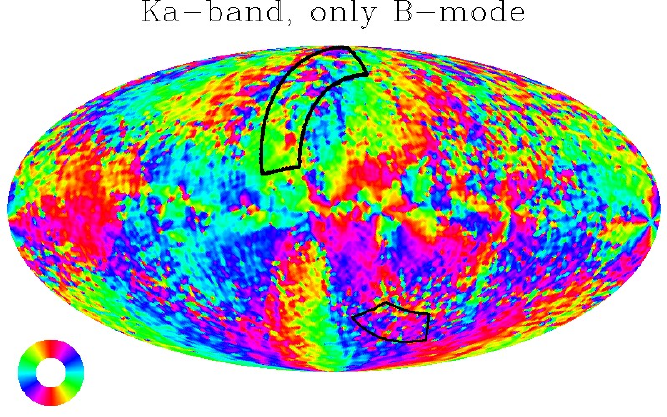} 

  \includegraphics[width=0.32\textwidth]{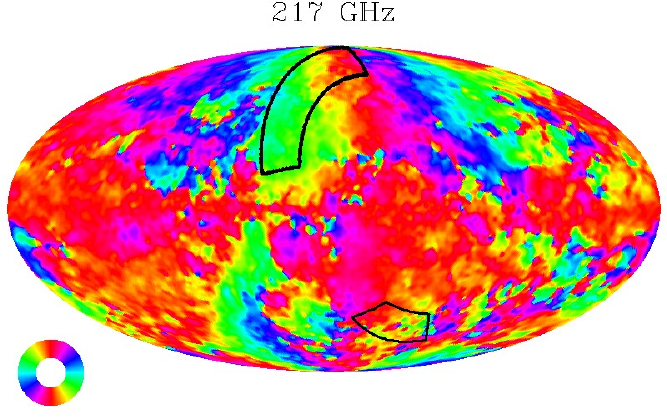} 
  \includegraphics[width=0.32\textwidth]{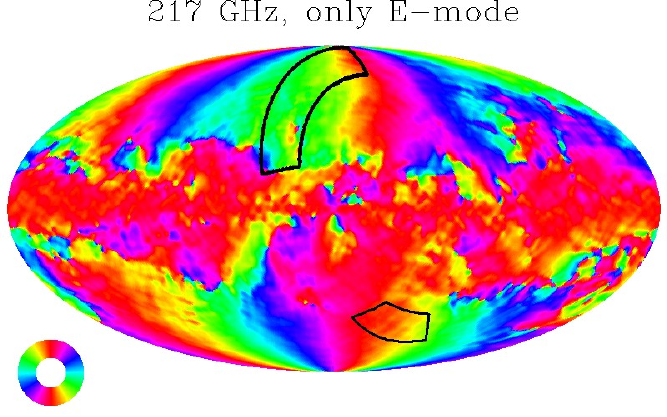}
  \includegraphics[width=0.32\textwidth]{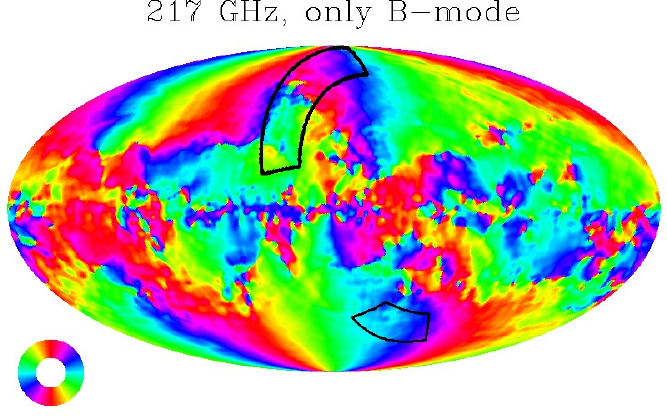}

  \includegraphics[width=0.32\textwidth]{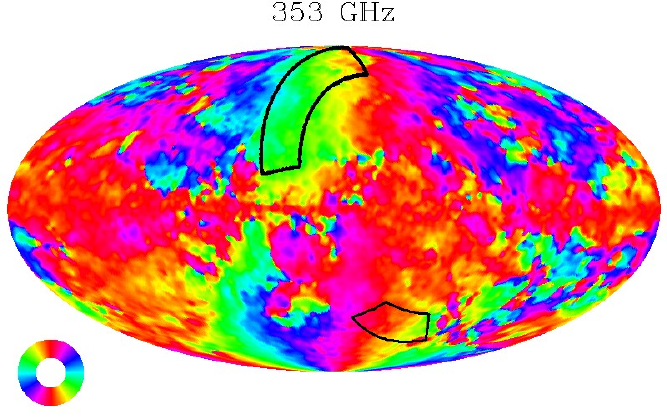} 
  \includegraphics[width=0.32\textwidth]{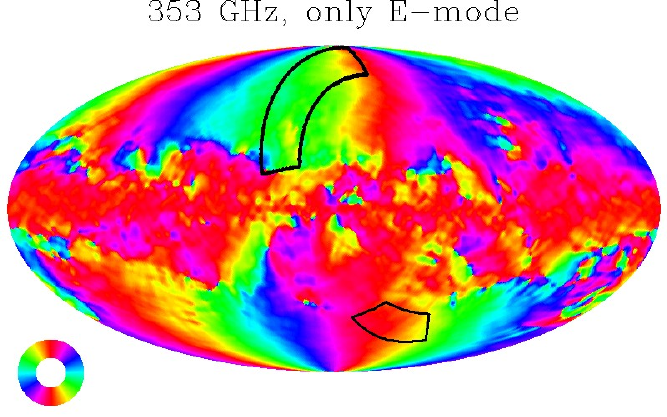}
  \includegraphics[width=0.32\textwidth]{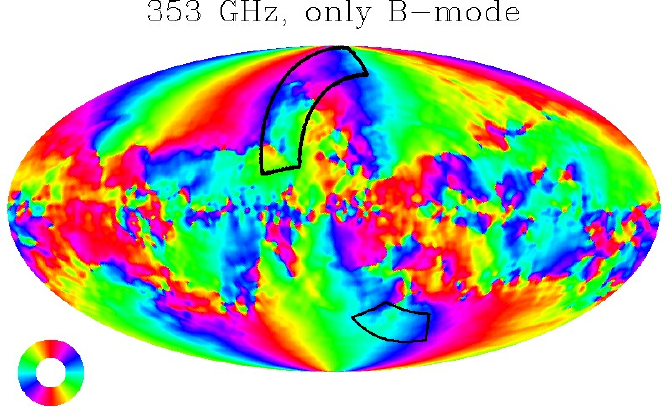}
  \caption{Maps of the polarization angles. From top to bottom: the WMAP
  K-band, Ka-band (synchrotron), and the Planck 217, 353 GHz (dust). From left
  to right: no E/B separation, the E family ($\theta_E$), and the B family
  ($\theta_B$). The color-to-angle mapping is given by the lower-left color
  disc. The black contour lines mark the North Polar Spur and the BICEP2
  zones.}
  \label{polangle}\label{plzones}
\end{figure*}

In this section we work with the polarization angles of the E and B families.
These polarization angles are plotted in figure~\ref{polangle}. They are shown
for the WMAP K and Ka bands (representing synchrotron emission) and the Planck
217 GHz and 353 GHz bands (representing TDE). For each band we calculate the
total polarization angle and also the polarization angles $\theta_E$ and
$\theta_B$ associated with only the E and B modes respectively (see
eq.~(\ref{angle})).

\subsection{Polarization angles of the E/B families: simple difference}

\begin{figure}[!hbtp]
 \centering
 \includegraphics[width=0.25\textwidth]{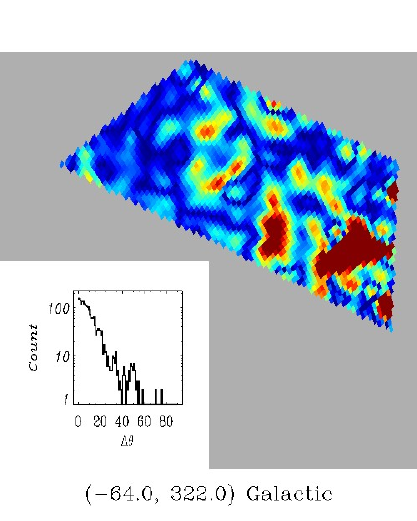}
 \includegraphics[width=0.25\textwidth]{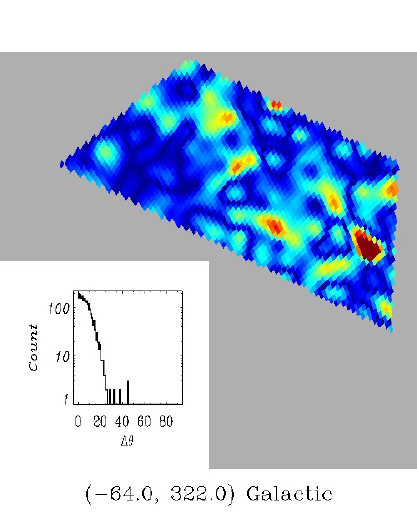}
 \includegraphics[width=0.25\textwidth]{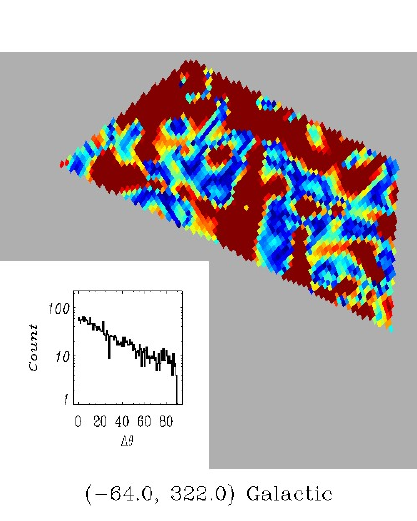}
 
 \includegraphics[width=0.25\textwidth]{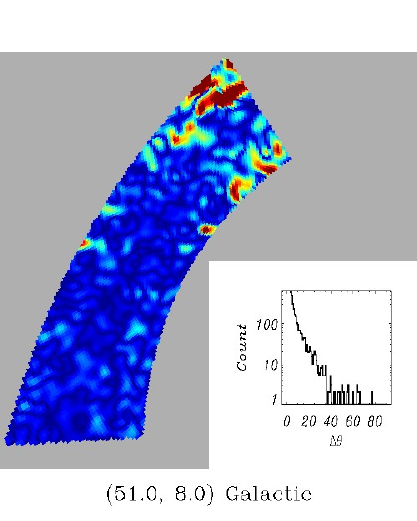}
 \includegraphics[width=0.25\textwidth]{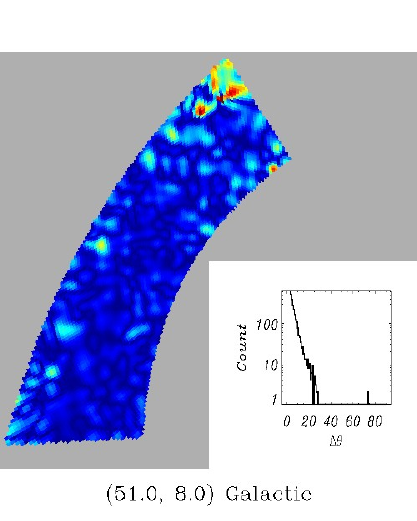}
 \includegraphics[width=0.25\textwidth]{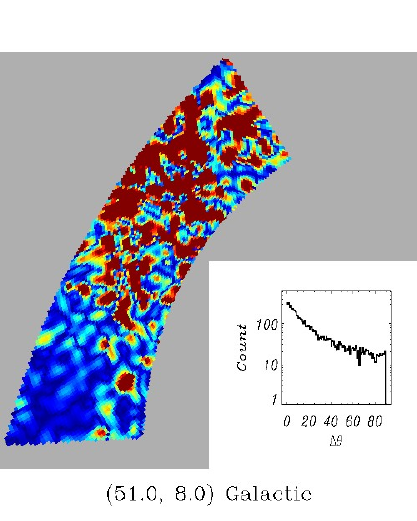}
 \caption{The differences of the polarization angles between the K and Ka
 bands for the BICEP2 zone (upper row) and the NPS zone (lower row). Note that
 the Ka band is apparently affected by systematics. From left to right: no
 separation for Q and U, only E family ($Q_E$ and $U_E$), and only B family
 ($Q_B$ and $U_B$). The inset histograms show the distributions of values for
 the pixels inside each zone.}
 \label{fig:cc_ang_diff_k_ka}
\end{figure}

\begin{figure}[!hbtp]
 \centering
 \includegraphics[width=0.25\textwidth]{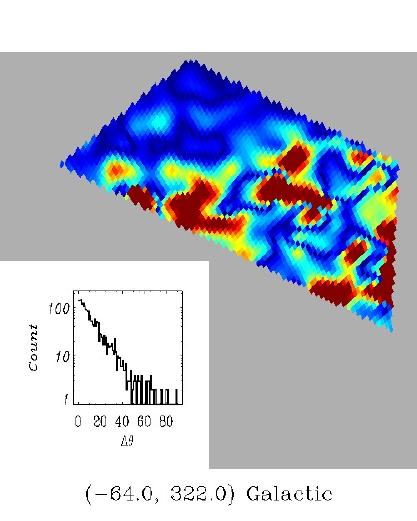}
 \includegraphics[width=0.25\textwidth]{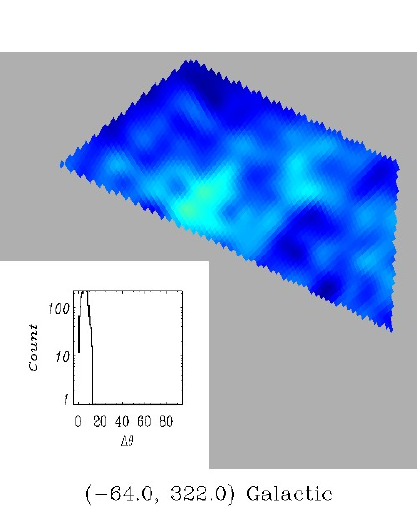}
 \includegraphics[width=0.25\textwidth]{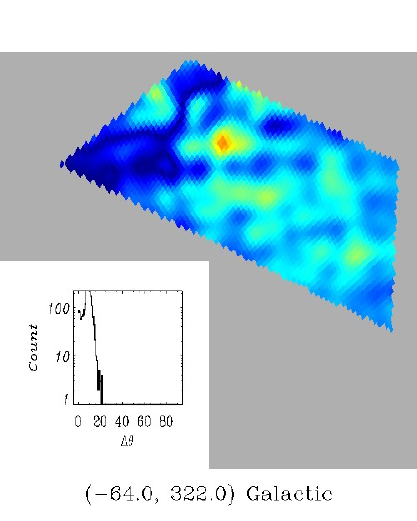}
 
 \includegraphics[width=0.25\textwidth]{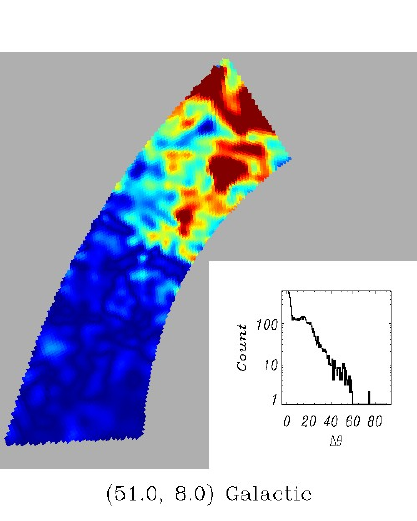}
 \includegraphics[width=0.25\textwidth]{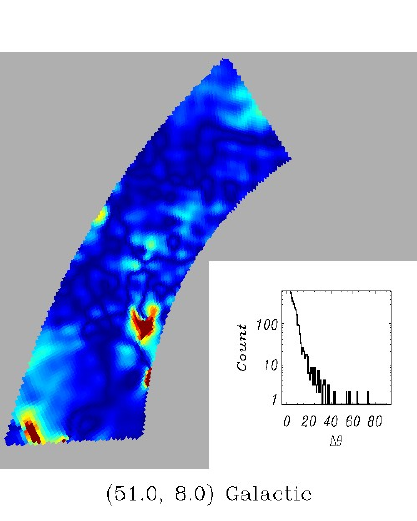}
 \includegraphics[width=0.25\textwidth]{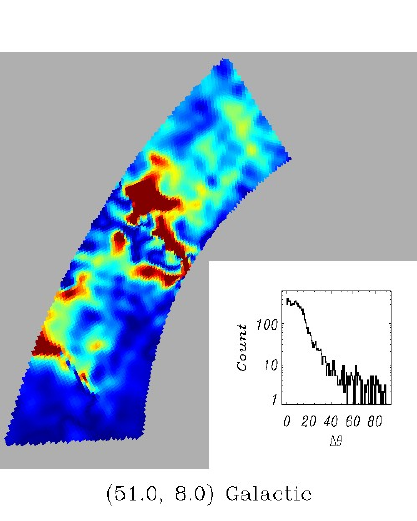}
 \caption{Similar to figure~\ref{fig:cc_ang_diff_k_ka} but for 217 and 353
 GHz. }
 \label{fig:cc_ang_diff_217_353}
\end{figure}

In order to characterize the residuals of the polarization angles between the
K, Ka and 217 GHz, 353 GHz bands we use the absolute polarization angle
difference, defined by
\begin{equation} \label{dif}
    \Delta\theta=|0.5\arctan[\sin(2\theta_1-2\theta_2),\cos(2\theta_1-2\theta_2)]|.
\end{equation}
The result of the calculation is presented in
figures~\ref{fig:cc_ang_diff_k_ka} and \ref{fig:cc_ang_diff_217_353} for K--Ka
and 217--353 GHz respectively. All the cases show remarkable similarity of the
polarization angles for the synchrotron and TDE bands (note that the upper
limit of the color map of the plots is only $30\degree$), which is consistent
with table~\ref{tab:cc ang} and figure~\ref{fig:hist_cc_ang}.

A flattening of the distribution of values, visible for example in the inset
histograms in the right panels of figure~\ref{fig:cc_ang_diff_k_ka}, indicates
the existence of higher amplitude residuals after subtraction of the
polarization angles.

From the upper panel of figure~\ref{fig:cc_ang_diff_217_353}, one can see that
for the 217 and 353 GHz bands, the polarization angles are highly consistent
for the E family and the B family individually, but less consistent for the
total angle without separation. This fact clearly indicates that the E and B
families have different spectral indices in this region.

\begin{figure}[!hbtp]
 \centering
 \includegraphics[width=0.32\textwidth]{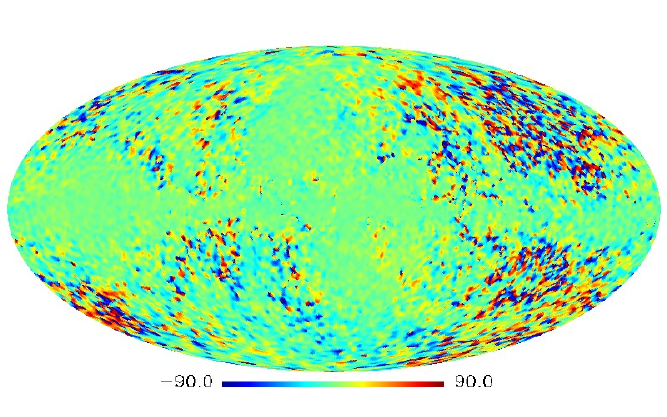}
 \includegraphics[width=0.32\textwidth]{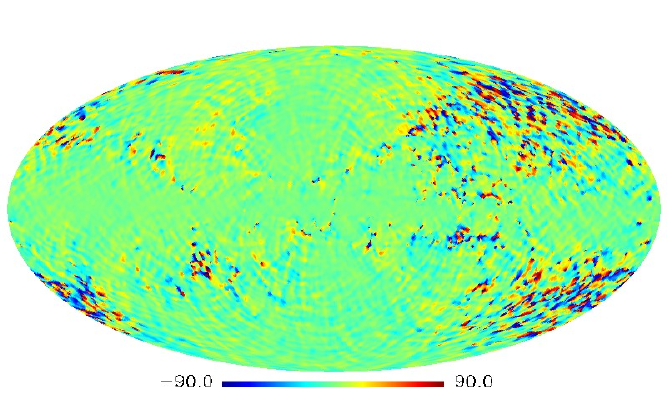}
 \includegraphics[width=0.32\textwidth]{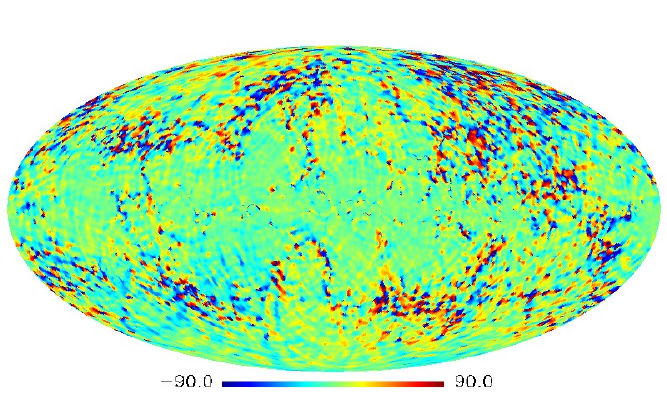}

 \includegraphics[width=0.32\textwidth]{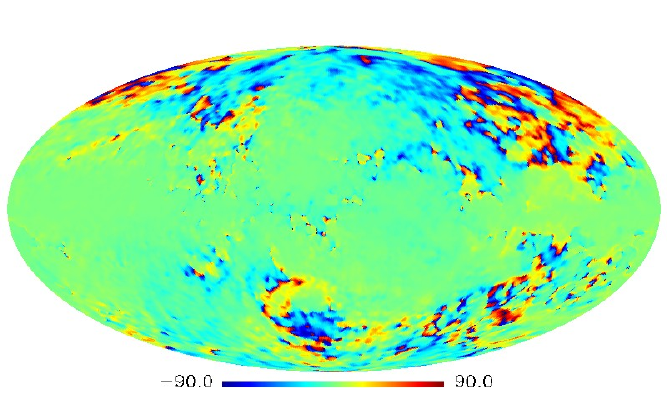}
 \includegraphics[width=0.32\textwidth]{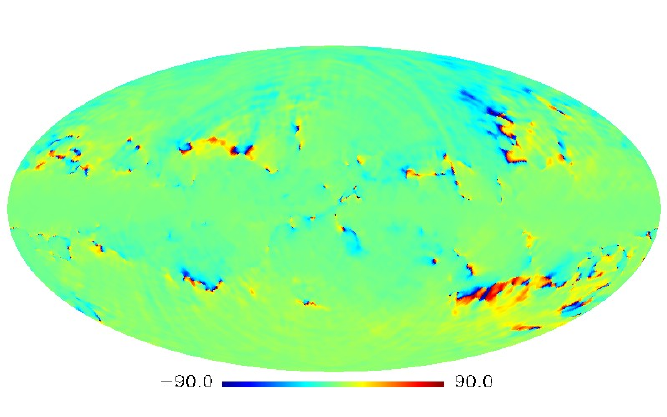}
 \includegraphics[width=0.32\textwidth]{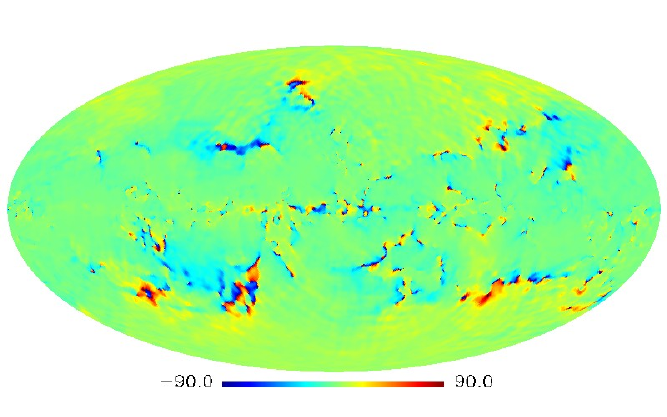}
 \caption{Simple differences of the polarization angles for K--Ka (upper) and
 353--217 GHz (lower). From left to right: No separation for Q and U, only
 E family ($\theta_E$), and only B family $\theta_B$. Apparently, in
 either case, the E mode gives higher band-to-band polarization angle
 consistency than the total polarization intensity.}
 \label{diff_217_353}
\end{figure}

Finally we would like to present the maps of differences for the polarization
angles $\theta_{\mathrm{dust}}=\theta_{353}-\theta_{217}$ and
$\theta_{\mathrm{sync}}=\theta_{\mathrm{K}}-\theta_{\mathrm{Ka}}$, both with
and without separation into the E and B families. From
figure~\ref{diff_217_353} we see the importance of the separation:
$\theta_{\mathrm{dust}}$ has very strong fluctuations in peripheral zones of
the full sky map outside the Loop I region, while for the E and B families
these differences reveal significantly lesser variations across the sky. This
result clearly illustrates the advantage of the method proposed over the
standard approaches. Due to its definition, the B family is acting as
``noise'' added to the E family, and in superposition it makes the
polarization angles $\theta_{353},\theta_{217}$ fluctate more than each
component. We have also confirmed that this result is not sensitive to the
removal of the best-estimated CMB signal.

\subsection{Polarization angles of the E/B families: angular correlation and significance}

The properties of the polarization angles can be understood if we consider eq.
(\ref{equ:Qe Qb Ue Ub}):
\begin{equation} \label{pan}
    \tan(2\theta)=\frac{P_E\sin(2\theta_E)+P_B\sin(2\theta_B)}{P_E\cos(2\theta_E)+P_B\cos(2\theta_B)}=\frac{\rho\sin(2\theta_E)+\sin(2\theta_B)}{\rho\cos(2\theta_E)+\cos(2\theta_B)}.
\end{equation}
Thus, if $\rho\gg 1$ (E family dominates), then $\theta\simeq \theta_E$, and
if $\rho\ll 1$ we have $\theta\simeq \theta_B$. In figure~\ref{polangle} we
can see strong similarity between $\theta$ and $\theta_E$, which indicates a
general dominance of the E family. For $\rho\gg 1$, the B family acts like a
perturbation of $\theta_E$:
\begin{equation} \label{pert}
    \theta \simeq \theta_E - \frac{1}{2\rho}\sin(2\theta_E-2\theta_B).
\end{equation}

In the domain of the North Polar Spur, the polarization angle of the total
signal is remarkably stable and homogeneous, and is dominated by the E family
of polarization. We will discuss this phenomenon in detail in a separate
paper.

In figure~\ref{plzones}, there is remarkable similarity between the two
synchrotron bands (K, Ka) and the two dust bands (217, 353 GHz) in the BICEP2
zone for $\theta_E$. For $\theta_B$ more differences can be seen in the same
zone. However, in both maps we can clearly see similarities in global
structure that cannot be reproduced by noise, which will be confirmed both by
Table~\ref{tab:cc ang} and in section~\ref{sub:test EB-family}. In the Ka
band, the structure consists of some stripe-like features across the Galactic
plane, which can be associated with the corresponding foregrounds and/or
residuals of systematic effects.

In order to characterize the similarity between these maps, we define the
cross-correlation coefficient between two polarization angle maps
$\theta_1(\mathbf{n})$ and $\theta_2(\mathbf{n})$ as follows:
\begin{equation}
    C(\theta_1, \theta_2) = \frac{1}{N_s} \sum_{i=0}^{N_s} \cos(2 \theta_1(\mathbf{n}_i) - 2 \theta_2(\mathbf{n}_i)),
\end{equation}
where $\theta_j(\mathbf{n}_i)$ is the polarization angle for $j$-th component
in the pixel $i$, $N_s$ is the size of the sample, and the factor 2 comes from
the definition of polarization angles. Therefore, $C(\theta_1,\theta_2)$ takes
values between $-1$ and $1$, and two $100\%$ correlated polarization angle
maps will give $C(\theta_1,\theta_2)=1$. The values of the cross-correlation
coefficient for different pairs of maps are shown in table~\ref{tab:cc ang},
for the full sky, BICEP2, and NPS regions.

\begin{table}[!htb]
 \caption{List of $C(\theta_1,\theta_2)$ for various 2-band combinations.
 Three regions are listed: full sky, Bicep2 and the NPS region.}
 \centering
 \begin{tabular}{|c|c|c|c|c|c|c|c|c|} \hline
       & K-Ka  & K-Ka (E)  & K-Ka (B) &K-217&K-217 (E)&K-217 (B)\\ \hline 
   Full sky & 0.73 &   0.85 &   0.67&0.50 &0.62 &0.37 \\ \hline
   NPS &   0.96 &   0.98 &   0.60 &0.89 & 0.81 &0.24\\ \hline
   BICEP2 &  0.87 &   0.95 &   0.45& 0.66 &0.93 &0.62  \\ \hline\hline
 &K-353&K-353 (E)&K-353 (B)&353-217&353-217 (E)&353-217 (B)   \\ \hline
   Full sky & 0.52 &   0.63 &   0.40& 0.79 & 0.94 &0.93 \\ \hline
  NPS  &   0.90 &   0.82 &   0.25& 0.88 &0.97 &0.87 \\ \hline
  BICEP2 &   0.85 &   0.88 &   0.42& 0.82 &0.97 &0.95 \\ \hline
 \end{tabular}
 \label{tab:cc ang}
\end{table}

\begin{figure}[!hbtp]
 \centering
 \includegraphics[width=0.48\textwidth]{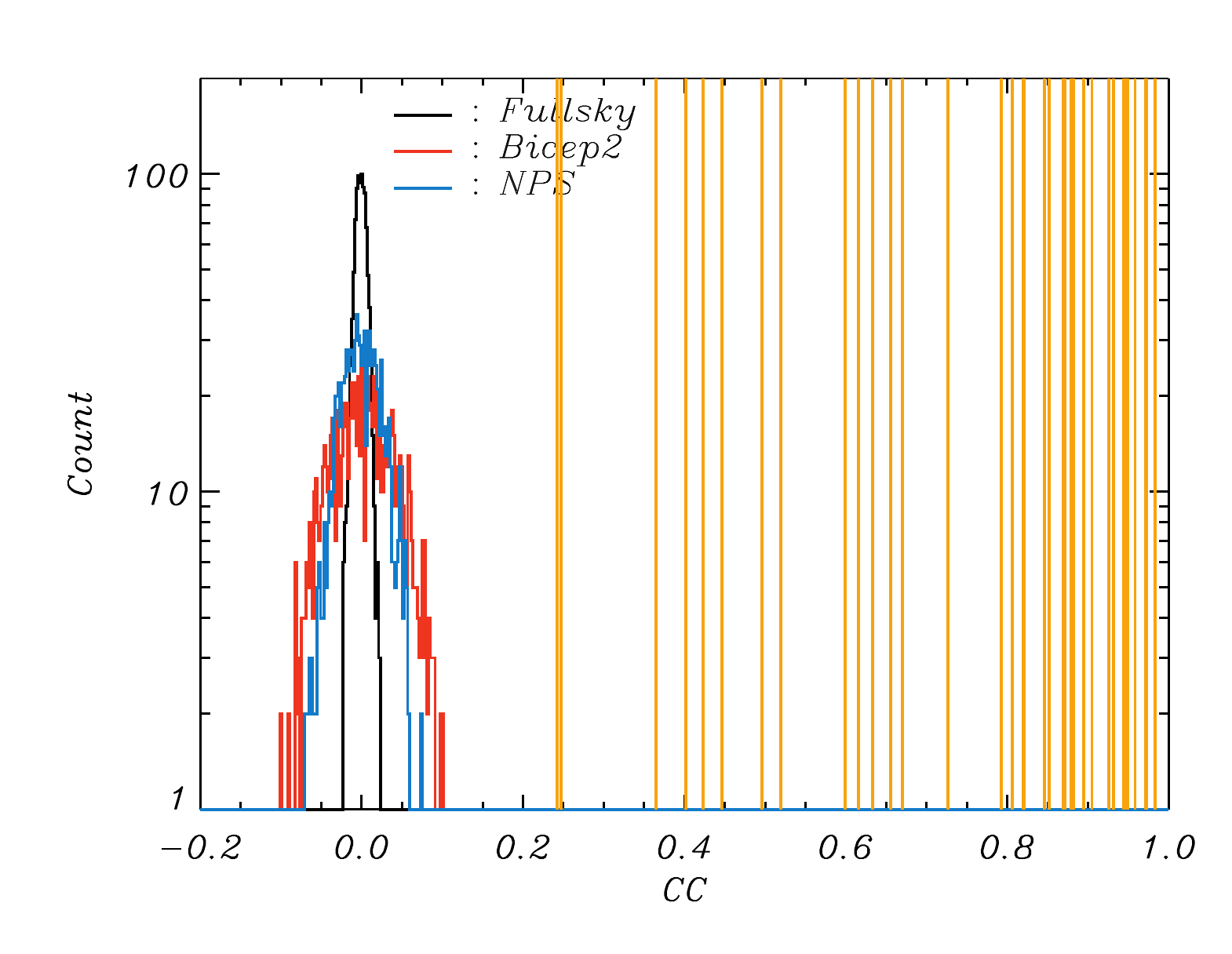}
 \includegraphics[width=0.48\textwidth]{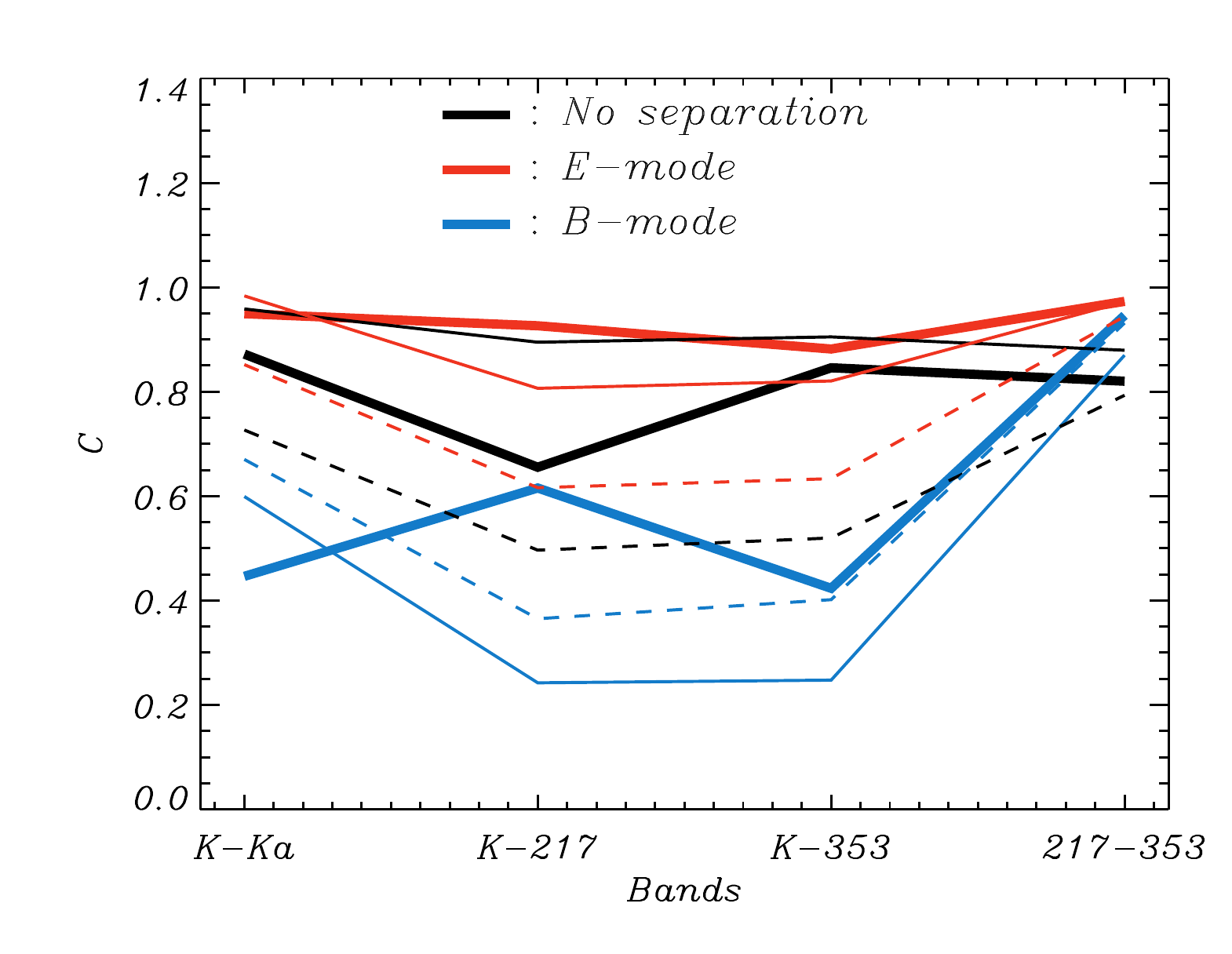}
 \caption{ \emph{Left}: the histogram of $C(\theta_1,\theta_2)$ from
 simulations assuming no band-to-band correlation. All values of
 $C(\theta_1,\theta_2)$ in table~\ref{tab:cc ang} are plotted as vertical
 lines, and they are all outside the range allowed by the simulations.
 \emph{Right}: the variation of the cross-correlation values as a function of
 the band combination. For the right panel: The BICEP2 region is shown with a
 thick solid line, the NPS region with a solid line, and the full sky with a
 dashed line.}
 \label{fig:hist_cc_ang}
\end{figure}

From table \ref{tab:cc ang} it is evident that for the K and Ka bands, the
polarization angles for the E family have very high cross-correlations in the
NPS and BICEP2 zones (0.98 and 0.95 respectively), and even the
synchrotron--TDE correlations (K--353 GHz) reach 0.82 and 0.88. For the TDE
bands 217-353 GHz these coefficients are both equal to 0.97 for E family, and
0.87, 0.95 for the B family. This clearly indicates that the structure of the
maps in figure~\ref{plzones} is associated with intrinsic properties of the
Stokes components $(Q_E,U_E)$ and $(Q_B,U_B)$ and is not an artifact of the
noise.

To estimate the significance of these values, we randomize the phases of the K
and Ka bands to get 1,000 simulated noise maps. The distribution of CC
coefficients is shown in figure~\ref{fig:hist_cc_ang}, together with vertical
lines for each value in table~\ref{tab:cc ang}. One can see that all values in
table~\ref{tab:cc ang} deviate significantly from the simulations. The
simulations assume no band-to-band correlation, yet the band-to-band
polarization angle correlation is very significant between synchrotron and
dust bands and for all selected sky regions.

\subsection{Polarization intensities of the E/B families} \label{sub:test EB-family}

\begin{figure*}[!htb]
 \centering
 \includegraphics[width=0.32\textwidth]{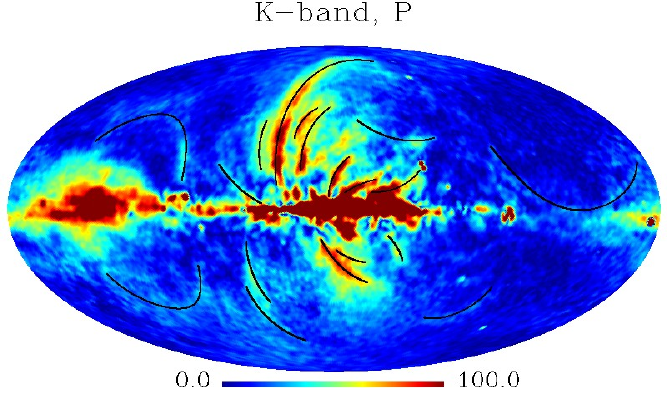}
 \includegraphics[width=0.32\textwidth]{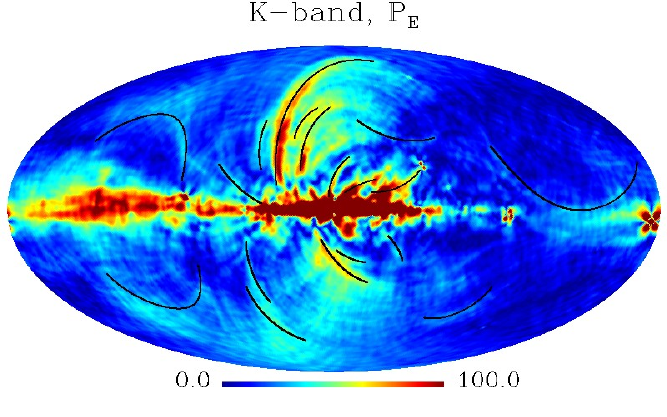}
 \includegraphics[width=0.32\textwidth]{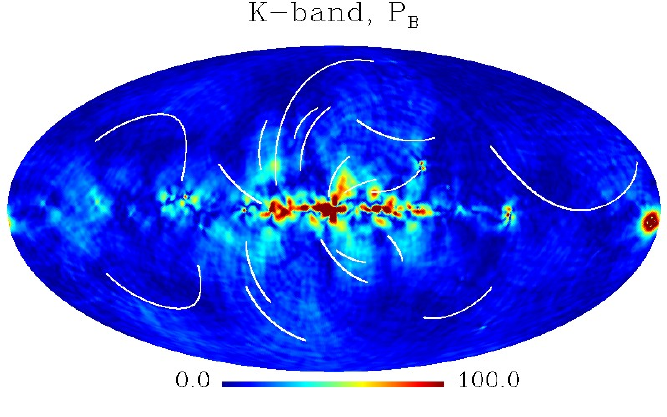}

 \includegraphics[width=0.32\textwidth]{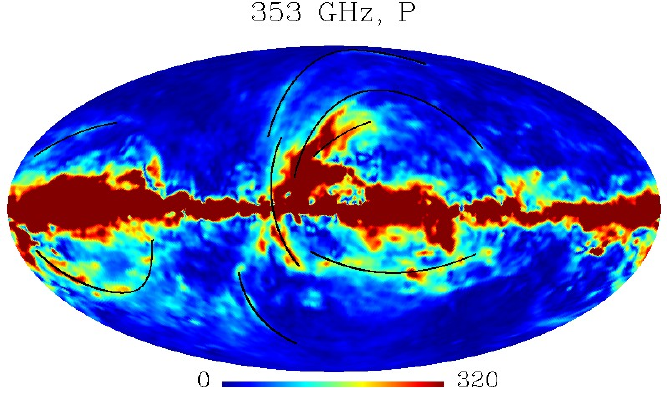}
 \includegraphics[width=0.32\textwidth]{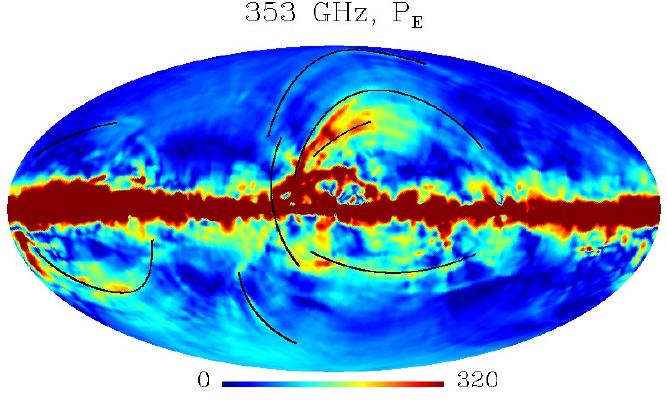}
 \includegraphics[width=0.32\textwidth]{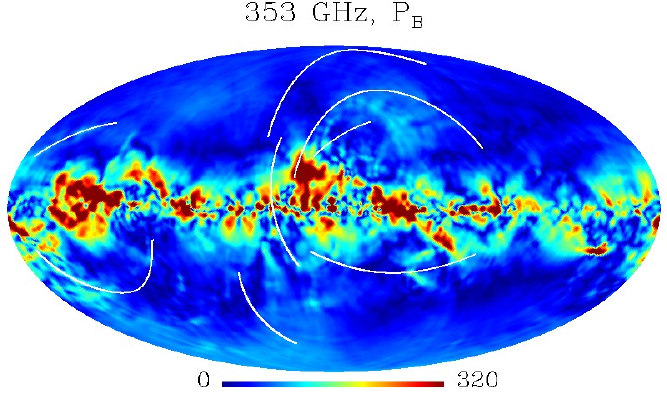}
 \caption{ The polarization intensity maps and its EB decomposition, in
 K-band (upper) and 353 GHz (lower). From left to right: total polarized
 intensity $P$, $P_E$, and $P_B$. The loop-like structures are marked by
 arches.}
 \label{fig:P Pe and Pb}
\end{figure*}

\begin{figure*}[!htb]
 \centering
 \includegraphics[width=0.32\textwidth]{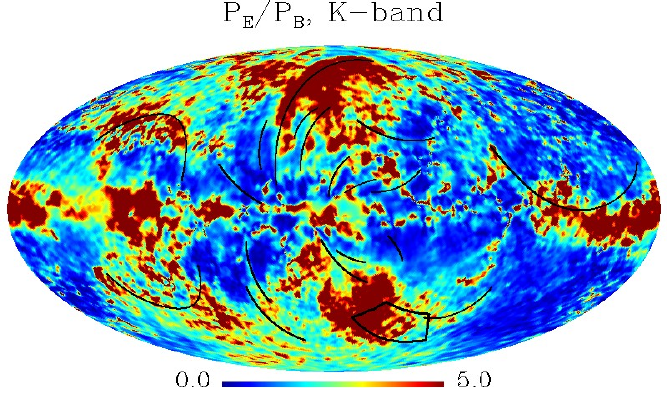}
 \includegraphics[width=0.32\textwidth]{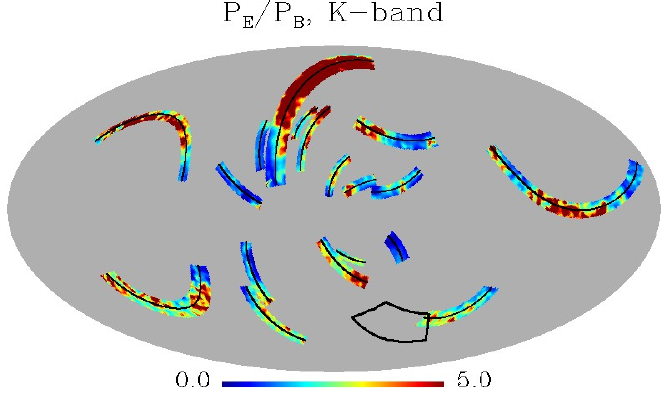}

 \includegraphics[width=0.32\textwidth]{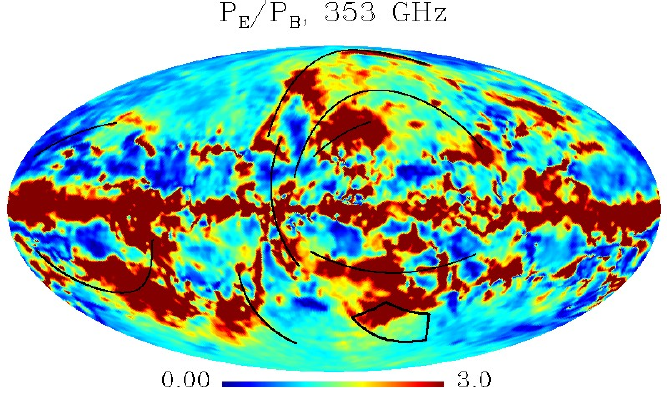}
 \includegraphics[width=0.32\textwidth]{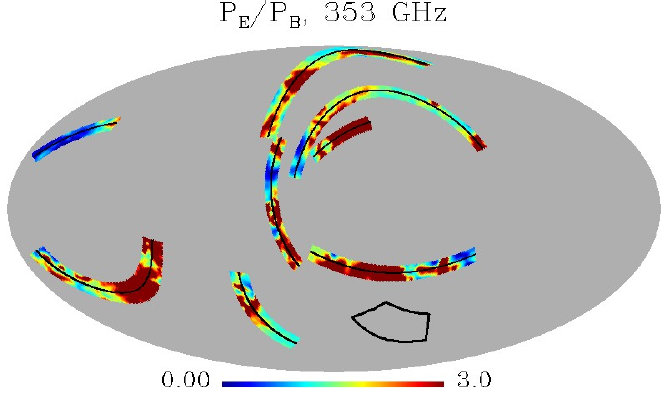}
 \caption{ The ratio $\rho=P_E/P_B$ in K-band (upper) and 353 GHz (lower),
 both full sky (left) and around the arches (right). The outline of the BICEP2
 zone is also plotted.}
 \label{fig:r K and 353}
\end{figure*}

In this section, using the definitions of $P_E$ and $P_B$ from eq.~\ref{equ:pe
pb} two foreground families are identified: one in which the E mode dominates
the polarization, and one in which the B mode dominates. For illustration of
our method we will use the WMAP K-band and the Planck 353 GHz maps, from which
$P_E$ and $P_B$ are derived and presented in figure~\ref{fig:P Pe and Pb}. The
signal in WMAP K-band map is mainly synchrotron emission, while the Planck 353
GHz map consists of thermal dust emission. The ratio between $P_E$ and $P_B$,
denoted by $\rho=P_E/P_B$, is presented in figure~\ref{fig:r K and 353}, with
the arches listed in~\citep{2015MNRAS.452..656V} marked for comparison.

In figure~\ref{fig:P Pe and Pb}, all loop-like structures are clearly visible
in the $P_E$ maps, but are missing in the $P_B$ maps. This tendency is further
confirmed in figure~\ref{fig:r K and 353} by higher amplitudes of $\rho$ near
the loops and arches. Furthermore, it can be seen from figure~\ref{fig:r K and
353} that the B mode tends to be subdominant ($\rho > 1$) in the BICEP2 zone,
NPS zone, and loops, which makes these parts of the sky peculiar.

\subsection{Significance of the EB families by the polarization intensities}\label{sub:EB family significance}

We define the cross-correlation coefficient for the $P$ and $\rho$ maps as
\begin{equation} \label{corr}
    C(P, \rho) = \frac{\sum_i (P_i - \left< P_i \right>) (\rho_i - \left< \rho_i \right>)}{\sqrt{\sum_i(P_i-\left< P_i \right>)^2 \sum_i(\rho_i-\left< \rho_i \right>)^2}},
\end{equation}
where $\left< X_i \right> = \frac{1}{N} \sum_i X_i$, the sums are taken over
all pixels $i$, and $N$ is the total number of pixels in the maps. As was
pointed out in~\cite{2016A&A...586A.133P}, at 353 GHz, the E mode is
stronger than the B mode, which is also the case for the WMAP K band (23 GHz).
For an input map with a stronger E mode, pixels with higher $\rho$ are more
affected by the E mode and thus tend to have higher $P_E$ in comparison to
$P_B$. Therefore we may expect to get positive cross correlation $C(P,\rho)$
for $\rho<1$ and $\rho>1$.

\begin{figure*}[!htb]
 \centering
 \includegraphics[width=0.22\textwidth]{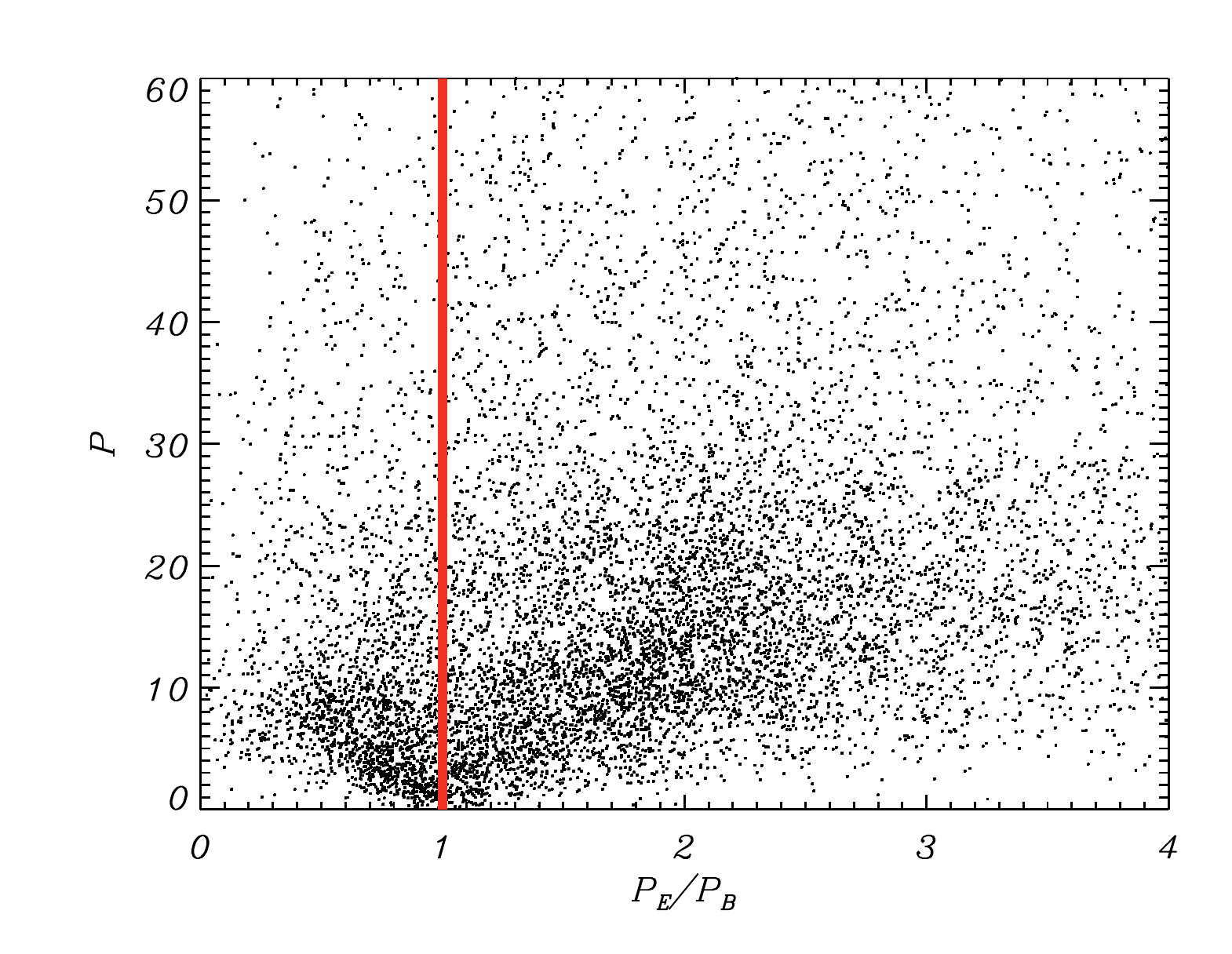}
 \includegraphics[width=0.22\textwidth]{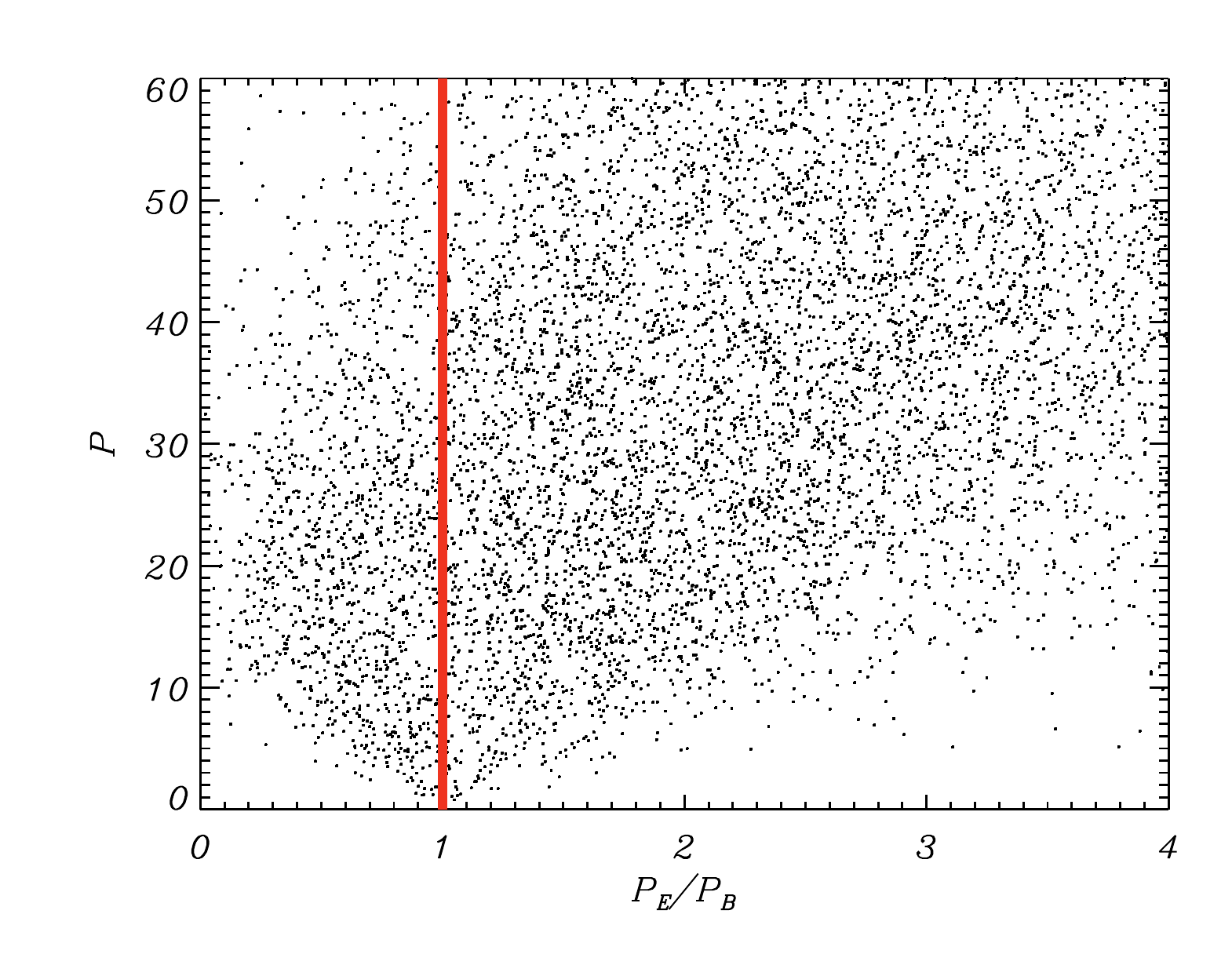}
 \includegraphics[width=0.22\textwidth]{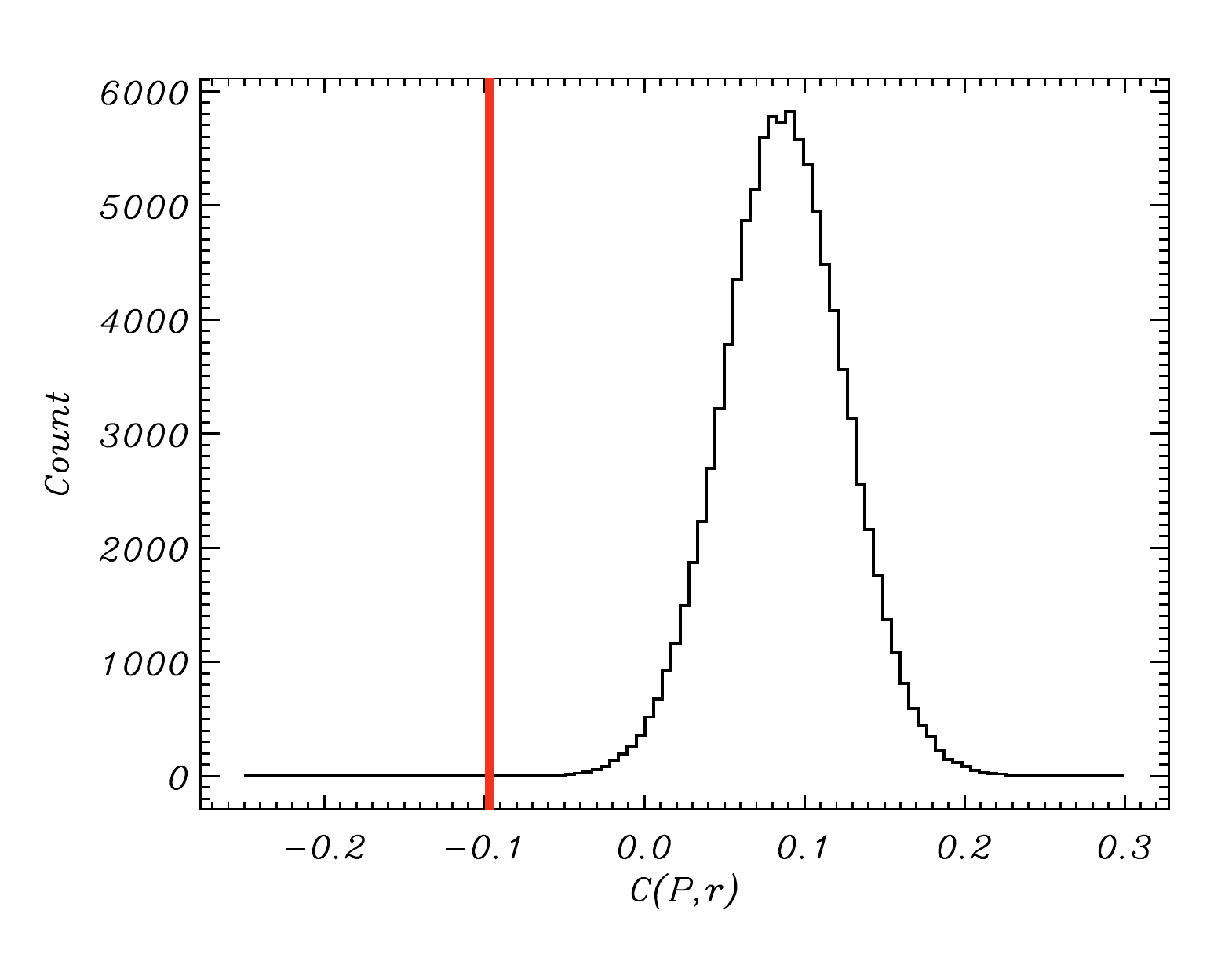}
\includegraphics[width=0.22\textwidth]{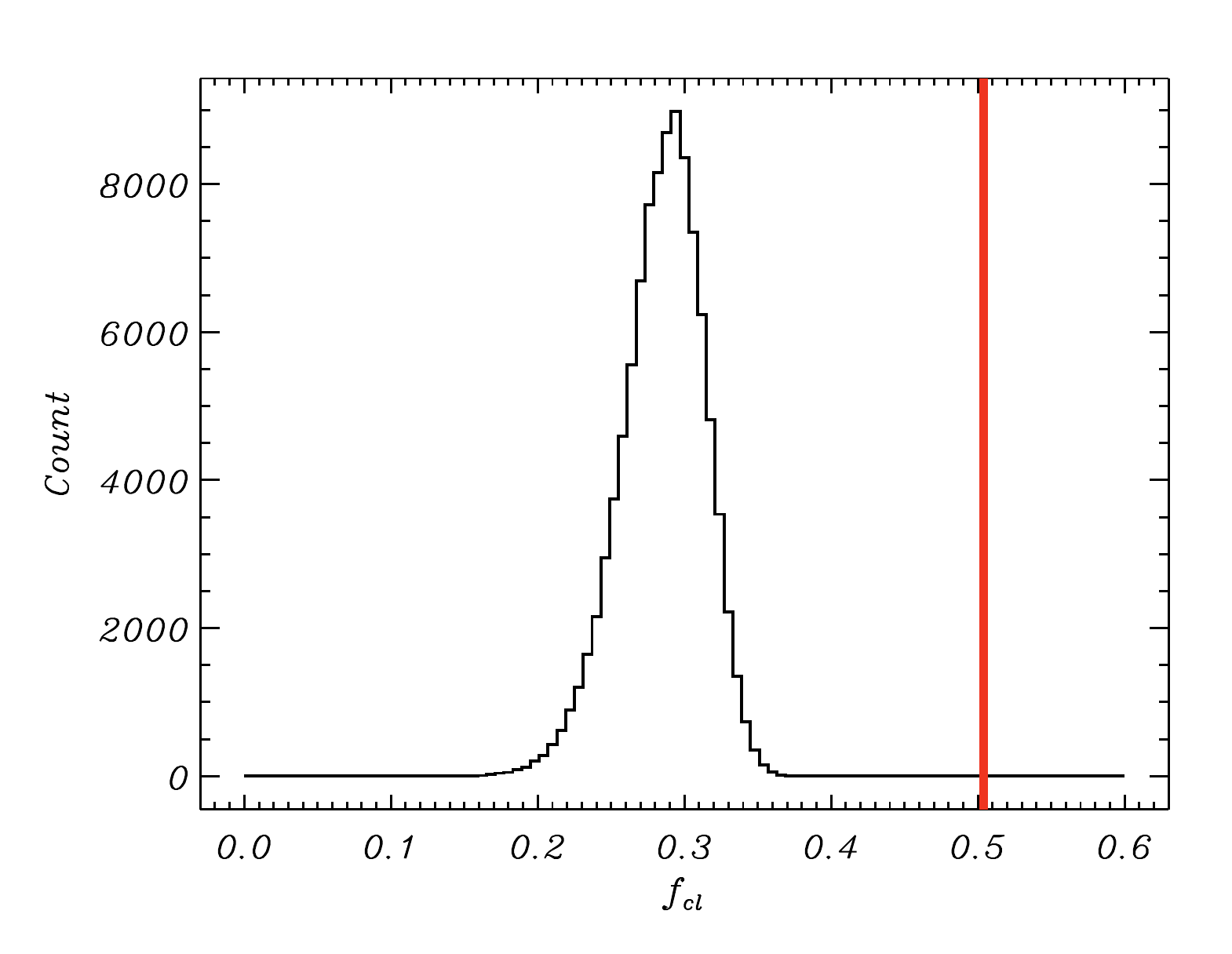}

 \includegraphics[width=0.22\textwidth]{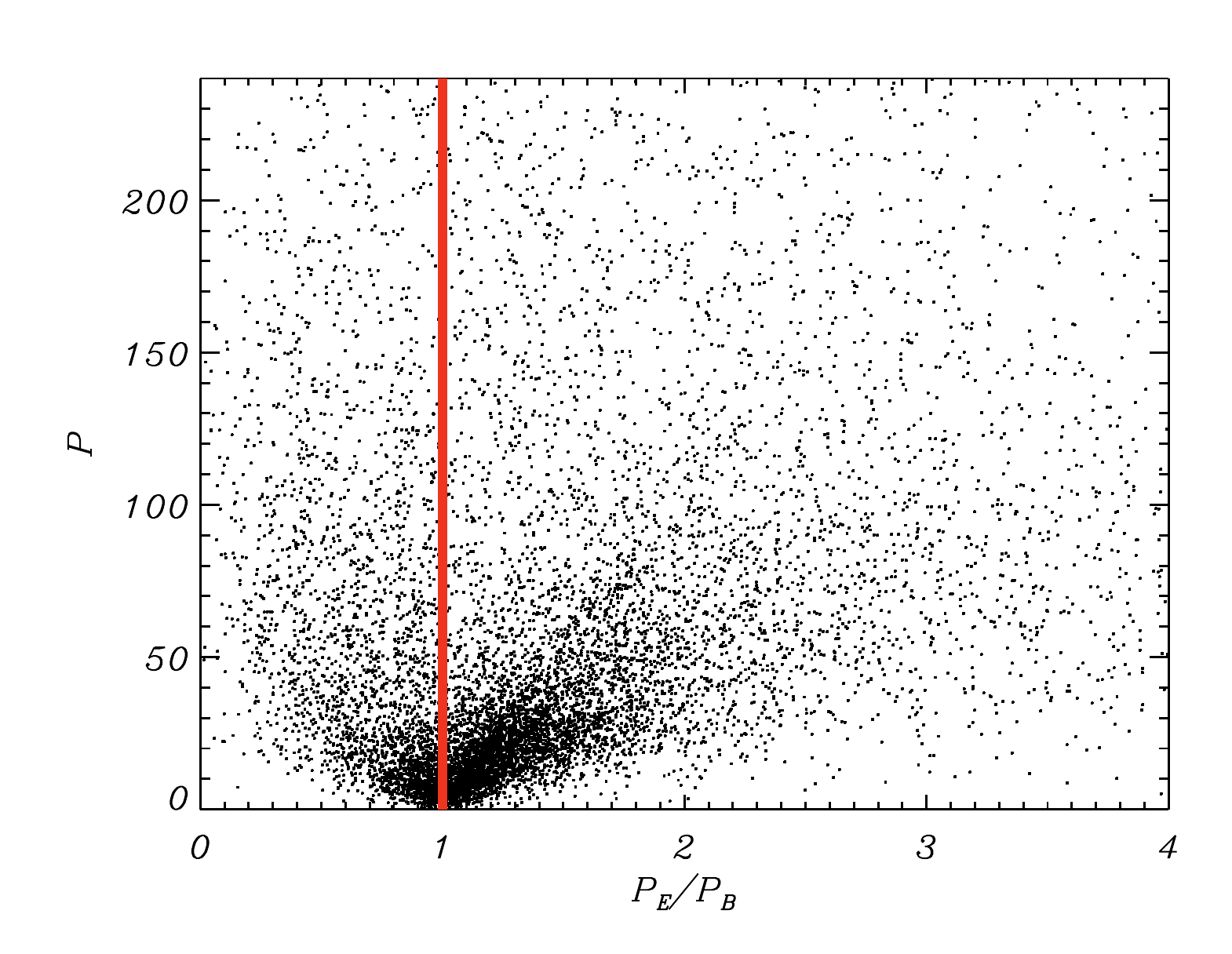}
 \includegraphics[width=0.22\textwidth]{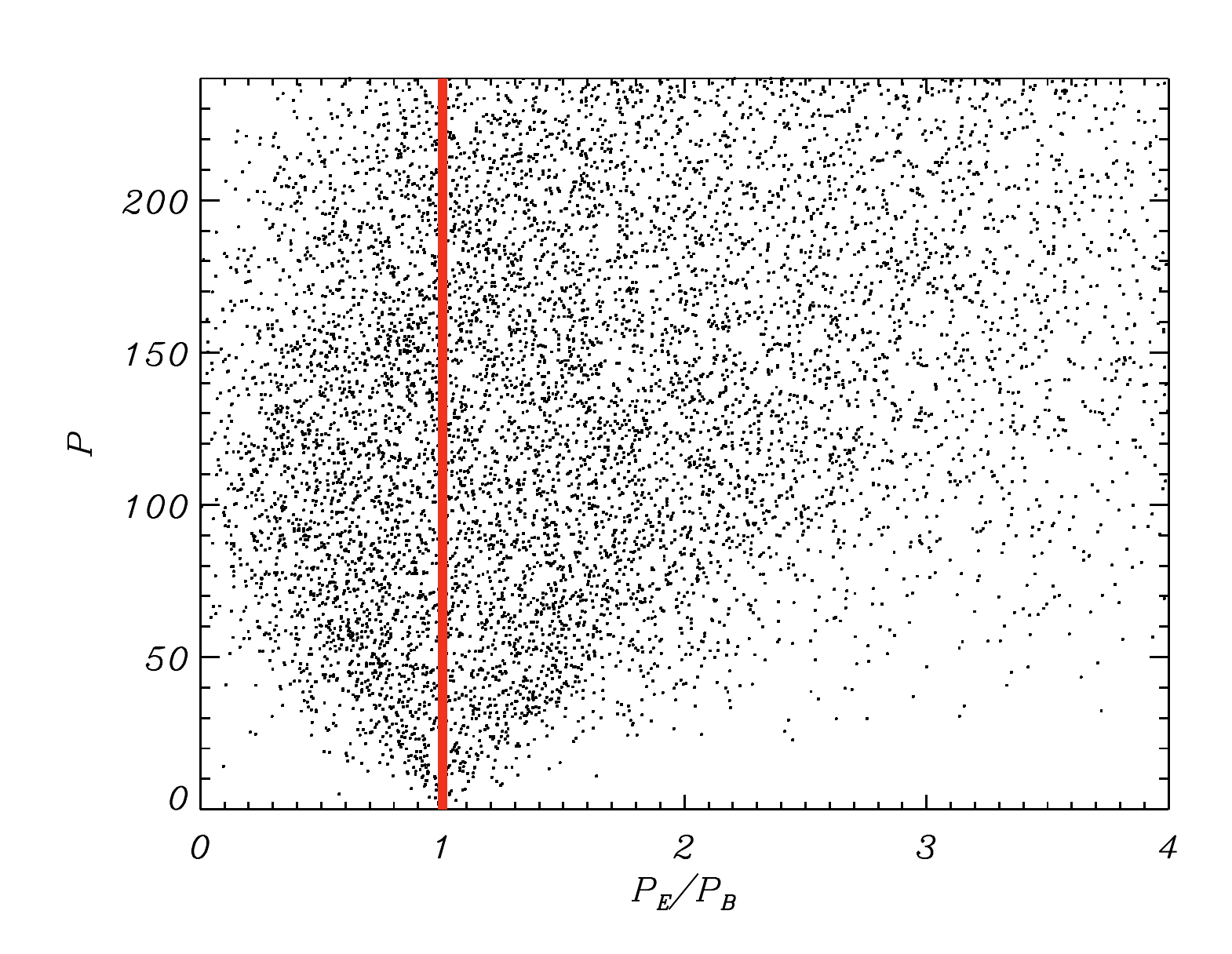}
 \includegraphics[width=0.22\textwidth]{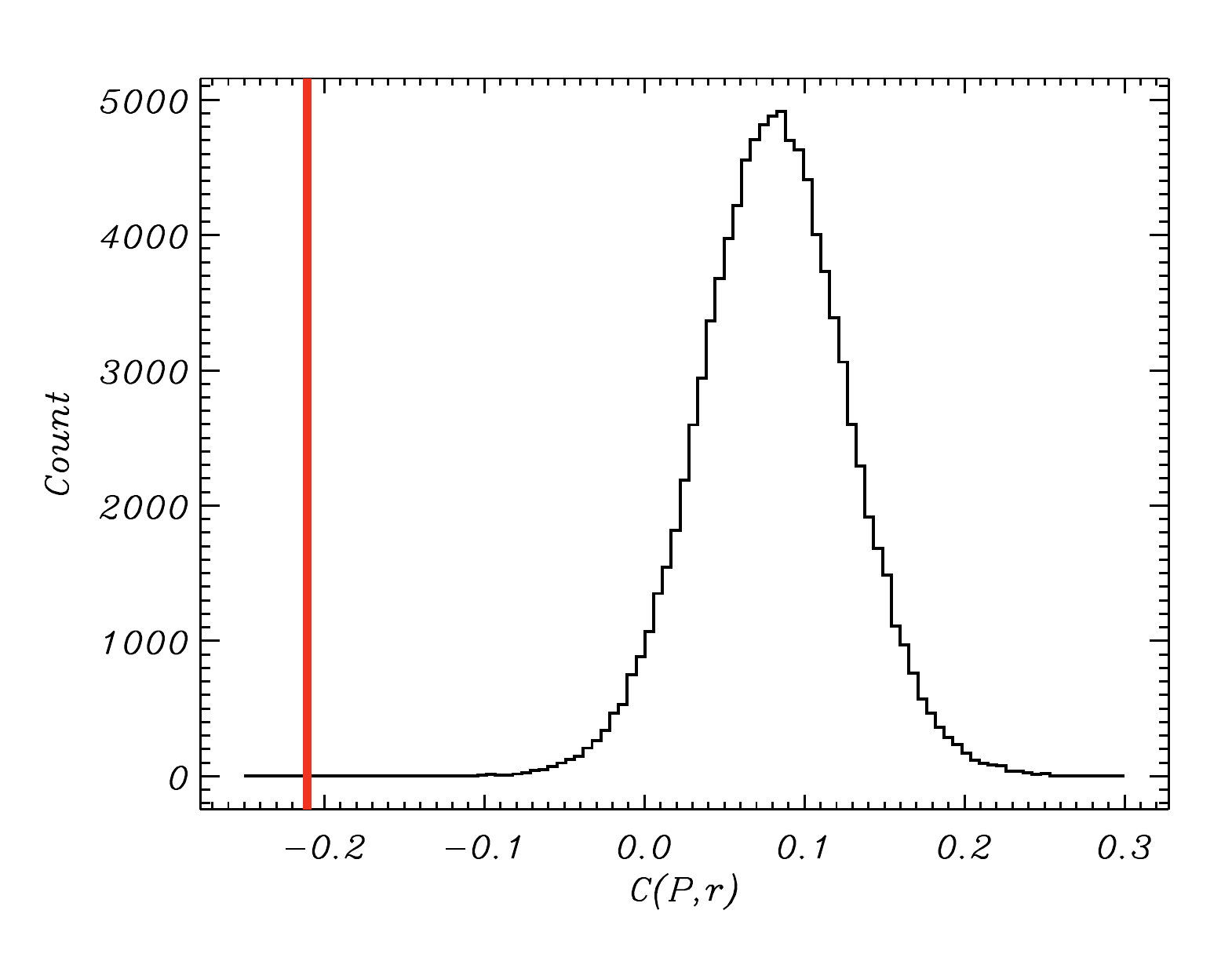}
 \includegraphics[width=0.22\textwidth]{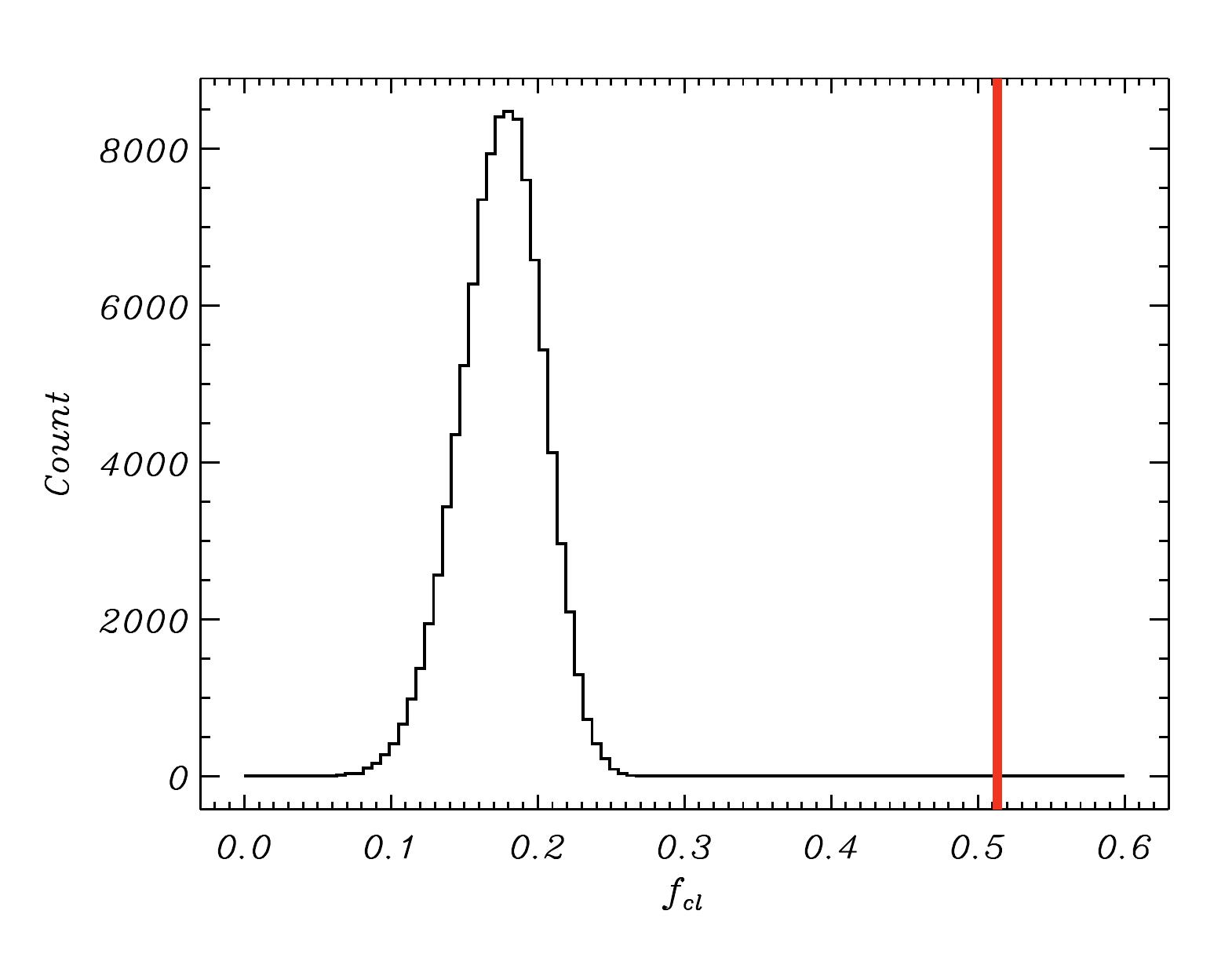}
 \caption{The relationship between total polarization intensity $P$ and the
 ratio $\rho=P_E/P_B$. Upper panels show the K-band and lower panels show 353
 GHz. From left to right: real data, one random realisation, the histograms of
 $C(P,\rho)|_{\rho<1}$ and for $\rho>1$ for $10^5$ simulations. The vertical
 lines for the two panels on the left mark $\rho=1$, and the vertical lines on
 the two right pannels mark $C(P,r)$ of the real data. A negative value of
 $C(P,\rho)|_{\rho<1}$ is an indicator for the B mode family. The positive
 $C(P,\rho)|_{\rho>1}$ indicate the E family. }
 \label{fig:P vs r}
\end{figure*}

However, the sign of correlations depends on the mean values of $P^2_B$ and
$\rho$, which makes the shape of $C(P,\rho)$ more complicated. In
figure~\ref{fig:P vs r} we plot the relationship between $P$ and $\rho$ which
shows anti-correlation in the $\rho<1$ region, and the histograms of
$C(P,\rho)$. To illustrate the significance of $C(P,\rho<1)<0$, we generate
$10^5$ realizations of the difference $\theta_E-\theta_B$ from a random
uniform distribution in $[0,2\pi]$ and plot the distribution $H(C_{rand})$ of
the corresponding values of $C_{rand}(P,\rho)$ in figure~\ref{fig:P vs r}.
Contrary to the actual distribution of $C(P,\rho)$, the result from the
simulation shows that $H(C_{rand})$ is consistent with $C_{rand}\ge 0$, and
strongly disfavoured at $\rho<1$. Thus, the anti-correlation
$C(P,\rho<1)<-(0.1-0.2)$ cannot be produced by noise.


In order to evaluate the significance of this B mode foreground family, $10^5$
random simulations are generated with the same angular power spectrum to the
WMAP K-band foreground map in use, and for each simulation, $P$ and $r$ are
calculated for the $|b|>10\degree$ and $\rho<1$ region. Under the same
conditions, the K-band map gives $C(P,\rho)=-0.1$, which is not reached by any
of the 100,000 simulations, as presented by the histogram in figure~\ref{fig:P
vs r}. Therefore, the existence of a B mode family is confirmed at a
$99.999\%$ confidence level. One possible candidate for this B mode family is
the asymmetry structure of the loops, which is briefly discussed in
section~\ref{sec:discussion}. The above analysis is also repeated for the 353
GHz band and the results are included in figure~\ref{fig:P vs r}, which are
similar to the K band. The existence of a B mode family with $\rho<1$ can be
an important source of contamination for future primordial gravitational
wave detection, which depends critically on the level of residual B mode
foreground.

In addition to the B family the existence of an E mode family can also be
identified from figure~\ref{fig:P vs r} by an apparent clustering of pixels in
the $\rho>1$ domain localized in the $P<40$ $\mu$K region for the K band. In
fact, this family was already discovered as the loops in figure~\ref{fig:P Pe
and Pb}.

The significance of detection of the E mode family is evaluated by the number
of pixels in the $\rho>1$ and $P<40$ $\mu$K region of the WMAP K band, which
covers 57$\%$ of the sky for the K band, whereas in 100,000 simulations, this
ratio is exceeds the threshold 45$\%$. This corresponds to a confidential
level of 99.999$\%$ against chance correlations. The same results are
obtained for the TDE in the Planck 353 GHz map.

\section{Frequency dependence of E and B families}\label{sec:EB ratio}

In the previous sections, we used the two most representative frequency
maps---the WMAP K-band as an indicator of synchrotron emission, and the Planck
353 GHz map as an indicator of thermal dust emission. Below we will extend
this analysis for all the WMAP and Planck frequency bands in order to trace
the properties of the E and B foreground families, and more importantly, to
investigate their frequency dependence. This analysis is very important for
future implementation of component separation tools, for verification of the
residuals of data cleaning (for instance, the bandpass leakage correction),
and for identification of different masks that can help to avoid peculiar
zones of the maps where there is strong variation of the spectral indices of
the components.

\subsection{Asymptotic of E mode excess for high E/B ratio} \label{sub:more about pe/pb} 

As we have shown in section~\ref{sub:test EB-family}--\ref{sub:EB family
significance}, the ratio $\rho=P_E/P_B$ is a very useful tool for determining
the morphological features of the synchrotron and TDE maps, especially because
it is unaffected by the absolute amplitude of the emission. In this section we
will extend this analysis for WMAP K, Ka and Planck 30-353 GHz maps in terms
of the E and B families. Needless to say, unlike the K-band and 353 GHz maps,
which are dominated by synchrotron emission and TDE respectively, for other
bands (44--217 GHz) the common signal is given by a superposition of these two
basic components, and also primordial CMB.
\begin{figure*}[!htb]
 \centering
 \includegraphics[width=0.32\textwidth]{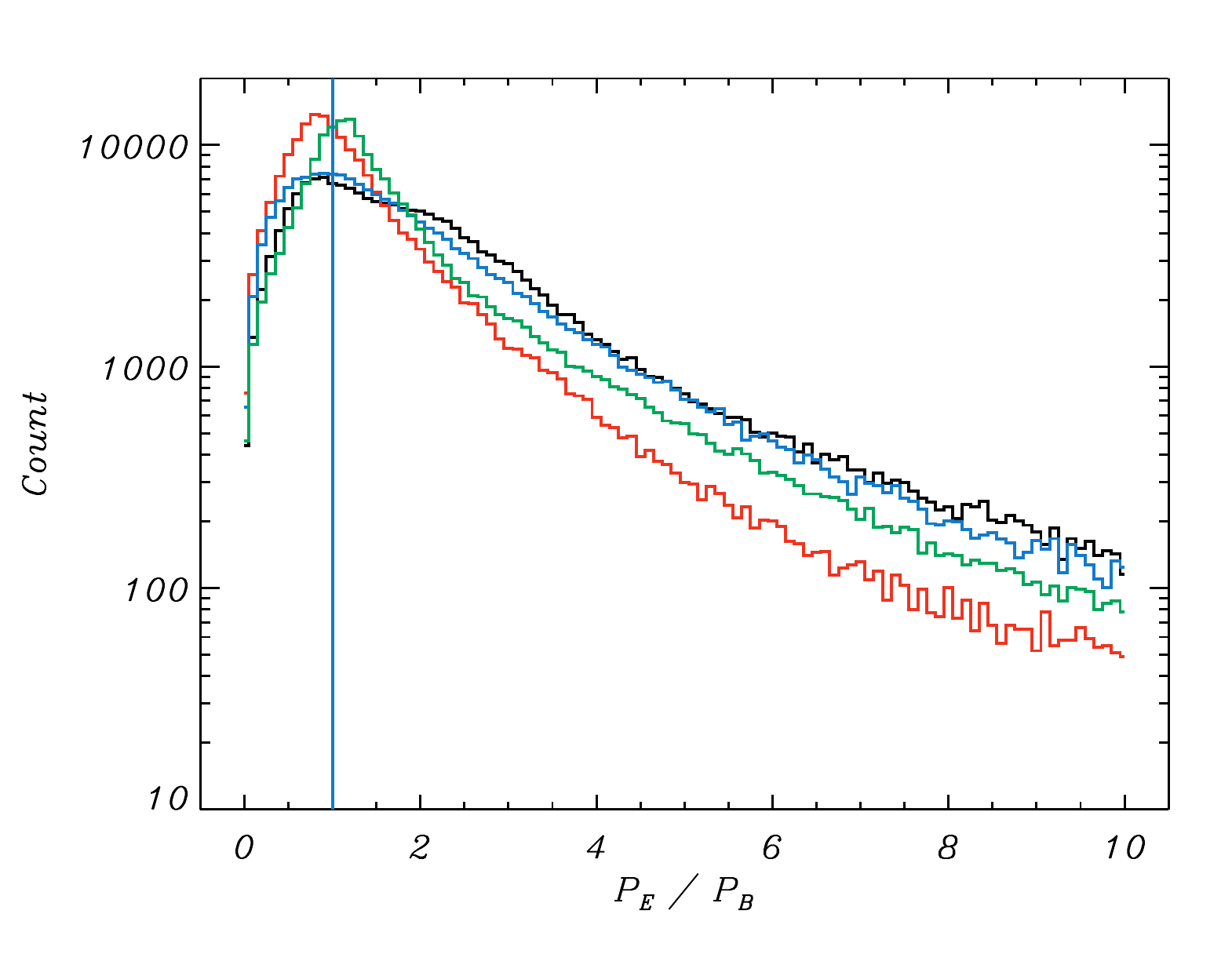}
 \includegraphics[width=0.32\textwidth]{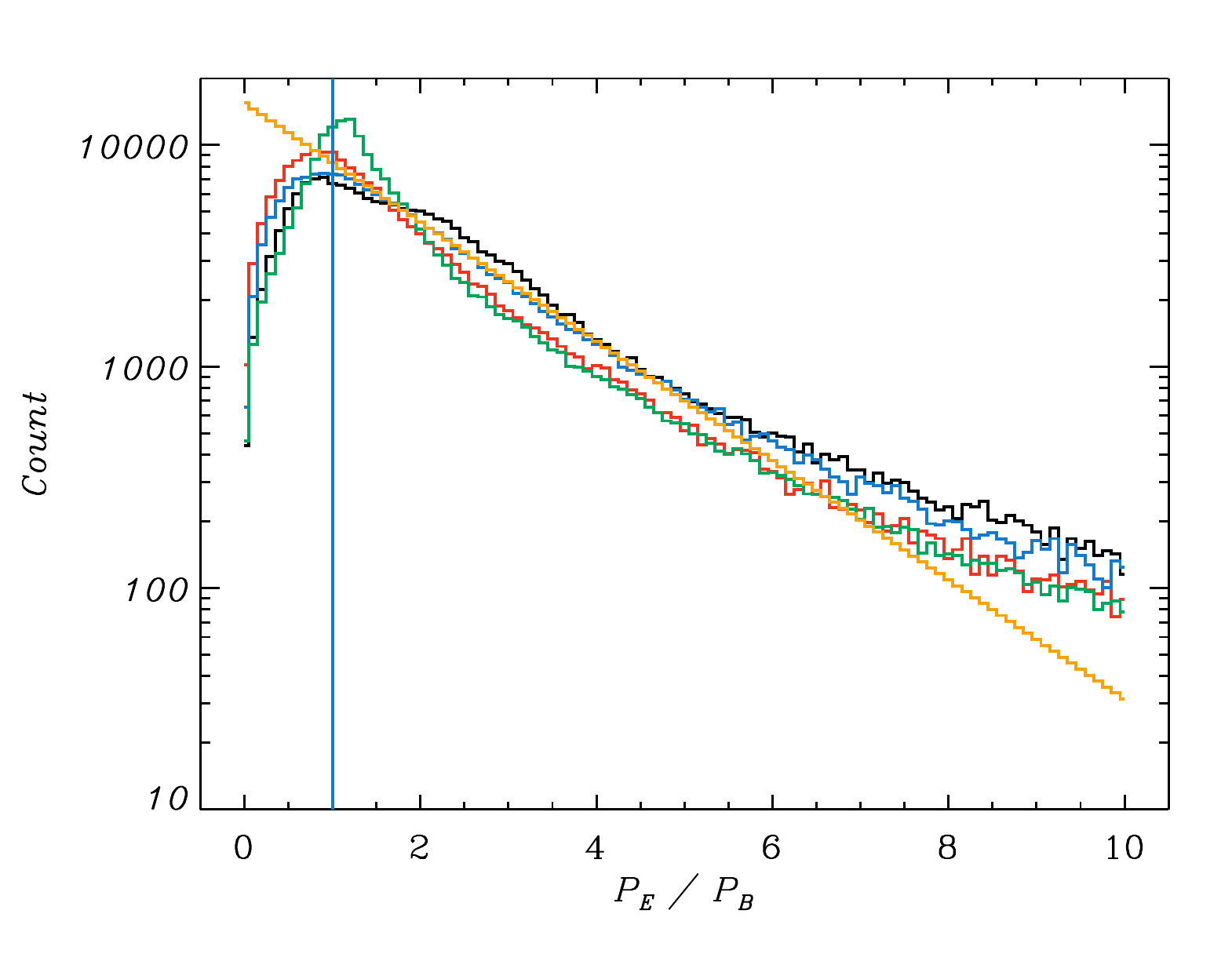}
 \includegraphics[width=0.32\textwidth]{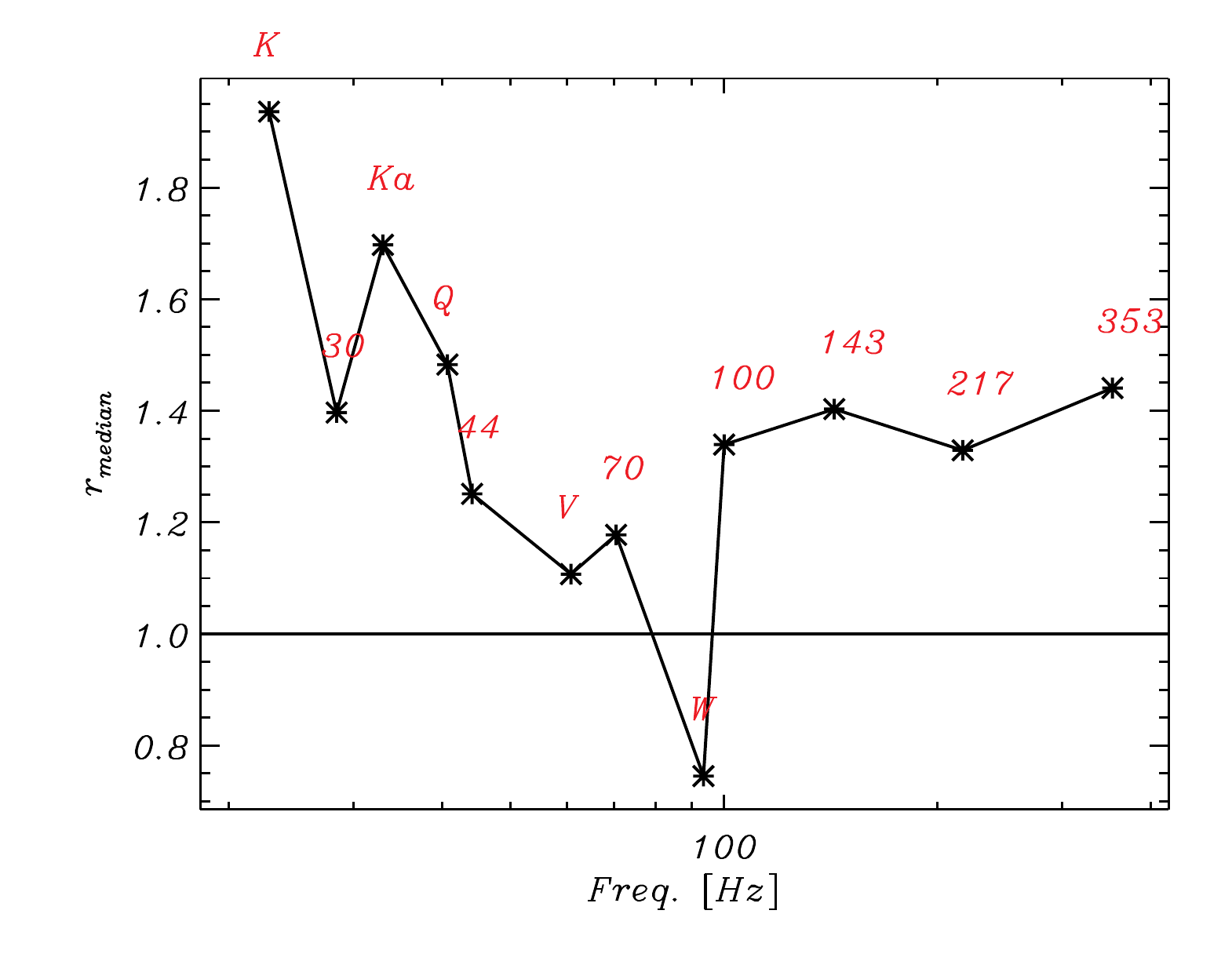}
 \caption{\emph{Left}: histogram of the ratio $\rho=P_E/P_B$ for 4 example
 frequency bands (K, Ka, 30 GHz and 353 GHz in black, blue, red, and green
 respectively.) without correction to 30 GHz (note that the uncorrected 30 GHz
 data is used \emph{only} in this panel). \emph{Middle}: similar to left but
 30 GHz is corrected. \emph{Right}: the median values of $\rho$ for each WMAP
 and Planck frequency band. The vertical lines in left and middle mark
 $\rho=1$.}
 \label{fig:EB ratio hist}
\end{figure*}

Our primary goal is to derive the corresponding maps of $\rho(\textbf n)$
for each frequency band and plot the histograms $H(\rho)$
for each map under investigation. In figure~\ref{fig:EB ratio
hist} we show these histograms (as a number of counts vs $\rho$) for the K, Ka,
30 and 353 GHz maps. From figure~\ref{fig:EB ratio hist} we come to the
very important conclusion,that in spite of very different physical mechanisms
underlying the synchrotron and polarized thermal dust generation, the corresponding
distributions of $\rho$ reveal remarkable similarity (convergence),
especially when $\rho>2$.

Surprisingly, the $H(\rho)$ estimator is very sensitive to the
systematic effects, namely a bandpass mismatch leakage in the Planck 30 GHz
band~\citep{2016A&A...594A...2P, 2016A&A...594A...3P}. On the left panel of
figure~\ref{fig:EB ratio hist} we show the $H(\rho)$ histograms
for the K, Ka and 30, 353 GHz maps, where the 30 GHz signal is not corrected for the
bandpass mismatch leakage. One can see that for $\rho>2$ the $H(\rho)$
histogram decays more rapidly than the other histograms. At the same time,
in the domain $0.7<\rho<1.5$, the uncorrected 30 GHz signal is more dominant than the 
WMAP K and Ka bands. Considering that the
Planck 30 GHz signal is close in frequency to the WMAP Ka (33 GHz) signal, it is
obvious that this very different asymptotic behavior is due to the bandpass
mismatch leakage in the Planck 30 GHz map.

In the middle plot of figure~\ref{fig:EB ratio hist} we show the same bands
as the left panel, but with correction of the 30 GHz map. Now we see the
convergence of the Planck 30 GHz signal to the 353 GHz distribution for
$\rho>2$. In this region the histograms closely follow an exponential
distribution, marked by an orange line in the same panel. Meanwhile, for
$0.7<\rho<1.5$, the amplitude of $H(\rho)$ for the 30 GHz bandpass leakage
corrected signal is closer to the corresponding distributions for K
and Ka bands, but not exactly equal to them. Although the 30 GHz band is
expected to give similar result to the K and Ka bands rather than the 353 GHz band,
this is not achieved even after correction, which confirms the conclusion
of~\citep{2018arXiv180101226W} that the bandpass mismatch leakage correction
needs further improvements, and reminds us that the 30 GHz map should always be
used with great caution.

The most pronounced part of the distributions in figure~\ref{fig:EB ratio
hist} is in the domain $0.7\le\rho\le 2$, where the synchrotron channels have
a plateau, and TDE channels have a point of maximum. To compare these
distributions we will use the median value of $\rho$ for all WMAP and Planck
23--353 GHz maps, which are plotted in the right panel of figure~\ref{fig:EB
ratio hist}. One can see that almost all bands show E mode excess with
the median value of $\rho$ between 1.1 and 1.8, except for the WMAP W band,
where the median $\rho_W\simeq 0.75 $. In addition to this feature, it is needless to
point out the convergence of all Planck 100--353 GHz signals to the asymptotic
$\rho\simeq 1.4$, which coincides with the Planck 30 GHz median (which might
still be abnormal).

\begin{figure*}[!htb]
 \centering
 \includegraphics[width=0.24\textwidth]{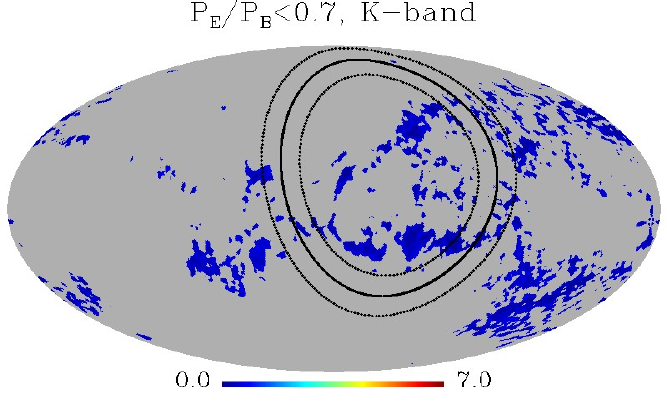}
 \includegraphics[width=0.24\textwidth]{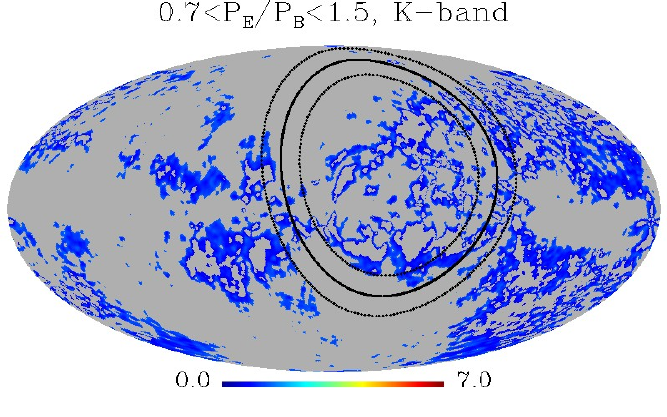}
 \includegraphics[width=0.24\textwidth]{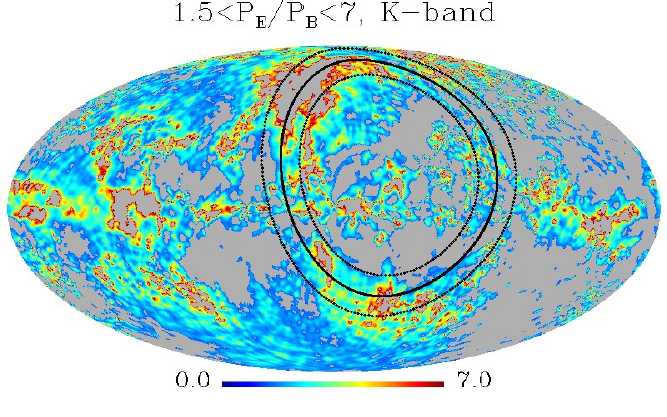}
 \includegraphics[width=0.24\textwidth]{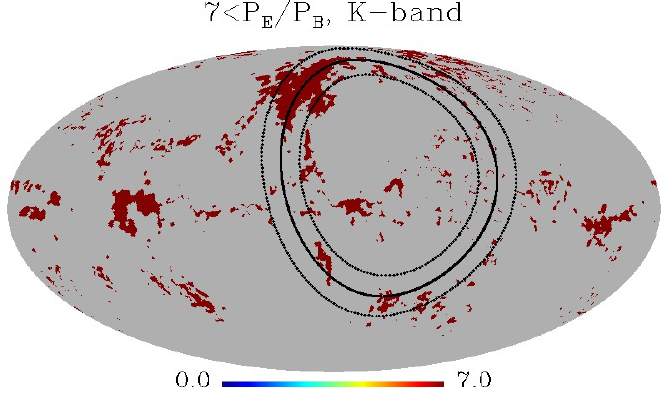}

 \includegraphics[width=0.24\textwidth]{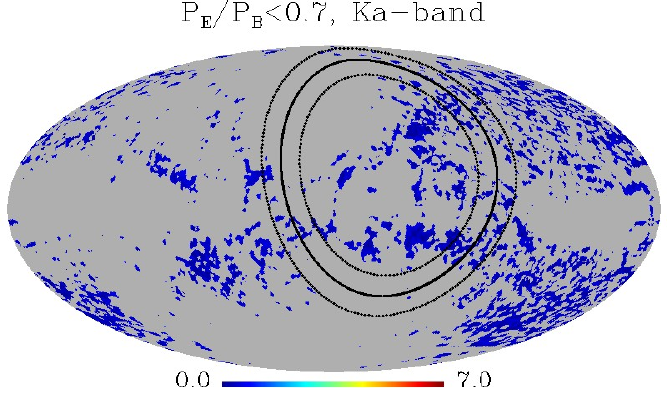}
 \includegraphics[width=0.24\textwidth]{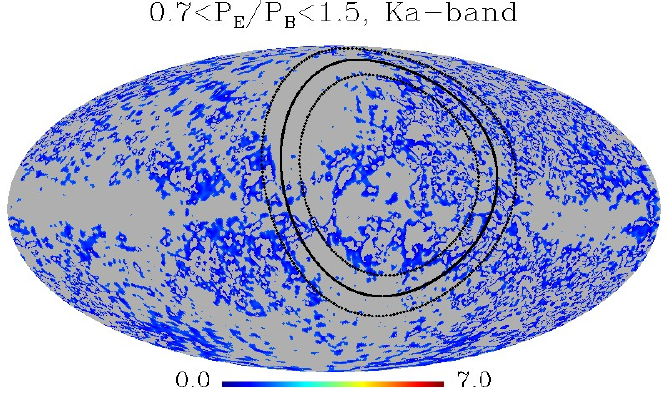}
 \includegraphics[width=0.24\textwidth]{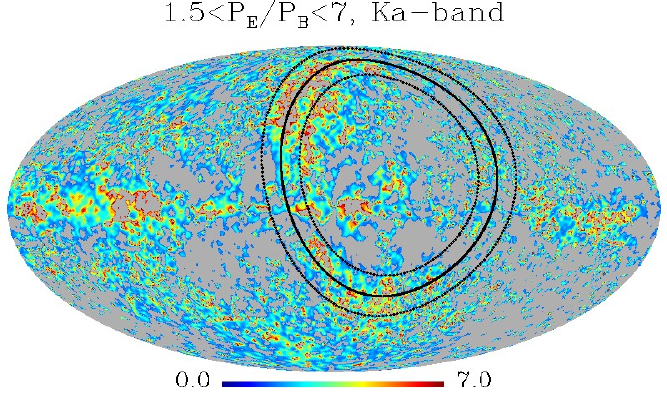}
 \includegraphics[width=0.24\textwidth]{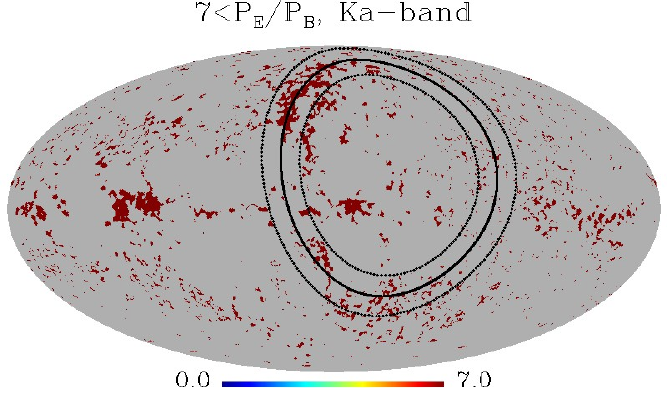}

 \includegraphics[width=0.24\textwidth]{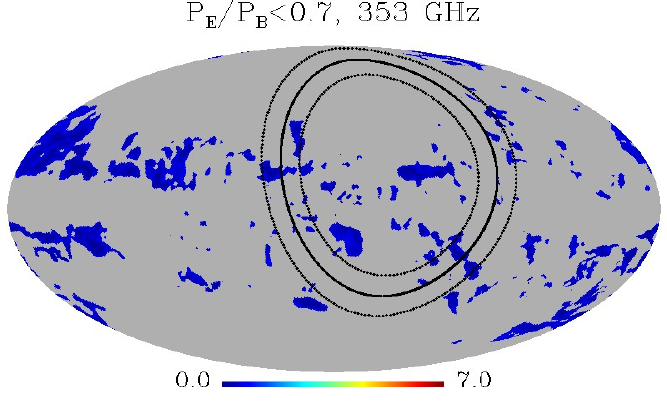}
 \includegraphics[width=0.24\textwidth]{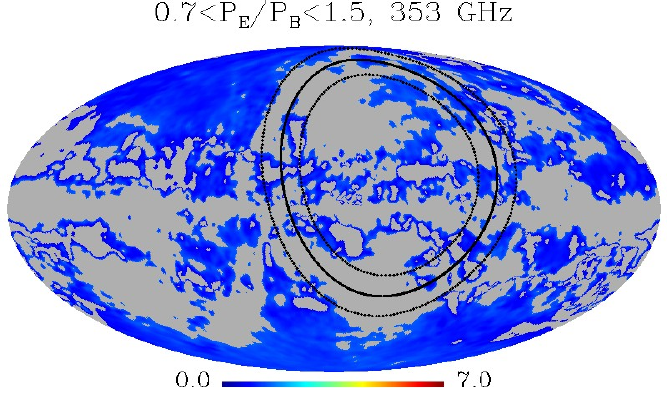}
 \includegraphics[width=0.24\textwidth]{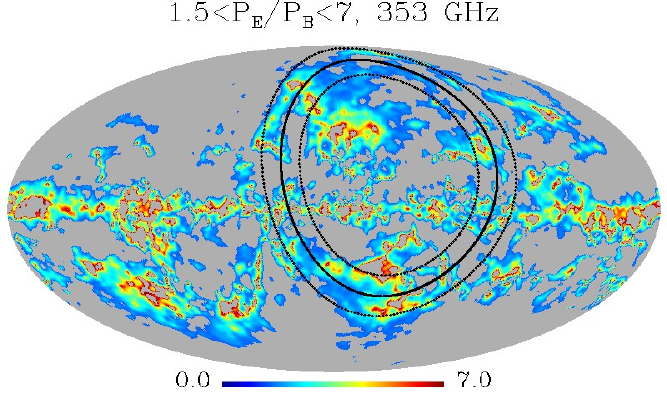}
 \includegraphics[width=0.24\textwidth]{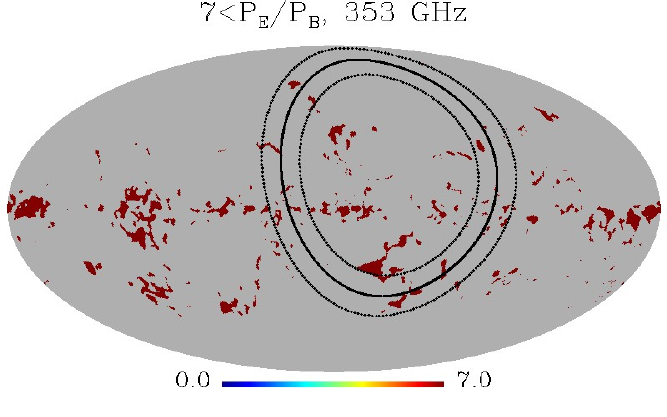}

 \caption{Maps of $\rho$ in various regions and frequency bands. From top to
 bottom: the WMAP K, Ka and Planck 353 GHz. From left to right: $0<\rho<0.7$
 (dominated by B mode), $0.7<\rho<1.5$ (moderate), $1.5<\rho<7$ (dominated by
 E mode), and $7<\rho$ (highly dominated by E mode).}
 \label{fig:EB ratio}
\end{figure*}

\subsection{Real space distribution of the E/B ratio}

The important question now arises: how much does the similarity of the
$H(\rho)$ distributions reflect similarity of the morphology of the
$\rho(\textbf n)$ sky maps? To answer this question, in figure~\ref{fig:EB
ratio} we show the maps of $\rho(\textbf n)$ for different thresholds of
$\rho$. The left column of this figure shows in blue those parts of the sky
where $\rho\le 0.7$ in the K, Ka and 353 GHz signals (from top to bottom).
There is well pronounced similarity between the K and Ka maps, but the 353 GHz
map has a different pattern. Note that this threshold indicates the angular
distribution of the B mode family. Thus, for synchrotron emission, this family
is largely localized in the right half of the map, while for TDE at 353 GHz
this family is closer to the Galactic plane (on the left part of the
map).

We would like to remind the reader that some of the most
interesting features of the polarization intensity maps are the NPS and
Loop I regions. In all the maps presented in figure~\ref{fig:EB ratio}, we
indicate the position of Loop I from~\cite{Berkhuijsen1971, Salter1983,
Wolleben2007} together with two others lines, which indicate $\pm 10^o$
circles around it. As one can see, for the synchrotron signals, the B family
of polarization is relatively very weak in the NPS zone.

The same tendency persists in column 2 in figure~\ref{fig:EB ratio}, showing
the zones where $0.7\le\rho\le 1.5$. Again, the NPS domain is empty in all the
synchrotron maps, and the only few spots are visible in the 353 GHz map. Note
that the threshold $0.7\le\rho\le 1.5$ includes the median value of
$\rho\simeq 1.4$ discussed above. The WMAP K and Ka maps significantly depart
from this value (see figure~\ref{fig:EB ratio hist}). Nevertheless, even for
these maps one can see that the NPS domain is empty or almost empty for the
353 GHz map.

Columns 3--4 of figure~\ref{fig:EB ratio} show the thresholds $1.5\le
\rho\le 7$ and $\rho>7$ respectively, which both correspond to the E family of
the polarization. An important feature of these maps is the significant
difference between the synchrotron and TDE signals. The most interesting
feature for the threshold $\rho>7$ occurs for the K band in the upper part of
the NPS region. There we find a wide zone dominated by the E family in the
synchrotron maps, which does not exist in the dust emission. In the BICEP2
zone we can see the peak of $\rho$ for the K-band, which is less visible at Ka
band, and most pronounced again in 353 GHz map. On the other hand, for the
$\rho>7$ threshold, the northern spot at NPS occurs for the K and Ka maps,
while the E family does not exist in the BICEP2 zone for this threshold.

\section{Spectral indices for the E and B families}\label{sec:EB-beta}

A critical point for extraction of the primordial CMB polarization is the
frequency dependence of the synchrotron and thermal dust
emission~\cite{2016A&A...594A..10P}. In addition to blind methods, such as
NILC~\cite{2016A&A...594A...1P}, others (Commander, SMICA) use well-formulated
assumptions about the transition of the linear combination of the foreground
templates from one frequency band to others. Assuming a power law for
synchrotron emission, we can convert the frequency dependence of the polarized
intensity to the corresponding spectral index, defined as
\begin{equation} \label{index}
    \beta(\textbf{n}) = \frac{\log(P_1(\textbf{n}) / P_2(\textbf{n}))}{\log(\nu_1 / \nu_2)},
\end{equation}
where $P_1$ and $P_2$ are the polarization intensities (either total or
associated with one of the E/B families) at frequncies $\nu_1$ and $\nu_2$
respectively. The spectral index is defined independently at each pixel
specified by $\textbf{n}$, and can vary across the sky.

In this section we will address the question of whether the E and B families
of polarized emission have the same frequency dependence, as measured by their
spectral indices $\beta_E(\textbf{n})$ and $\beta_B(\textbf{n})$. If they are
different, then the transition of Q and U, or E and B, in the frequency domain
will be more complicated.

\subsection{Spectral indexes for the full sky}\label{sub:beta_eb_fullsky}

\begin{figure}[!htb]
 \centering
 \includegraphics[width=0.24\textwidth]{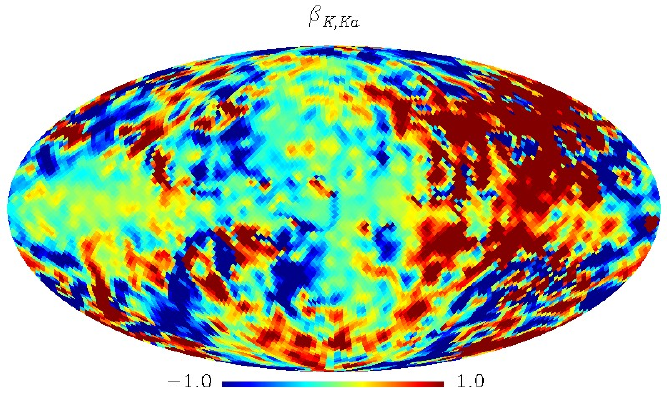}
 \includegraphics[width=0.24\textwidth]{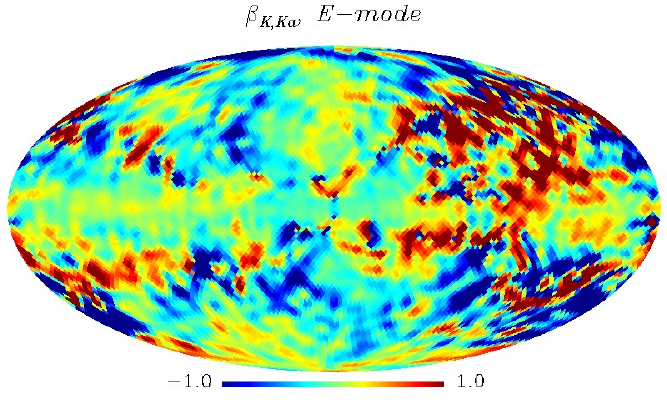}
 \includegraphics[width=0.24\textwidth]{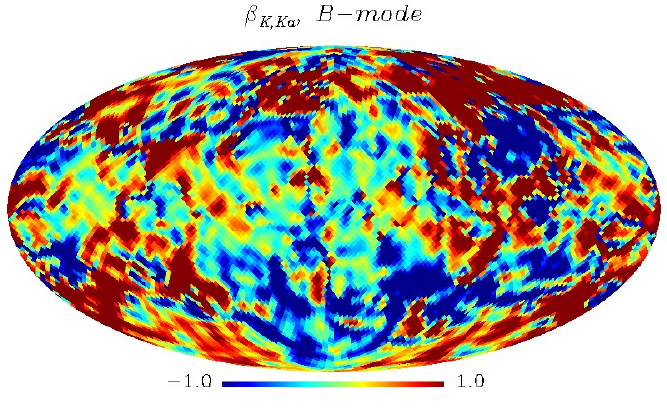}
 \includegraphics[width=0.24\textwidth]{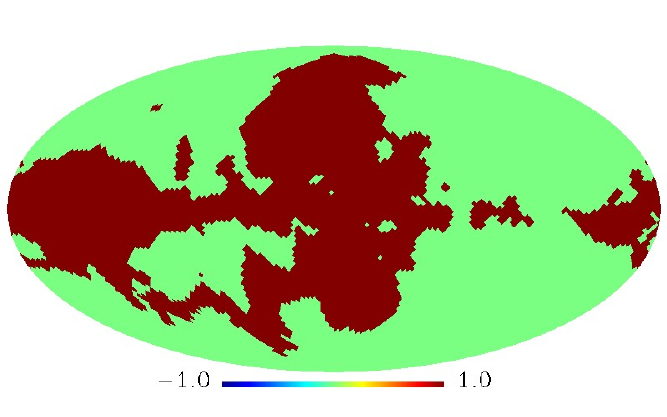}

 \includegraphics[width=0.24\textwidth]{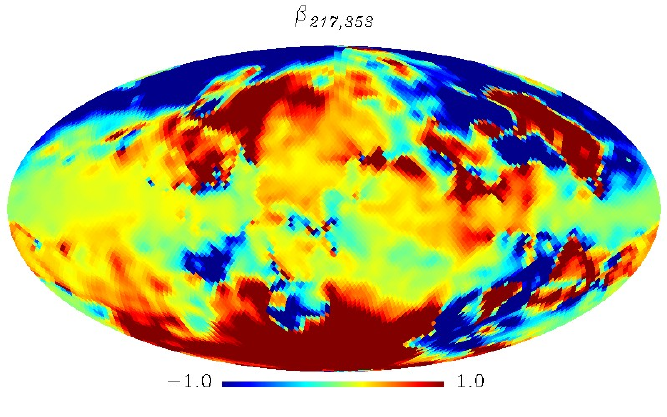}
 \includegraphics[width=0.24\textwidth]{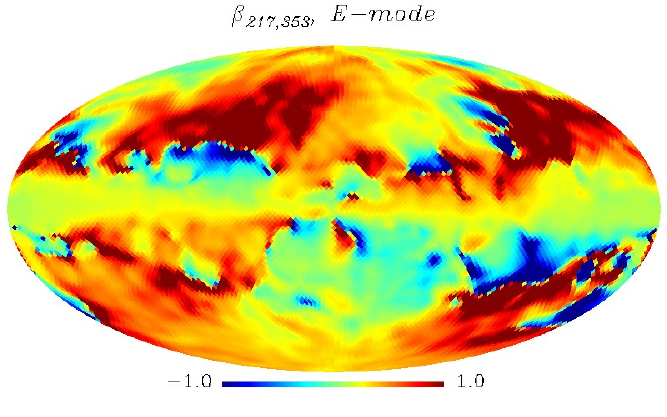}
 \includegraphics[width=0.24\textwidth]{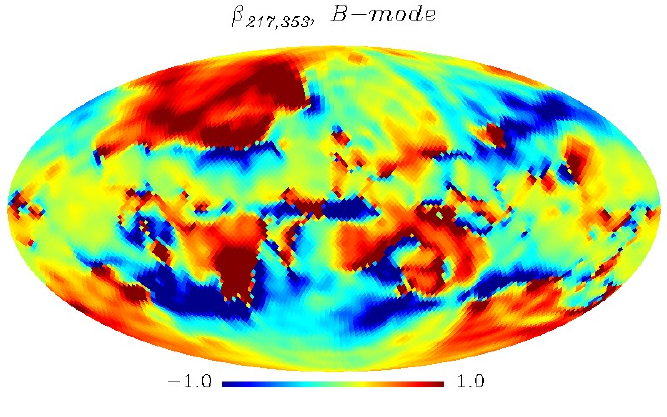}
 \includegraphics[width=0.24\textwidth]{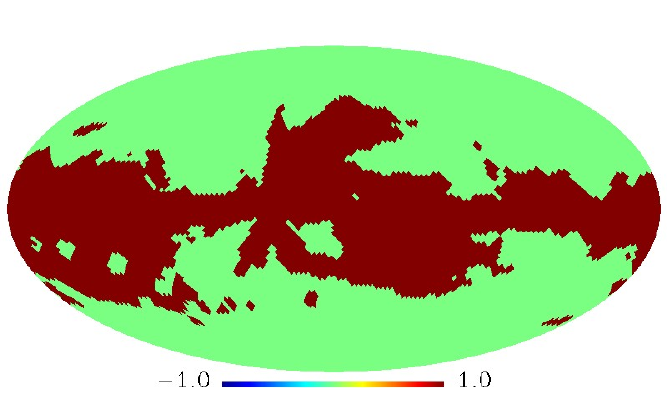}
 \caption{The spectral index maps derived from K and Ka bands (upper panels)
 and 217, 353 GHz (lower panels). From left to right: the total polarization
 intensity, only E-mode family, only B mode family, and the mask for
 statistics that covers only the highest 40$\%$ polarization intensities (the
 red zone is used).}
 \label{fig:beta_EB}
\end{figure}

\begin{figure}[!htb]
  \centering
  \includegraphics[width=0.42\textwidth]{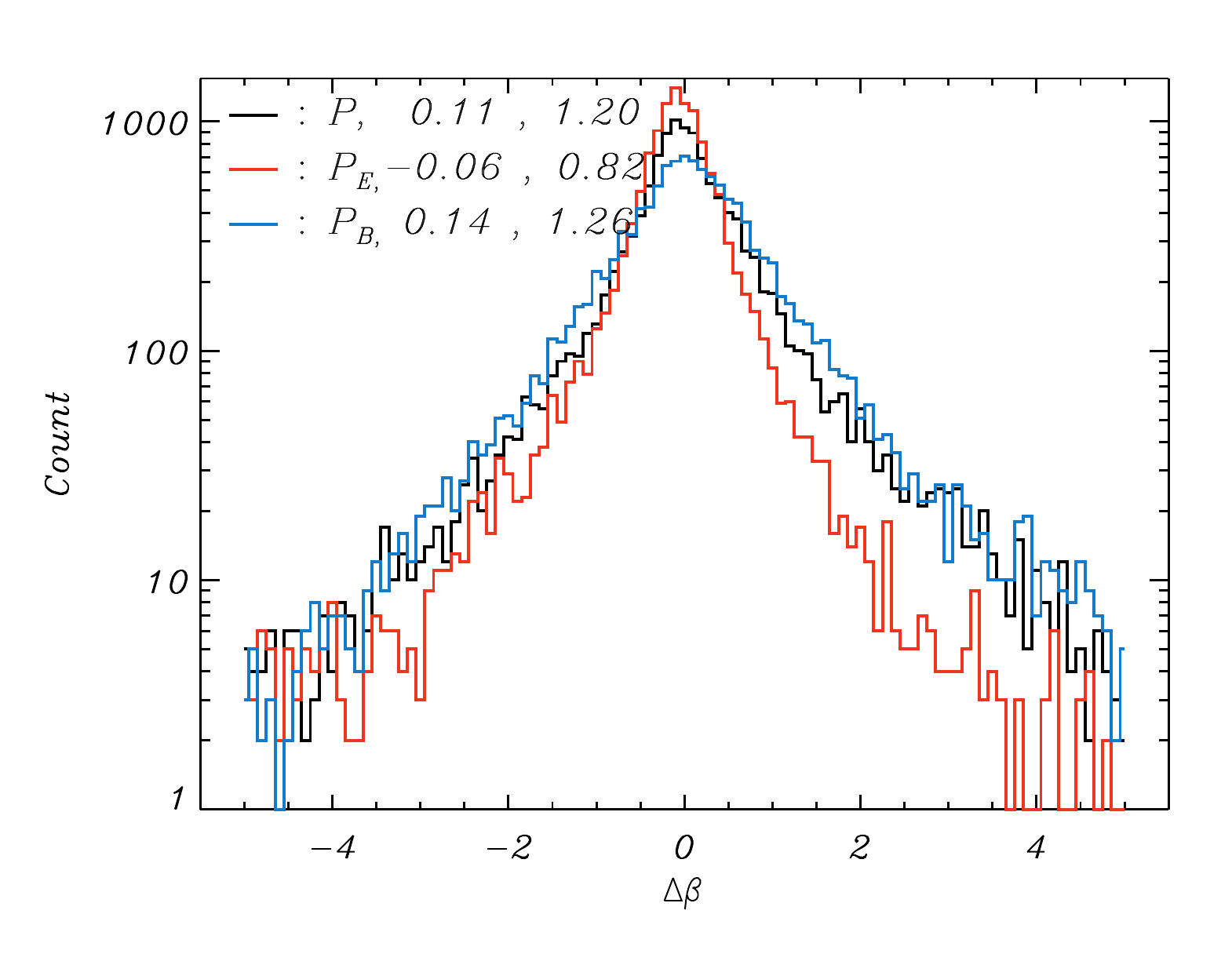}
  \includegraphics[width=0.42\textwidth]{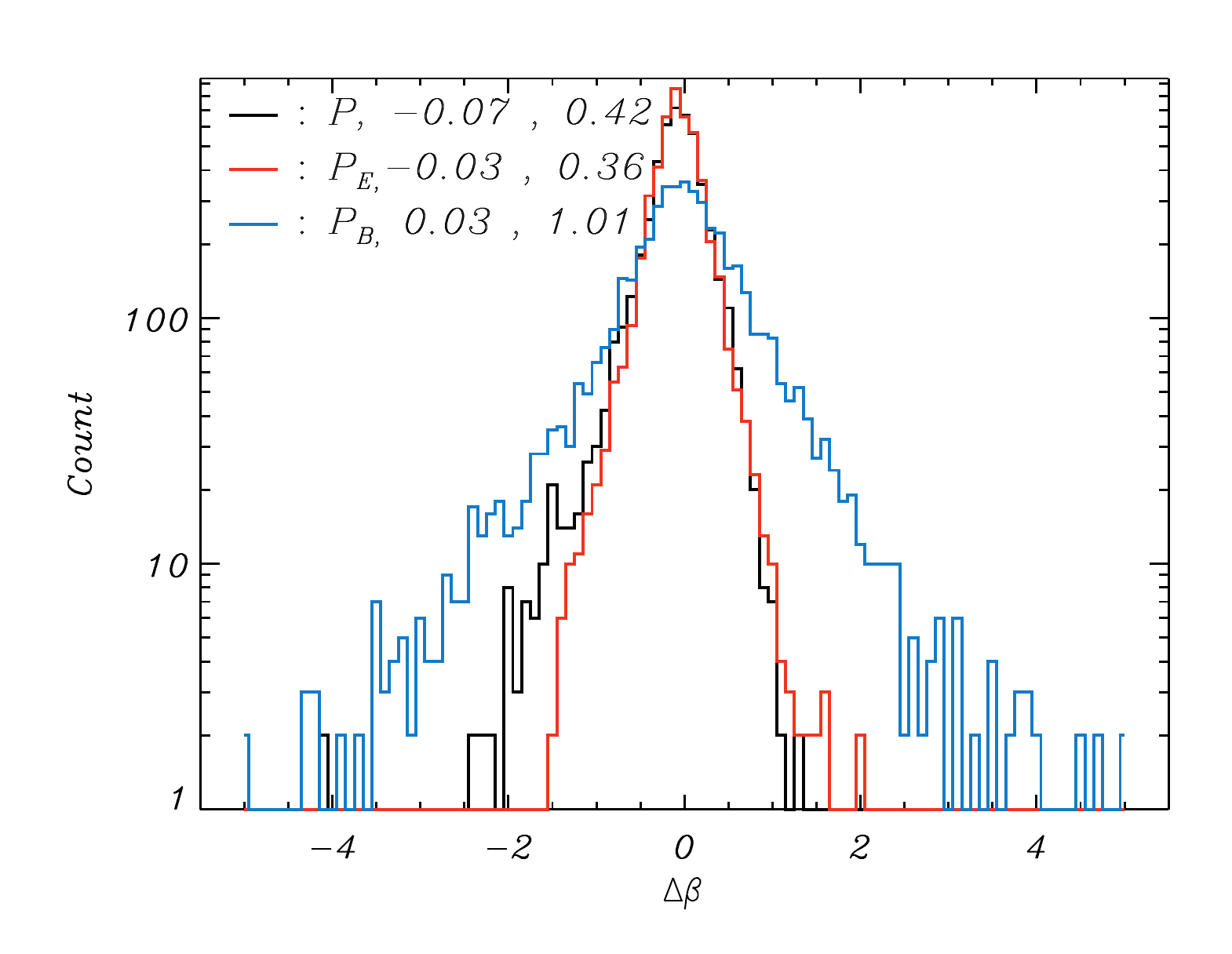}

  \includegraphics[width=0.42\textwidth]{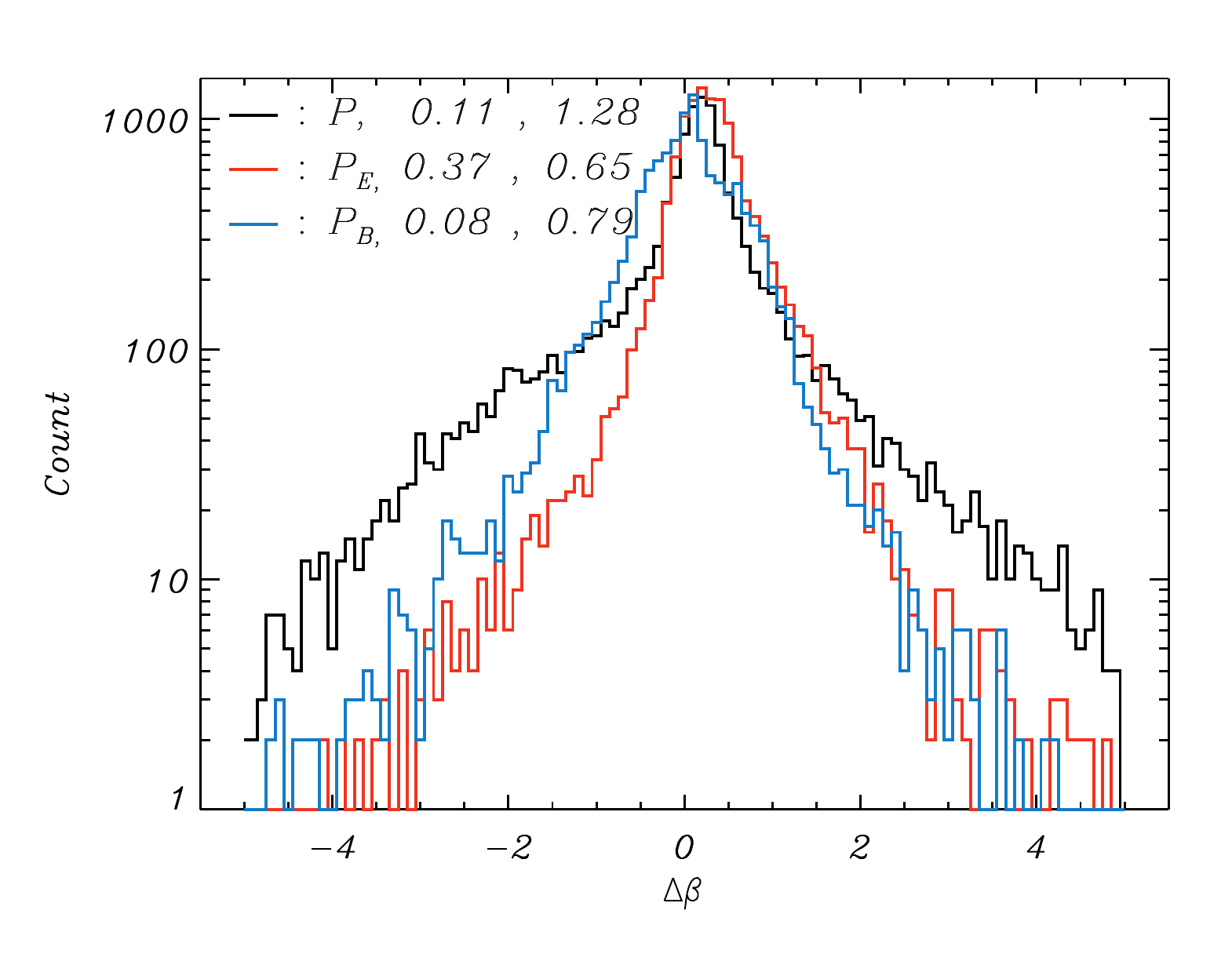}
  \includegraphics[width=0.42\textwidth]{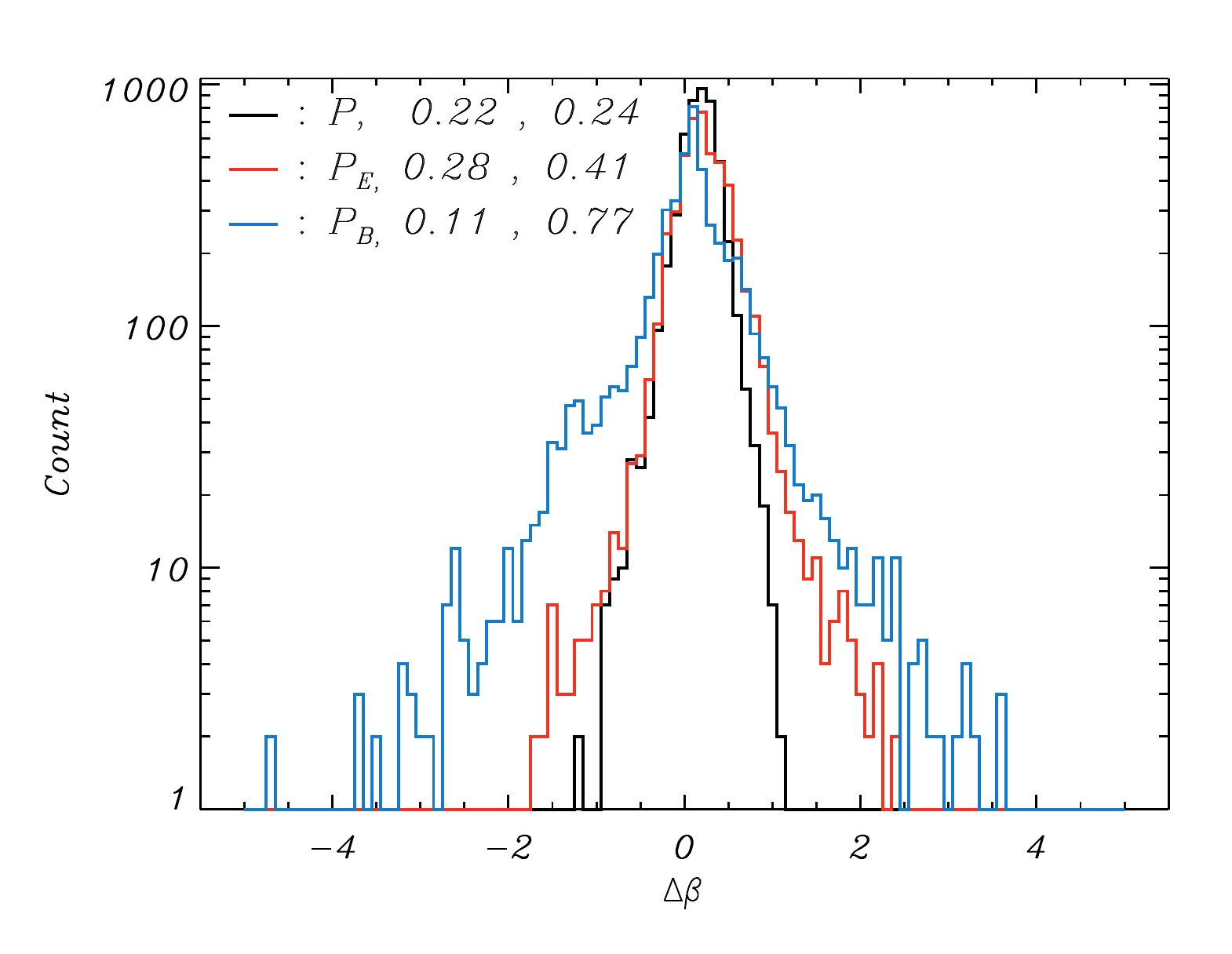}
  \caption{The histograms of the spectrum index maps shown in
  figure~\ref{fig:beta_EB}. \emph{Upper}: K and Ka bands. \emph{Lower}:
  217, 353 GHz. \emph{Left}: full sky. \emph{Right}: only the
  high-intensity zone shown in figure~\ref{fig:beta_EB}. }
  \label{fig:beta_EB_hist}
\end{figure}

We consider the low frequency WMAP K and Ka bands for synchrotron emission,
and the high frequency Planck 217 and 353 GHz bands for TDE. The spectral
indices for these map pairs are presented in figure~\ref{fig:beta_EB}, and the
corresponding histograms $H(\beta)$ are shown in
figure~\ref{fig:beta_EB_hist}. Note that a center value of $\beta=-3$ is
subtracted from the synchrotron band results and $\beta=4$ is subtracted from
the dust band results, and for the dust emission a power law is assumed for
simplification (see Appendix~\ref{app:simp dust beta} for details). From
figure~\ref{fig:beta_EB} we can see that variation of the synchrotron spectral
index is strong in the upper right corner of the polarized intensity map
$P(\textbf n)$. For the E family we can see significant homogenization of the
spectral index $\beta_E$, and great non-uniformity for $\beta_B$. We
illustrate the corresponding distributions $H(\beta)$ in
figure~\ref{fig:beta_EB_hist} for the full sky and for selected high
polarization intensity zones, and in Table~\ref{tab:pe/pb} we show the mean
values and standard deviations for spectral indices in each case.

\begin{table}[!htb]
 \caption{The mean values and standard deviation $\sigma$ for $P$, $P_E$ and
 $P_B$ spectral indexes after monopole subtraction (the first, the second
 and the last values in the boxes) for figure~\ref{fig:beta_EB_hist}. }
 \centering
 \begin{tabular}{|c|c|c|c|c|c|} \hline
 & K/Ka, no mask & K/Ka with mask& 217/353, no mask & 217/353 with mask \\ \hline
 Mean & $0.14$,~$-0.03$,~$0.16$ &$-0.06$,~$-0.02$,~$0.04$& $0.17$,~$0.34$,~$0.13$&$0.22$,~$0.26$,~$0.16$\\ \hline
 $\sigma$ &$1.21$,~ $0.81$,~$1.31$&$0.43$,~ $0.37$,~$0.98$& $1.14$,~$0.59$,~$0.71$&$0.19$,~$0.36$,~$0.70$\\ \hline
 \end{tabular}
 \label{tab:pe/pb}
\end{table}

Apparently, for the sky region with higher polarization intensity, included in
the masks in figure~\ref{fig:beta_EB}, the B mode foreground has much wider
distribution of the spectral index; however, the mean values of the spectral
indices do not deviate significantly from the center value. This rules out
noise as an explanation for the broadening of the distribution, because the
noise will broaden and shift the distributions at roughly same level.

\subsection{Spectral indices for BICEP2 and NPS zones} 

In section \ref{sec:EB families in foreground} we have shown that the E family
dominates in the NPS and BICEP2 zones of the synchrotron K band map and TDE
353 GHz map (see figure~\ref{fig:r K and 353}). The BICEP2 zone has attracted
very serious attention after BICEP2 experiment~\cite{2014PhRvL.112x1101B,
PhysRevLett.107.091101} as well as the corresponding comments
in~\cite{2015PhRvL.114j1301B}. Following the analysis presented above, we
would like to discuss here the properties of the spectral indices of the E and
B families in the NPS and BICEP2 zones.

\begin{figure}[!htb]
 \centering
 \includegraphics[width=0.42\textwidth]{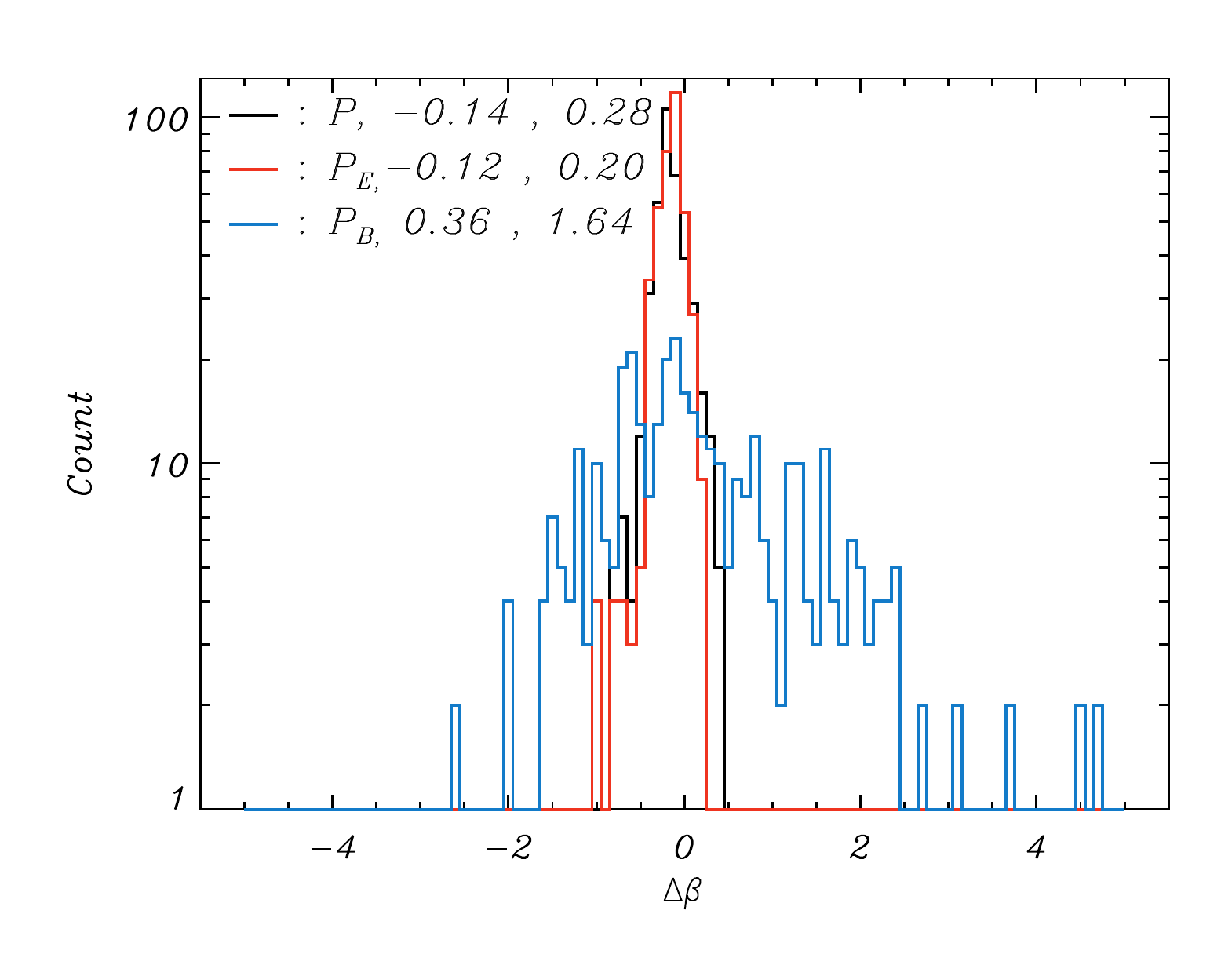}
 \includegraphics[width=0.42\textwidth]{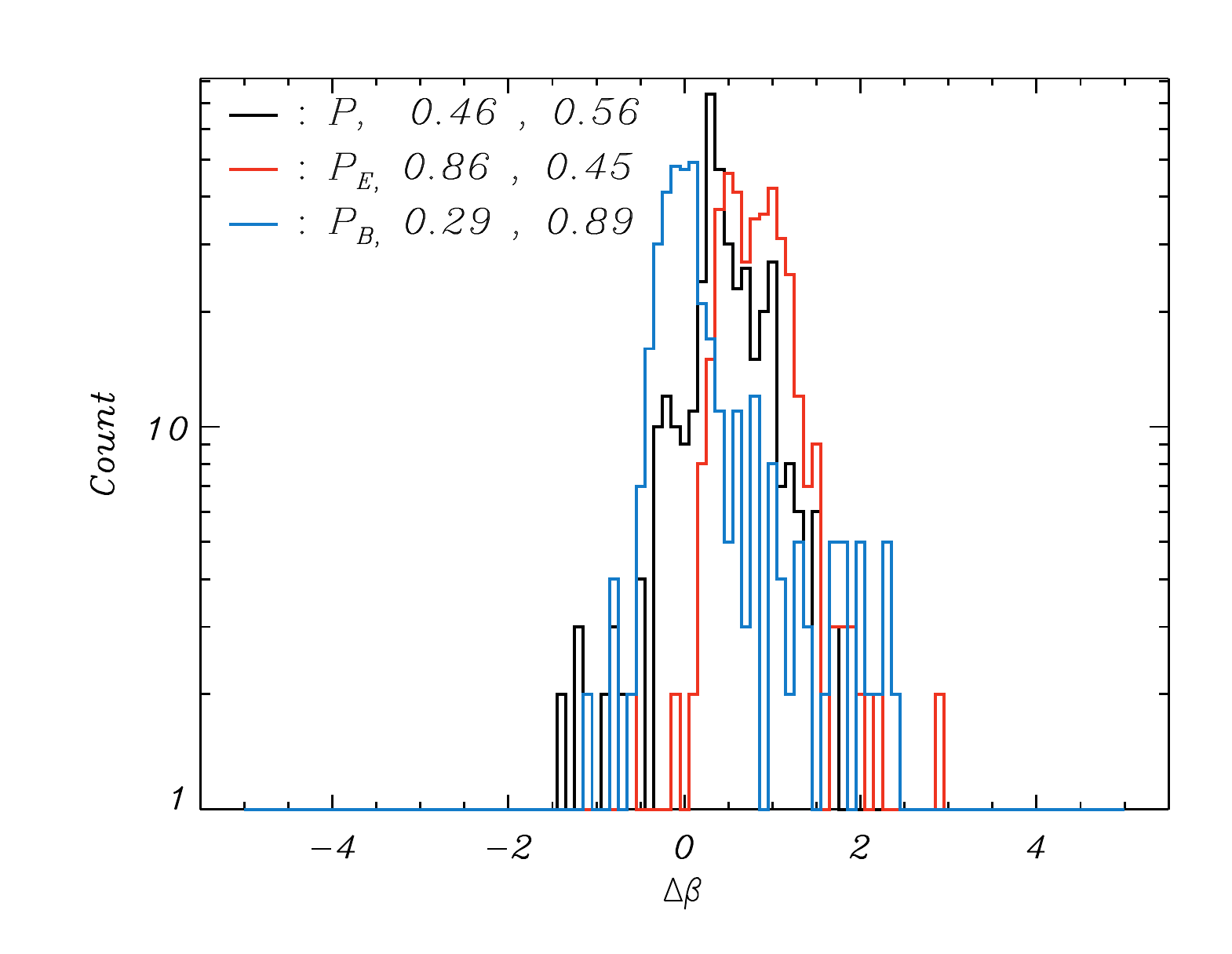}

 \includegraphics[width=0.42\textwidth]{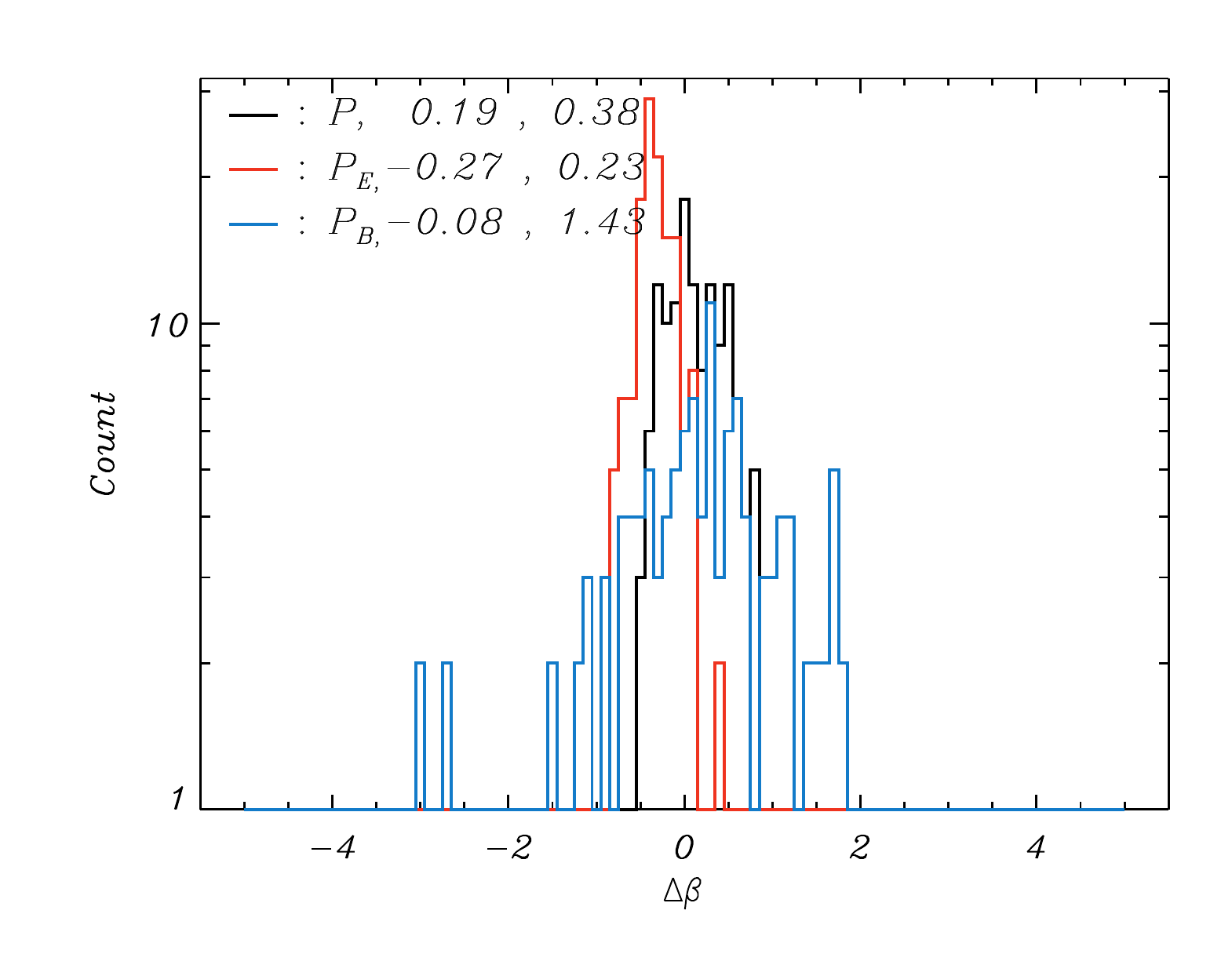}
 \includegraphics[width=0.42\textwidth]{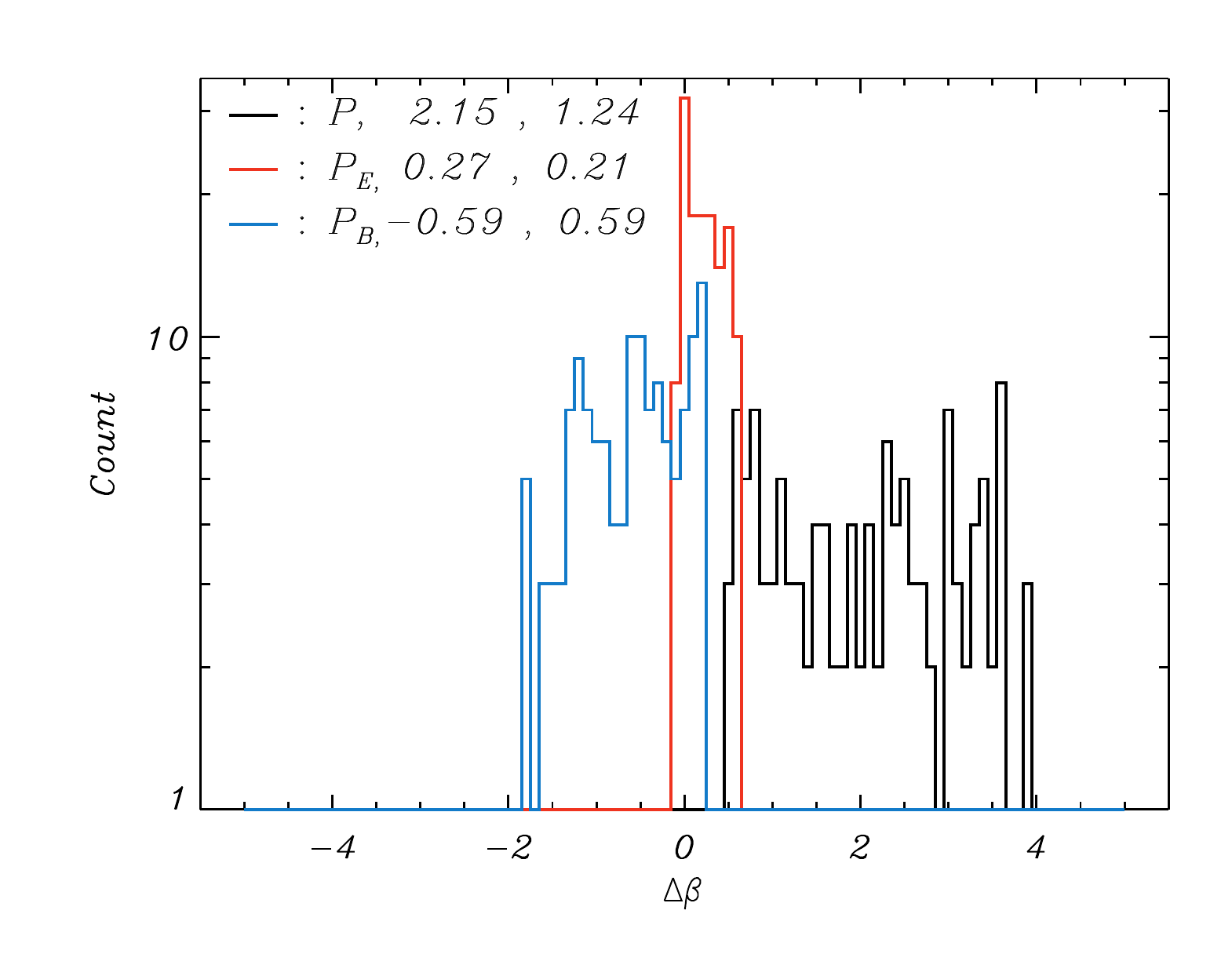}
 \caption{The histograms of the spectrum index variations for the NPS zone
 (upper) and the BICEP2 zone (lower). Left: for K-Ka. Right: for 217-353 GHz.}
 \label{hist_zones}
\end{figure}

In figure~\ref{hist_zones} we show the histograms $H(\beta)$ for the NPS and
BICEP2 zones for $P$, $P_E$ and $P_B$, similar to figure~\ref{fig:beta_EB}.
From this figure we can see that the spectral index for the E family deviates
significantly from that of the B family. In the NPS zone, the B family is
sub-dominant with respect to the E family, and the spectral index for the
total polarization intensity $P$ is given mainly by the E-component. This
result is consistent with figure~\ref{fig:r K and 353}. At the same time, for
the NPS zone we can see significant differences in variation of the B family
synchrotron spectral index with respect to the $\beta$ for the total
polarization intensity $P$, and the E family.

\begin{figure}[!htb]
 \centering
 \includegraphics[width=0.43\textwidth]{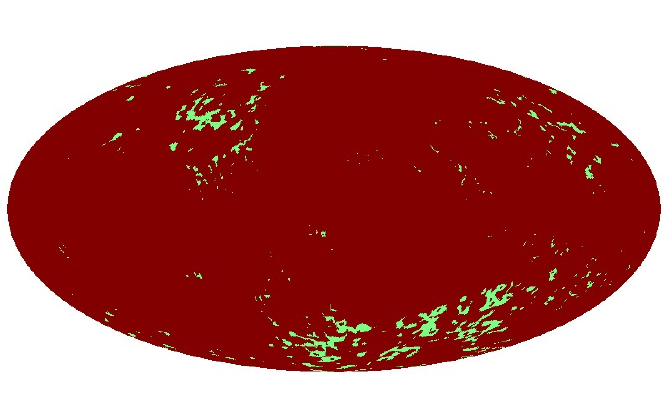}
 \includegraphics[width=0.43\textwidth]{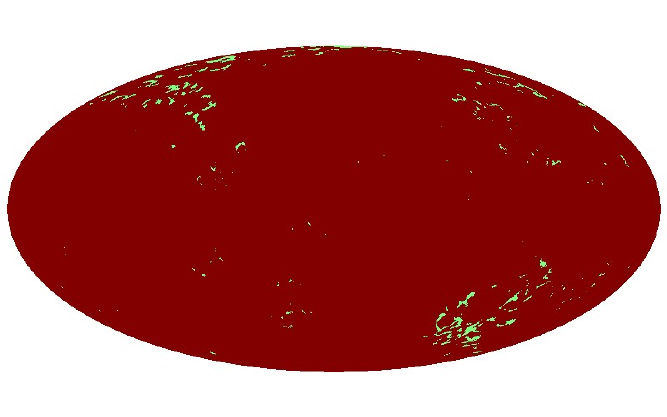}

 \includegraphics[width=0.43\textwidth]{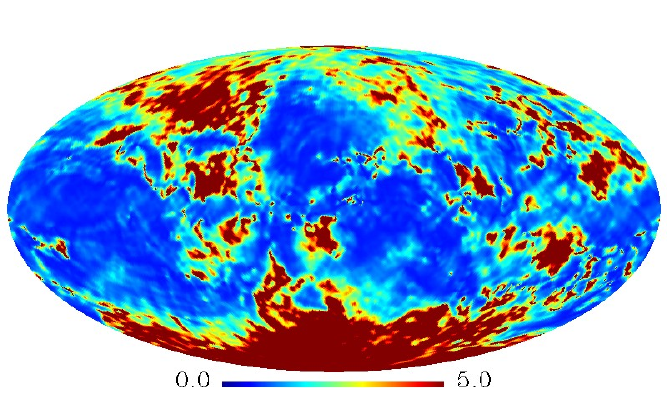}
 \includegraphics[width=0.43\textwidth]{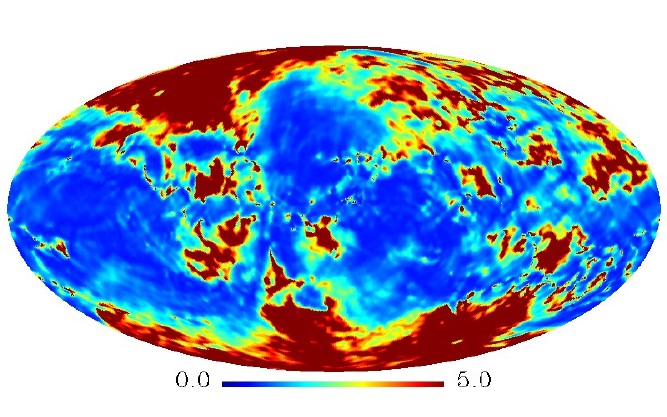}
 \caption{\emph{Upper}: The regions with SNR greater than 1 (in red) for 217
 (left) and 353 GHz (right) without separation. \emph{Lower}: The ratio of
 $\Gamma=\sqrt{P_E^2+P_B^2}/P$ for  217 GHz (left) and 353 GHz (right). }
 \label{zones1}
\end{figure}

We also calculate the noise level for the $Q$ and $U$ maps with
$N_{\mathrm{side}}=128$ and 2$\degree$ smoothing using simulations based on
the released Planck $QQ$ and $UU$ covariances, which gives about
$n_1\approx0.6 \; \mu \mathrm{K}$ for the 217 GHz band and $n_2\approx2.4 \;
\mu\mathrm{K}$ for the 353 GHz band. In figure~\ref{zones1}, we show the
regions with total polarization intensity $P>\sqrt{2}n_{1,2}$ for the two
bands (without separation). We can see that most of the region has a
reasonably high signal-to-noise ratio. We also note that the region with
$P>\sqrt{2}n_{1,2}$ is even larger---covering almost the full sky---for the E
and B families separately than for the total signal, which indicates an E--B
anti-correlation. We have also confirmed that there are no apparent
differences with or without CMB removal.

The properties of the $\Gamma$-parameter, presented in the last row of
figure~\ref{zones1}, reflect the coupling between the total intensity
$P(\textbf n)$, $P_E(\textbf n)$, $P_B(\textbf n)$, and the polarization
angles $\theta_E, \theta_B$ (see eq. (\ref{215}). We have
\begin{equation} \label{zone2}
    \Gamma^2=\frac{P_E^2+P_B^2}{P^2}=\frac{1+\rho^2}{1+2\rho\cos(2\theta_E-2\theta_B)+\rho^2}.
\end{equation}
In all zones in the second row of figure~\ref{zones1} with $\Gamma>1$, we have
$\cos(2\theta_E-2\theta_B)<0$, which indicates E--B anti-correlation.

In the BICEP2 zone the variation of the B family synchrotron spectral index
has significantly different behavior than the NPS zone. Firstly, the mean
value of $\beta$ is $3.27$ with standard deviation $\sigma\simeq \pm 0.38$.
However, for the E family we get $\beta_E\simeq 2.76\pm 0.24$, and
$\beta_B\simeq 2.77\pm 1.1$. The strongest variations of $\beta_B$ occurs in
TDE, where $\beta\simeq 5.48\pm 0.73$, $\beta_E\simeq 4.26\pm 0.18$, but
$\beta_B\simeq 3.56\pm 0.53$. In spite of very narrow distribution of
$\beta_E$, the distributions of $\beta$ and $\beta_B$ are almost opposites of
each other (see figure~\ref{hist_zones}). We summarize these results for
spectral index variations in NPS and BICEP2 zones in Table~\ref{tabbicep1}.
\begin{table}[!htb]
 \caption{The mean values and standard deviations $\sigma$ for the $P$, $P_E$
 and $P_B$ spectral indices (the 1st, 2nd and 3rd values in the boxes), after
 subtraction of the spectrum index offset  for NPS and BICEP2 zones. }
 \centering
 \begin{tabular}{|c|c|c|c|c|c|} \hline
 & NPS sync. & NPS dust& BICEP2 sync.&BICEP2 dust \\ \hline
 Mean & -0.08,~-0.09,~0.39 &0.39,~0.75,~0.32& 0.27,~-0.24,~-0.23&1.48,~0.26,~-0.44\\ \hline
 $\sigma$ &0.28,~ 0.20,~1.64&0.48,~0.38,~0.84&0.38,~ 0.24,~1.10&0.73,~0.18,~ 0.53\\ \hline
 \end{tabular}
 \label{tabbicep1}
\end{table}

An important question related to figure~\ref{hist_zones} is the morphology of
fluctuation causing the corresponding departure of spectral indexes
$\beta,\beta_E$ and $\beta_B$ from their central (mean) values. The most
obvious (but not only) explanation of that effect is the presence of
noise or systematic effects in the corresponding K, Ka, 217 and 353 GHz maps.
If the instrumental noise is a source of spectral index variation, we may
expect small scale fluctuations of the polarization angles, coinciding with
noise patterns at North and South ecliptic poles. For systematic effects
we should see some peculiarities adjusted to the Galactic and ecliptic
planes~\cite{Wehus:2017mwo}.

\section{Conclusion}\label{sec:discussion}

The separation of the Stokes parameters $Q$ and $U$ into E and B families
reveals very important properties of these components, valuable for
determination of peculiarities of the synchrotron and TDE polarization. Since
the E/B family representation is linear, all information stored in $Q$ and $U$
is preserved in $(Q_E,U_E)$ and $(Q_B,U_B)$, but redistributed between them.
This redistribution allows to detect the following features of the foreground
polarization:\\
\begin{enumerate}
\item We have shown that for the synchrotron emission the local features of
   polarization in the loops and arches are mainly associated with the E
   family with much less influence from the B family.
  
\item We have applied the E/B family decomposition for determination of the
   polarization angles for each component: $\theta$, $\theta_E$ and
   $\theta_B$. We have found remarkable stability of $\theta_E$ for
   synchrotron emission and TDE in the whole sky, and in the NPS and BICEP2
   zones as well. We have also discovered well pronounced large sky patterns
   at 217--353 GHz for the B family well above and below the Galactic plane,
   highly contaminated by the residuals of systematic effects. The origin of
   this structure needs further investigation.

\item We have shown that the ratio of the corresponding intensities
   $\rho=P_E/P_B$ is very useful tool for identification of the synchrotron
   and TDE dominated zones, which is important for construction of the
   corresponding masks for future CMB analysis. For the synchrotron emission
   and TDE, the asymptotic of $\rho$ with $\rho>7$ indicate the NPS and BICEP2
   zones with low amplitude of the B family. Both these zones are anomalous in
   respect to other zones and the properties of the derived B mode of
   polarization (not the B family!) needs to be taken with care. As we have
   shown, the $\rho$-test is very sensitive to the residuals of systematics.
   The number of pixels with $\rho>7$ is significantly different for the
   Planck 30 GHz map with and without the bandpass leakage correction,
   relative to the WMAP K and Ka distribution of $\rho$.
 
\item We have investigated the dependency of the spectral indices over
    frequency for synchrotron emission and TDE, using the K and Ka bands for
    synchrotron and 217 and 353 GHz bands for dust emission. We have detected
    different patterns for variation of $\beta$, $\beta_E$ and $\beta_B$ for
    total intensity $P$, $P_E$ and $P_B$. The important feature of the
    $\beta_E$ index is that it is significantly less fluctuated across the
    full sky, and especially in the NPS and BICEP2 zones. For the B family the
    variations of the spectral index are much more pronounced in the same
    zones. Taking into account, that $(Q,U)\equiv(Q_E,U_E)+(Q_B,U_B)$, for
    total intensity $P$ the variation of the TDE spectral index $\beta$ is
    mainly associated with B family. This effect dominates the BICEP2 zone,
    where the $\beta$ spectral index variation is given by variation of
    $\beta_B$. This means the removal of the E family in this domain by
    multi-frequency methods will be extremely effective, while for the B
    family it will be quite problematic. For future ground based and space CMB
    experiments, devoted to determination of the primordial B mode from
    cosmological gravitational waves, these variations of $\beta_B$ will act
    as an additional ``noise''. From the ground the highest frequency bands
    are limited by the opacity of the Earth's atmosphere (200--250 GHz). At
    these frequencies synchrotron emission is already sub-dominant to cosmic
    thermal dust emission. However, since we are interested in very small
    tensor-to-scalar ratio $r\le 0.01-0.0001$, even small synchrotron
    components require more specific attention.
\end{enumerate}

\appendix

\section{Simplification of the dust spectrum index}\label{app:simp dust beta}

Note that in Section~\ref{sub:beta_eb_fullsky}, the definition of $\beta$ for
TDE is different in respect to the gray-body model for non-polarized spectral
energy density $I_d$: In the optically thin limit, the model of spectral
energy distribution $ I_d(\nu)$ (SED) in the direction $\textbf{n}$ and the
frequency $\nu$ for the dust grains is given by~\cite{PlanckXI}:
\begin{eqnarray}
I(\nu)=\tau(\textbf{n})\left(\frac{\nu}{\nu_o}\right)^{\beta_d(\textbf{n})}B_{\nu}\left(T_d(\textbf{n})\right)
\label{eq1}
\end{eqnarray}
where: $\tau_{\nu}=\tau(\textbf{n})\left(\frac{\nu}{\nu_o}\right)^{\beta_d(\textbf{n})}=\sigma_d(\nu)N_H(\textbf{n})$ is the dust optical depth,
$\sigma_d(\nu)$ is the dust opacity, $N_H$ is the gas column density,
\begin{eqnarray}
B_{\nu}\left(T_d(\textbf{n})\right)=\frac{2h\nu^3}{c^2}\left(e^{\frac{h\nu}{kT_d(\textbf{n})}}-1\right)^{-1}
\label{eq2}
\end{eqnarray}
is the Planck spectral function with the temperature $T_d(\textbf{n})$,
$\beta(\textbf{n})$ is the spectral index, and $\nu_o$ is a reference
frequency. For any two frequency bands , $\nu_1$ and $\nu_2$ the ratio
$I(\nu_1)/I(\nu_2)$ is given by
\begin{eqnarray}
\frac{I(\nu_1)}{I(\nu_2)}=\left(\frac{\nu_1}{\nu_2}\right)^
{\beta_d(\textbf n)}\frac{B_{\nu_1}(T_d)}{B_{\nu_2}(T_d)}
\label{eq3}
\end{eqnarray}
Thus, the spectral index $\beta$ related to $\beta_d$ through the following
equation:
\begin{eqnarray}
\beta(\textbf{n})=\beta_d(\textbf{n})+
\ln\left[\frac{(e^{\frac{h\nu_2}{kT_d(\textbf{n})}}-1}{e^{\frac{h\nu_1}{kT_d(\textbf{n})}}-1}\right]/\ln(\frac{\nu_1}{\nu_2}) +3
\label{eq4}
\end{eqnarray}
As one can see from eq. (\ref{eq4}), for $\nu_1,\nu_2\ll kT_d/h$, we have
$\beta\simeq \beta_d+2$. In this case the variation of the spectral index for
$\beta$ and $\beta_d$ is almost identical. If we take $\nu_1 = 353 \,
\mathrm{GHz}$ and $\nu_2 = 217 \, \mathrm{GHz}$, and use $T_d = 20 \,
\mathrm{K}$, then eq.~(\ref{eq4}) simplifies to $\beta =
\beta_d + 1.62$.

\Ack{

This research has made use of data/product from the
WMAP~\citep{WMAPdata:online} and Planck~\citep{Planckdata:online}
collaborations. Some of the results in this paper are derived using the
HEALPix~\citep{2005ApJ...622..759G}) package. This work was partially funded
by the Danish National Research Foundation (DNRF) through establishment of the
Discovery Center and the Villum Fonden through the Deep Space project. Hao Liu
is also supported by the Youth Innovation Promotion Association, CAS.

}


\begin{thebibliography}{10}

\bibitem{2011PhRvD..83h3003B}
E.~F. {Bunn}, \emph{{Efficient decomposition of cosmic microwave background
  polarization maps into pure E, pure B, and ambiguous components}},
  \href{https://doi.org/10.1103/PhysRevD.83.083003}{\emph{\prd} {\bfseries 83}
  (Apr., 2011) 083003}, [\href{https://arxiv.org/abs/1008.0827}{{\ttfamily
  1008.0827}}].

\bibitem{2012SPIE.8442E..19H}
M.~{Hazumi}, J.~{Borrill}, Y.~{Chinone}, M.~A. {Dobbs}, H.~{Fuke}, A.~{Ghribi}
  et~al., \emph{{LiteBIRD: a small satellite for the study of B-mode
  polarization and inflation from cosmic background radiation detection}},  in
  \emph{Space Telescopes and Instrumentation 2012: Optical, Infrared, and
  Millimeter Wave}, vol.~8442 of \emph{\procspie}, p.~844219, Sept., 2012,
  \href{https://doi.org/10.1117/12.926743}{DOI}.

\bibitem{2016arXiv161002743A}
K.~N. {Abazajian}, P.~{Adshead}, Z.~{Ahmed}, S.~W. {Allen}, D.~{Alonso}, K.~S.
  {Arnold} et~al., \emph{{CMB-S4 Science Book, First Edition}}, {\emph{ArXiv
  e-prints} (Oct., 2016) }, [\href{https://arxiv.org/abs/1610.02743}{{\ttfamily
  1610.02743}}].

\bibitem{2011arXiv1110.2101K}
B.~{Keating}, S.~{Moyerman}, D.~{Boettger}, J.~{Edwards}, G.~{Fuller},
  F.~{Matsuda} et~al., \emph{{Ultra High Energy Cosmology with POLARBEAR}},
  {\emph{ArXiv e-prints} (Oct., 2011) },
  [\href{https://arxiv.org/abs/1110.2101}{{\ttfamily 1110.2101}}].

\bibitem{quijote2012}
J.~A. {Rubi\~no-Mart\'{\i}n}, R.~{Rebolo}, M.~{Aguiar}, R.~{G\'enova-Santos},
  F.~{G\'omez-Re\~nasco}, J.~M. {Herreros} et~al., \emph{The quijote-cmb
  experiment: studying the polarisation of the galactic and cosmological
  microwave emissions},
  \href{https://doi.org/10.1117/12.926581}{\emph{Proc.SPIE} {\bfseries 8444}
  (2012) 8444 -- 8444 -- 11}.

\bibitem{2015PhRvL.114j1301B}
{BICEP2/Keck Collaboration}, {Planck Collaboration}, P.~A.~R. {Ade},
  N.~{Aghanim}, Z.~{Ahmed}, R.~W. {Aikin} et~al., \emph{{Joint Analysis of
  BICEP2/Keck Array and Planck Data}},
  \href{https://doi.org/10.1103/PhysRevLett.114.101301}{\emph{Physical Review
  Letters} {\bfseries 114} (Mar., 2015) 101301},
  [\href{https://arxiv.org/abs/1502.00612}{{\ttfamily 1502.00612}}].

\bibitem{WMAPdata:online}
{The WMAP data release}. \url{https://lambda.gsfc.nasa.gov/product/map/dr5/ },
  2013.

\bibitem{Planckdata:online}
{The Planck data release}. \url{https://www.cosmos.esa.int/web/planck/pla },
  2015.

\bibitem{2016A&A...586A.133P}
{Planck Collaboration}, R.~{Adam}, P.~A.~R. {Ade}, N.~{Aghanim}, M.~{Arnaud},
  J.~{Aumont} et~al., \emph{{Planck intermediate results. XXX. The angular
  power spectrum of polarized dust emission at intermediate and high Galactic
  latitudes}}, \href{https://doi.org/10.1051/0004-6361/201425034}{\emph{\aap}
  {\bfseries 586} (Feb., 2016) A133},
  [\href{https://arxiv.org/abs/1409.5738}{{\ttfamily 1409.5738}}].

\bibitem{2017ApJ...839...91C}
R.~R. {Caldwell}, C.~{Hirata} and M.~{Kamionkowski}, \emph{{Dust-polarization
  Maps and Interstellar Turbulence}},
  \href{https://doi.org/10.3847/1538-4357/aa679c}{\emph{\apj} {\bfseries 839}
  (Apr., 2017) 91}, [\href{https://arxiv.org/abs/1608.08138}{{\ttfamily
  1608.08138}}].

\bibitem{2003moco.book.....D}
S.~{Dodelson}, \emph{{Modern cosmology}}.
\newblock 2003.

\bibitem{PhysRevD.55.1830}
M.~Zaldarriaga and U.~c.~v. Seljak, \emph{All-sky analysis of polarization in
  the microwave background},
  \href{https://doi.org/10.1103/PhysRevD.55.1830}{\emph{Phys. Rev. D}
  {\bfseries 55} (Feb, 1997) 1830--1840}.

\bibitem{0004-637X-503-1-1}
M.~Zaldarriaga, \emph{Cosmic microwave background polarization experiments},
  {\emph{The Astrophysical Journal} {\bfseries 503} (1998) 1}.

\bibitem{1997PhRvD..55.7368K}
M.~{Kamionkowski}, A.~{Kosowsky} and A.~{Stebbins}, \emph{{Statistics of cosmic
  microwave background polarization}},
  \href{https://doi.org/10.1103/PhysRevD.55.7368}{\emph{\prd} {\bfseries 55}
  (June, 1997) 7368--7388},
  [\href{https://arxiv.org/abs/astro-ph/9611125}{{\ttfamily
  astro-ph/9611125}}].

\bibitem{1997PhRvL..78.2058K}
M.~{Kamionkowski}, A.~{Kosowsky} and A.~{Stebbins}, \emph{{A Probe of
  Primordial Gravity Waves and Vorticity}},
  \href{https://doi.org/10.1103/PhysRevLett.78.2058}{\emph{Physical Review
  Letters} {\bfseries 78} (Mar., 1997) 2058--2061},
  [\href{https://arxiv.org/abs/astro-ph/9609132}{{\ttfamily
  astro-ph/9609132}}].

\bibitem{2016ARA&A..54..227K}
M.~{Kamionkowski} and E.~D. {Kovetz}, \emph{{The Quest for B Modes from
  Inflationary Gravitational Waves}},
  \href{https://doi.org/10.1146/annurev-astro-081915-023433}{\emph{\araa}
  {\bfseries 54} (Sept., 2016) 227--269},
  [\href{https://arxiv.org/abs/1510.06042}{{\ttfamily 1510.06042}}].

\bibitem{2010A&A...519A.104K}
J.~{Kim} and P.~{Naselsky}, \emph{{E/B decomposition of CMB polarization
  pattern of incomplete sky: a pixel space approach}},
  \href{https://doi.org/10.1051/0004-6361/201014739}{\emph{\aap} {\bfseries
  519} (Sept., 2010) A104}, [\href{https://arxiv.org/abs/1003.2911}{{\ttfamily
  1003.2911}}].

\bibitem{2015MNRAS.452..656V}
M.~{Vidal}, C.~{Dickinson}, R.~D. {Davies} and J.~P. {Leahy}, \emph{{Polarized
  radio filaments outside the Galactic plane}},
  \href{https://doi.org/10.1093/mnras/stv1328}{\emph{\mnras} {\bfseries 452}
  (Sept., 2015) 656--675}, [\href{https://arxiv.org/abs/1410.4438}{{\ttfamily
  1410.4438}}].

\bibitem{2016A&A...594A...2P}
{Planck Collaboration}, P.~A.~R. {Ade}, N.~{Aghanim}, M.~{Ashdown},
  J.~{Aumont}, C.~{Baccigalupi} et~al., \emph{{Planck 2015 results. II. Low
  Frequency Instrument data processings}},
  \href{https://doi.org/10.1051/0004-6361/201525818}{\emph{\aap} {\bfseries
  594} (Sept., 2016) A2}, [\href{https://arxiv.org/abs/1502.01583}{{\ttfamily
  1502.01583}}].

\bibitem{2016A&A...594A...3P}
{Planck Collaboration}, P.~A.~R. {Ade}, J.~{Aumont}, C.~{Baccigalupi}, A.~J.
  {Banday}, R.~B. {Barreiro} et~al., \emph{{Planck 2015 results. III. LFI
  systematic uncertainties}},
  \href{https://doi.org/10.1051/0004-6361/201526998}{\emph{\aap} {\bfseries
  594} (Sept., 2016) A3}, [\href{https://arxiv.org/abs/1507.08853}{{\ttfamily
  1507.08853}}].

\bibitem{2018arXiv180101226W}
J.~L. {Weiland}, K.~{Osumi}, G.~E. {Addison}, C.~L. {Bennett}, D.~J. {Watts},
  M.~{Halpern} et~al., \emph{{Effect of Template Uncertainties on the WMAP and
  Planck Measures of the Optical Depth Due To Reionization}}, {\emph{ArXiv
  e-prints} (Jan., 2018) }, [\href{https://arxiv.org/abs/1801.01226}{{\ttfamily
  1801.01226}}].

\bibitem{Berkhuijsen1971}
E.~M. {Berkhuijsen}, C.~G.~T. {Haslam} and C.~J. {Salter}, \emph{{Are the
  galactic loops supernova remnants?}}, {\emph{\aap} {\bfseries 14} (Sept.,
  1971) 252--262}.

\bibitem{Salter1983}
C.~J. {Salter}, \emph{{Loop-I the North Polar Spur - a Major Feature of the
  Local Interstellar Environment}}, {\emph{Bull. Astron. Soc. India} {\bfseries
  11} (Mar., 1983) 1}.

\bibitem{Wolleben2007}
M.~Wolleben, \emph{{A New Model For The Loop-I (The North Polar Spur) Region}},
  \href{https://doi.org/10.1086/518711}{\emph{Astrophys.J.} {\bfseries 664}
  (2007) 349--356}, [\href{https://arxiv.org/abs/0704.0276}{{\ttfamily
  0704.0276}}].

\bibitem{2016A&A...594A..10P}
{Planck Collaboration}, R.~{Adam}, P.~A.~R. {Ade}, N.~{Aghanim}, M.~I.~R.
  {Alves}, M.~{Arnaud} et~al., \emph{{Planck 2015 results. X. Diffuse component
  separation: Foreground maps}},
  \href{https://doi.org/10.1051/0004-6361/201525967}{\emph{\aap} {\bfseries
  594} (Sept., 2016) A10}, [\href{https://arxiv.org/abs/1502.01588}{{\ttfamily
  1502.01588}}].

\bibitem{2016A&A...594A...1P}
{Planck Collaboration}, R.~{Adam}, P.~A.~R. {Ade}, N.~{Aghanim}, Y.~{Akrami},
  M.~I.~R. {Alves} et~al., \emph{{Planck 2015 results. I. Overview of products
  and scientific results}},
  \href{https://doi.org/10.1051/0004-6361/201527101}{\emph{\aap} {\bfseries
  594} (Sept., 2016) A1}, [\href{https://arxiv.org/abs/1502.01582}{{\ttfamily
  1502.01582}}].

\bibitem{2014PhRvL.112x1101B}
{BICEP2 Collaboration}, P.~A.~R. {Ade}, R.~W. {Aikin}, D.~{Barkats}, S.~J.
  {Benton}, C.~A. {Bischoff} et~al., \emph{{Detection of B-Mode Polarization at
  Degree Angular Scales by BICEP2}},
  \href{https://doi.org/10.1103/PhysRevLett.112.241101}{\emph{Physical Review
  Letters} {\bfseries 112} (June, 2014) 241101},
  [\href{https://arxiv.org/abs/1403.3985}{{\ttfamily 1403.3985}}].

\bibitem{PhysRevLett.107.091101}
P.~Mertsch and S.~Sarkar, \emph{Fermi gamma-ray ``bubbles'' from stochastic
  acceleration of electrons},
  \href{https://doi.org/10.1103/PhysRevLett.107.091101}{\emph{Phys. Rev. Lett.}
  {\bfseries 107} (Aug, 2011) 091101}.

\bibitem{Wehus:2017mwo}
I.~K. Wehus and H.~K. Eriksen, \emph{{Bayesian component separation: The Planck
  experience}},  2017, \href{https://arxiv.org/abs/1712.10223}{{\ttfamily
  1712.10223}},
  \href{http://inspirehep.net/record/1645398/files/arXiv:1712.10223.pdf}{http://inspirehep.net/record/1645398/files/arXiv:1712.10223.pdf}.

\bibitem{PlanckXI}
{\scshape Planck Collaboration} collaboration, A.~Abergel et~al., \emph{{Planck
  2013 results. XI. All-sky model of thermal dust emission}},
  \href{https://arxiv.org/abs/1312.1300}{{\ttfamily 1312.1300}}.

\bibitem{2005ApJ...622..759G}
K.~M. {G{\'o}rski}, E.~{Hivon}, A.~J. {Banday}, B.~D. {Wandelt}, F.~K.
  {Hansen}, M.~{Reinecke} et~al., \emph{{HEALPix: A Framework for
  High-Resolution Discretization and Fast Analysis of Data Distributed on the
  Sphere}}, \href{https://doi.org/10.1086/427976}{\emph{\apj} {\bfseries 622}
  (Apr., 2005) 759--771},
  [\href{https://arxiv.org/abs/astro-ph/0409513}{{\ttfamily
  astro-ph/0409513}}].

\end{thebibliography}

\providecommand{\href}[2]{#2}\begingroup\raggedright\endgroup

\end{document}